\documentclass[11pt,dvips,twoside,letterpaper]{article}
\usepackage{graphicx}
\usepackage{times}
\usepackage{fancyhdr}
\usepackage{geometry}
\usepackage{epsfig}
\usepackage{epsf}

\newcommand{\msol}{M$_\odot$}
\newcommand{\kms}{km~s$^{-1}$}
\newcommand{\lsim}{{\, \lower2truept\hbox{
${< \atop\hbox{\raise4truept\hbox{$\sim$}}}$}\,}}
\newcommand{\gsim}{{\, \lower2truept\hbox{
${> \atop\hbox{\raise4truept\hbox{$\sim$}}}$}\,}}

\newcommand{\feiiq}{\rm Fe{\sc ii }$\lambda$4570\/}
\newcommand{\feiiopt}{{Fe \sc{ii}}$_{\rm opt}$\/}
\newcommand{\feiir}{\rm Fe{\sc ii }$\lambda$5250\/}
\def\cmq{cm$^{-2}$\/}
\def\l{$\lambda$}
\def\lm{$L_{\rm bol}/M_{\rm BH}$}
\def\lbol{$L_{\rm bol}$\/}
\def\lledd{$L_{\rm bol}/L_{\rm Edd}$}
\def\ne{$n_{\rm e}$\/}

\def\rfe{$R_{\rm FeII}$}

\def\msol{M$_\odot$\/}
\def\msoly{M$_\odot~\rm yr^{-1}$}
\def\rg{$R_{\rm g}$\/}
\def\md{$\dot{m}$}
\def\ltsima{$\; \buildrel < \over \sim \;$}
\def\simlt{\lower.5ex\hbox{\ltsima}}            
\def\gtsima{$\; \buildrel > \over \sim \;$}

\def\simgt{\lower.5ex\hbox{\gtsima}}            

\def\ha{{\sc H}$\alpha$}

\def\civ{{\sc{Civ}}$\lambda$1549\/}
\def\civnc{{\sc{Civ}}$\lambda$1549$_{\rm NC}$\/}
\def\civbc{{\sc{Civ}}$\lambda$1549$_{\rm BC}$\/}

\def\cm3{cm$^{-3}$\/}
\def\hb{{\sc{H}}$\beta$\/}
\def\hg{{\sc{H}}$\gamma$\/}
\def\hbbc{{\sc{H}}$\beta_{\rm BC}$\/}
\def\hbvbc{{\sc{H}}$\beta_{\rm VBC}$\/}
\def\habc{{\sc{H}}$\alpha_{\rm BC}$\/}
\def\hbnc{{\sc{H}}$\beta_{\rm NC}$\/}
\def\mgii{{Mg\sc{ii}}$\lambda$2800\/}

\def\ciii{{\sc{Ciii]}}$\lambda$1909\/}
\def\oiiiopt{{\sc{[Oiii]}}$\lambda\lambda$4959,5007\/}
\def\o4363{{\sc{[Oiii]}}$\lambda$4363\/}

\def\siiii{{Si}{\sc iii}]$\lambda$1892\/}

\def\aliii{{Al}{\sc iii}$\lambda$1808\/}
\def\nv{{\sc Nv}$\lambda$1240\/}

\def\feiiuv{{{Fe\sc{ii}}}$_{\rm UV}$\/}
\def\feiiopt{{Fe\sc{ii}}$_{\rm opt}$\/}
\def\feii{{Fe\sc{ii}}\/}

\def\fevii6087{{\sc [Fe vii]}$\lambda$6087\/}
\def\oiii{{\sc [Oiii]}$\lambda$5007}

\def\kms{km~s$^{-1}$}
\def\ergss{ergs s$^{-1}$\/}
\def\mbh{$M_{\rm BH}$\/}
\def\gs{$\Gamma_{\rm soft}$\/}
\def\mb{$ M_{\rm B}$}

\def\hi{H{\sc i}\/}
\def\rk{{$R{\rm _K}$}\/}
\def\heii{{{\sc H}e{\sc ii}}$\lambda$4686\/}
\def\hei{{{\sc H}e{\sc i}}\/}
\def\rb{$r_{\rm BLR}$\/}
\def\rs{$r_{\rm SF}$\/}
\def\ledd{$L_{\rm Edd}$\/}

\begin{document}
\title{
~\\[-1.2in]
{\normalsize\noindent
\begin{picture}(0,0)(208,-3.15)
\begin{tabular}{l p{0.7in} r}
In: New Developments in Black Hole Research  & & ISBN 1-59454-641-X \\
Editor: Paul V. Kreitler, pp. 123-183 & & \copyright 2006 Nova Science Publishers, Inc.\\
\end{tabular}
\end{picture}}
{\begin{flushleft} ~\\[0.63in]{\normalsize\bfseries\textit{Chapter~4}} \end{flushleft} ~\\[0.13in]
\bfseries\scshape Accretion onto Supermassive Black Holes in Quasars:
Learning from Optical/UV Observations}}
\author{
\bfseries\itshape Paola Marziani$^1$, Deborah Dultzin-Hacyan$^2$ and Jack W.
Sulentic$^3$\\
$^1$Istituto Nazionale di Astrofisica, Osservatorio
Astronomico di Padova, \\Vicolo dell'Osservatorio 5, I--35122
Padova, Italia\\
$^2$Instituto de
Astronom\'{\i}a, Universidad Nacional Autonoma de M\'exico,\\
Apartado postal 70-264, 4510 M\'exico, D.F.,  M\'exico,\\
$^3$Department of Physics \& Astronomy,
University of Alabama, \\Tuscaloosa, AL 35487, USA}
\date{}
\maketitle \thispagestyle{empty} \setcounter{page}{123}

\pagestyle{fancy}
\fancyhead{} \fancyhead[EC]{Paola Marziani, Deborah Dultzin-Hacyan
and Jack W. Sulentic} \fancyhead[EL,OR]{\thepage}
\fancyhead[OC]{Accretion onto Supermassive Black Holes in Quasars}
\fancyfoot{}
\renewcommand\headrulewidth{0.5pt}
\addtolength{\headheight}{2pt} 
\headsep=9pt

\begin{abstract}

Accretion processes in quasars and active galactic nuclei are
still poorly understood, especially as far as the connection
between observed spectral properties and physical parameters is
concerned. Quasars show an additional degree of complexity
compared to stars that is related to anisotropic
emission/obscuration influencing the observed properties in most
spectral ranges. This complicating factor has hampered efforts to
define the equivalent of an Hertzsprung-Russel diagram for
quasars. Even if it has recently become possible to estimate black
hole mass and Eddington ratio for sources using  optical and UV
broad emission lines, the results are still plagued by large
uncertainties. Nevertheless, robust trends are emerging from
multivariate analysis of large spectral datasets of quasars. A
firm observational basis is being laid out by accurate
measurements of broad emission line properties especially when the
source rest-frame is known. We consider the most widely discussed
correlations (i.e. the so-called ``eigenvector 1 parameter space"
and the ``Baldwin effect") and analyze how they can be explained
in terms of accretion properties, broad line region structure, and
source evolution. We critically review recent estimates of black
hole mass, accretion rate, spin and possible orientation
indicators, stressing that any improvement in these parameters
will provide a much better understanding of the physics and
dynamics of the region producing the optical and UV broad emission
lines. More accurate measurements of Eddington ratio and black
hole mass may have a significant impact on our ideas about
evolution of quasar properties with redshift and luminosity as
well as on broader cosmological issues.


\end{abstract}

\section{Introduction}
Tremendous data gathering advancements make now possible to see a
distant solution for most of the deepest conundrums concerning
quasar research. There has been an exponential growth in the
number of known quasars lasting since the 1970s. Presently, the
11$^{th}$\ edition of the V\'eron-Cetty \& V\'eron Catalogue of
AGNs and quasars \cite{veroncettyveron03} lists $\approx$50000
quasars.  The {\it Sloan Digital Sky Survey} (SDSS) plans to
compile a sample of $\sim$ 100000 quasars. The SDSS Third Data
Release provides {\it Charge Coupled Device} (CCD) spectra for
51000 quasars found over 4200 deg$^2$\ \cite{sdss}. The
improvement in observational capabilities has been possible by the
introduction and spread of CCDs and other linear, high detective
quantum efficiency devices as detectors for astronomical
observations since the late 1980s. At a second stance comes the
increase in access to large light-gathering power telescopes of
aperture $\simgt$ 4m,  crucial for spectroscopic observations. The
ability to carry out multi-frequency observations with increasing
spectral resolution and sensitivity (most notably provided by {\it
Hubble Space Telescope} (HST) and by the {\it Far Ultraviolet
Spectroscopic Explorer} (FUSE) for the optical/UV) through the
1990s and early 2000s has been a third factor of relevance. The
data gathering improvements -- which are the foundation of every
astronomical advancement -- have led to the discovery of
systematic trends in quasar properties. These suggest that quasars
are {\em not} well described by an average spectrum. We can
amplify this statement to say that the spectra of quasars seen at
a fixed viewing angle  are also not the same -- the basic tenet of
Unification Schemes \cite{antonucci93}. Yet, data have not been
fully digested in physical terms. Observational constraints have
been only very recently organized in meaningful ways. We still
lack the ability to derive important physical information from
observational parameters on an object-by-object basis. The first
aim of this paper is to set the point of our present knowledge on
the diversity of quasar spectral properties, especially in the
optical and UV spectral ranges (\S\ \ref{e1}). We then discuss
much needed improvement to ensure accurate measurements of the
main physical parameters (\S\ \ref{mass}) as well as structural
constraints on the line emitting regions (\S\ \ref{blr}) and the
physical basis of quasar diversity (\S\ \ref{spin}, \S\ \ref{edd},
\S\ \ref{end}).

\subsection{Basic Accretion Parameters}

We regard quasars and active galactic nuclei as systems accreting matter onto a massive ($\simgt 10^5$ \msol) black
hole (see \S \ref{arebhneeded} for possible caveats). The main parameters are, as for any accreting system, black hole
mass (\mbh, \S \ref{mass}), Eddington ratio (\S \ref{edd}), and mass accretion rate $\dot{M}$ in \msoly. The Eddington
ratio \lledd\ is defined as the ratio between the AGN bolometric luminosity \lbol\ and the Eddington luminosity \ledd,
i.e., the limiting luminosity beyond which radiation pressure overcomes gravitational attraction if accretion is
spherical \cite{peterson97}. The Eddington luminosity is directly proportional to  \mbh, and can be written as

$$L_{\rm Edd} \approx 1.3 \times 10^{38} (M_{\rm BH}/M_\odot)~ {\rm ergs~ s}^{-1}.$$

Since the power emitted by an active nucleus through conversion of mass into energy can be expressed as \lbol$ = \eta
\dot{M} c^2$, it is possible to define an Eddington accretion rate $\dot{M}_{\rm Edd} = L_{\rm Edd}/(\eta c^2)$, and
hence a dimensionless accretion rate $\dot{m}={\dot{M}}/\dot{M}_{\rm Edd}$.  We stress that \lledd $\propto$ \lm, and
that \lm\ is a quantity that can be derived from observations. On the other hand, the value of the efficiency $\eta$
($\sim 0.1$) depends on the accretion mode which may not be the same for all sources: if matter is confined in an
accretion disk as it is customarily assumed \cite{shakurasunyaev73}, $\eta$\ depends  on the geometry and radiative
properties of the disk, which in turn may depend on $\dot{M}$). Therefore we will not always confuse the dimensionless
accretion rate $\dot{m}$ and the Eddington ratio as it is frequently done in literature: for a fixed efficiency, there
could be well super-Eddington accretion even if the source is radiating at, or below, the Eddington limit
\cite{collinetal02}. Black hole spin (\S \ref{spin}) and an orientation angle (defined as the angle between the line
of sight and the axis of the accretion disk around the black hole $\theta$, \S \ref{theta})  probably matter in the
context of UV/optical properties although they have turned out to be very elusive to measure.

\subsection{Nomenclature and Samples \label{samples}}

We use the word quasar here in a generic fashion which means all
classes of extragalactic objects that show {\em broad} (full width
at half maximum, FWHM $\simgt$ 1000 \kms) optical and UV emission
lines. This includes the nuclei of Seyfert and broad-line radio
galaxies (BLRG) as well as radio-quiet (RQ) and radio-loud (RL)
optically unresolved sources. They are often referred to under the
umbrella of {\it Active Galactic Nuclei} (AGNs) which reenforces
the paradigm that they are driven by the same physics differing
only in their redshift-implied distances and, hence, luminosity.
At the same time, there is no divide in AGN occurrence at the
canonical boundary (absolute B magnitude \mb\ $\approx$ 22.5 for
$H_0 \approx$ 65 \kms Mpc$^{-1}$) that separates Seyfert nuclei
from quasars \cite{veroncettyveron03}, so that nomenclature may be
kept luminosity-independent. For instance, we keep using the word
Narrow Line Seyfert 1 (\S \ref{e1extremes}) for sources that are
actually luminous quasars.

Most sources considered in this review were ``classically" selected through color criteria. Major color-based surveys
include the {\it Palomar-Green} (PG), the {\it Large Bright Quasar Survey} (LBQS), the {\it ESO-Hamburg} (HE) quasar
survey and the SDSS. The SDSS photometric system was designed to allow quasars at 0 $\simlt z \simlt$ 6 to be
identified with multicolor selection techniques \cite{schneideretal02}. Other high-quality data-sets, even if  more
heterogeneous and less complete, are also considered
\cite{borosongreen92,corbinboroson96,grupeetal99,marzianietal03a,shangetal04,willsetal93}, especially if spectral
resolution $\lambda/\Delta\lambda \sim 1000$ and continuum $S/N \simgt 20$. These samples often include $\sim 100$
objects of ``good" spectra; studies on the LBQS involve $\sim 10^3$\ sources, while the SDSS accounts for another
order of magnitude leap in sample size, with $\sim 10^4$\ sources.

No consideration to obscured AGNs will be given since the very possibility of accretion parameter estimation from the
observed optical/UV quasar spectra turns out to be related to the measurement of quasar broad emission line shifts,
profile widths, and equivalent widths. Quasars showing broad lines are often referred to as ``type-1" AGNs to
distinguish them from Seyfert 2 nuclei and type 2 quasars (see e.g., Refs. \cite{antonucci93,padovanietal04}) which
will not be considered here.

There are more cumbersome, borderline AGNs which could be grouped
into three classes:

\begin{enumerate}
\item sources that show very large broad Balmer intensity
decrement \ha/\hb $\simgt$ 7 \cite{dongetal04}. Objects of this
kind often show dramatic differences between the broad component
of \ha\ and \hb, and are  not always found in color-based surveys.
RL AGNs hosted in early-type galaxies like Pictor A
 are among them \cite{sulenticetal95}.

\item  An analysis of spectra of quasars in the {\it Half-Jansky Parkes Survey} \cite{francisetal01} suggests that the
wide majority ($\approx$ 80\%) are sources whose continuum is dominated by a pure power-law spectrum ($f_\nu \propto
\nu^{0 \div -2}$) ascribed to synchrotron radiation. These sources show broad lines and are bona-fide type-1 sources
although their equivalent width is somewhat lower than that expected for color-selected quasars (see also \S
\ref{edd}). Sources with rather low W(\hbbc) and very broad \heii\ (most notably NGC 1275, indeed a core-dominated
BRLG) may be seen as somewhat peculiar with respect to the other AGNs, especially if an important goal of the spectral
analysis is
 \mbh\ measurements.

\item A third class of sources may show contamination in the
nuclear spectrum of emission components associated to strong star
formation. At the very least, the bolometric luminosity can be
significantly affected as in the case of Mkn 231
\cite{braitoetal04,crescietal04,sulenticetal05}. Circumnuclear
starbursts do occur in Seyfert 1 sources and quasars
\cite{guetal01a,imanishiwada04,daviesetal04}, and are usually {\em
not} resolved. The emission line spectra of luminous Seyfert 1
nuclei may well be affected by the absorption/emission features
observed in strong wind galaxies (e.g. NGC 4691;
\cite{garciabarretoetal95}).  {\it Broad Absorption Lines} (BAL)
observed in quasars could be associated to nova stars
\cite{shields96}.
\end{enumerate}

We mention these three AGN optical/UV spectral typologies since they may be extremes of AGNs accretion parameters,
especially of \lledd\ (\S \ref{mass}, \S \ref{edd}): sources in the third class are likely to be ``young" quasars
radiating at high \lledd; the first class may include ``dying"  quasars radiating at very low \lledd\   with little
reservoir of matter (\S \ref{evolution}).

\subsection{Rationale for a Black Hole in Active
Galactic Nuclei \label{arebhneeded}}

Do we have a definitive proof that the central massive object at the center of quasar is a black hole?  From the
definition of event horizon (\rg $ = 2 G M /c^2$\ for a non-rotating black hole, where $G$\ is the gravitational
constant), the mass of the black hole must satisfy the condition \mbh/\rg $\simgt 6.7 \cdot 10^{27}$ g cm$^{-1}$.
Although this criterion has not been satisfied yet by observations of extragalactic sources, circumstantial evidence
in favor of a black hole is now considered overwhelming \cite{blandford90}. From the time delay in the response of
emission lines to continuum variations (``reverberation mapping") it is possible to estimate the central mass assuming
that the line broadening is due to Doppler effect, and to test whether the emitting gas motion is virial. In luminous
Seyfert 1 galaxies, the central mass can be $\sim 10^8$ \msol\ within the emitting region of the broad lines (the
``Broad Line Region", BLR) size, of the order of 1 light month. The resulting \mbh/\rb $\sim$ 10$^{24}$ g cm$^{-1}$\
is far less than the \mbh/$r$\ requirement for a black hole. However, time responses for lines of different width are
roughly consistent with a Keplerian trend i.e., with a ``point-like" mass concentration with respect to the BLR size
\cite{krolik01}. It is therefore legitimate to talk at the very least about ``compact massive objects" at the center
of AGNs. This would be a proper, observation-bounded definition.   We will use the term black hole with the
reassurance of a large body of cirumstantial evidence as well as with the perspective of decisive tests. In addition,
most of the reasoning presented in this paper is largely independent on the exact nature of the ``compact massive
object," provided that it is really compact and massive.

\section{Eigenvector 1 \label{e1}}

\subsection{Quasar Surveys and Quasar Spectral Diversity}

\begin{figure}
\hspace{-0.5cm} \epsfxsize=7cm \epsfbox{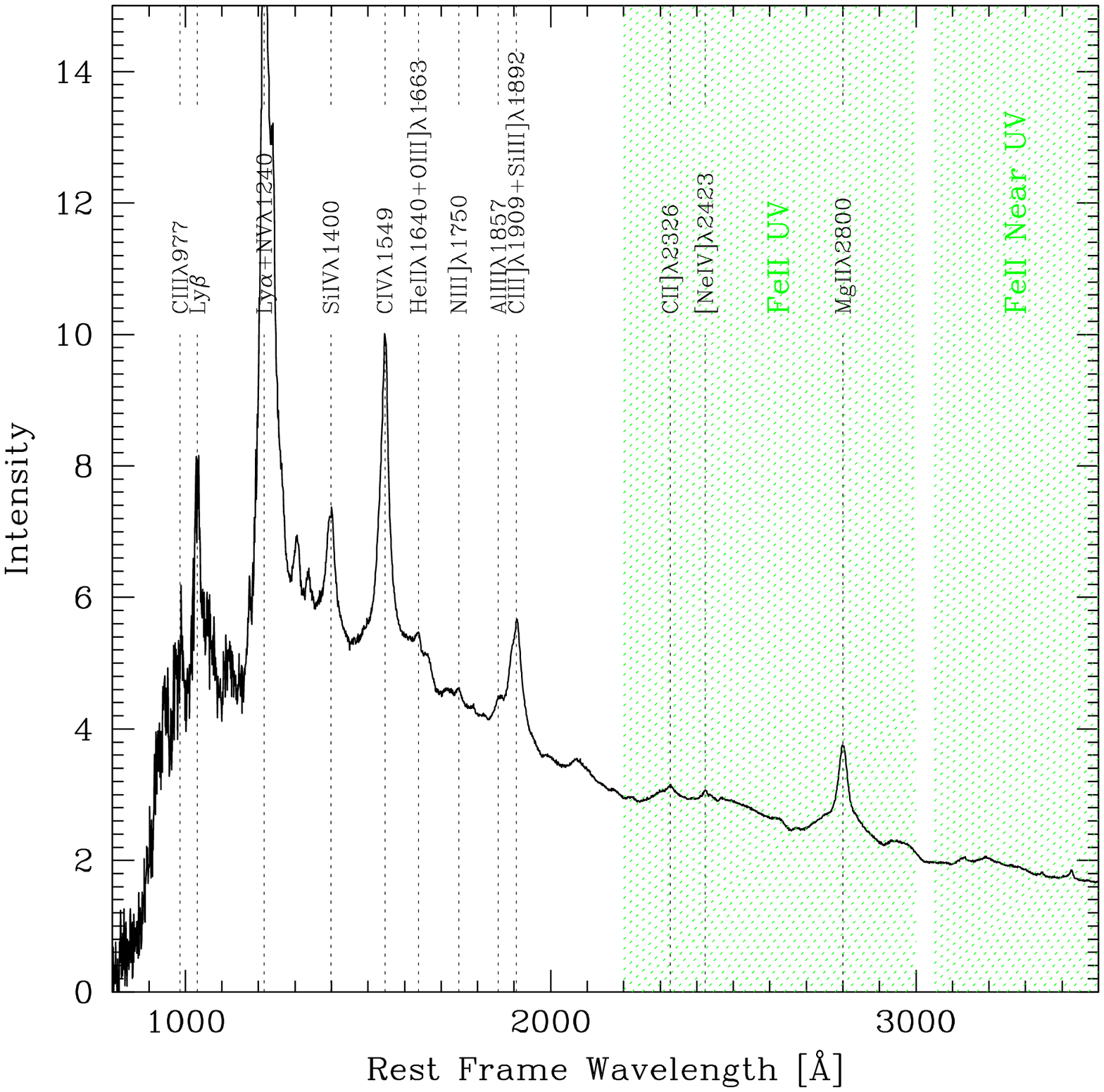} \epsfxsize=7cm \epsfbox{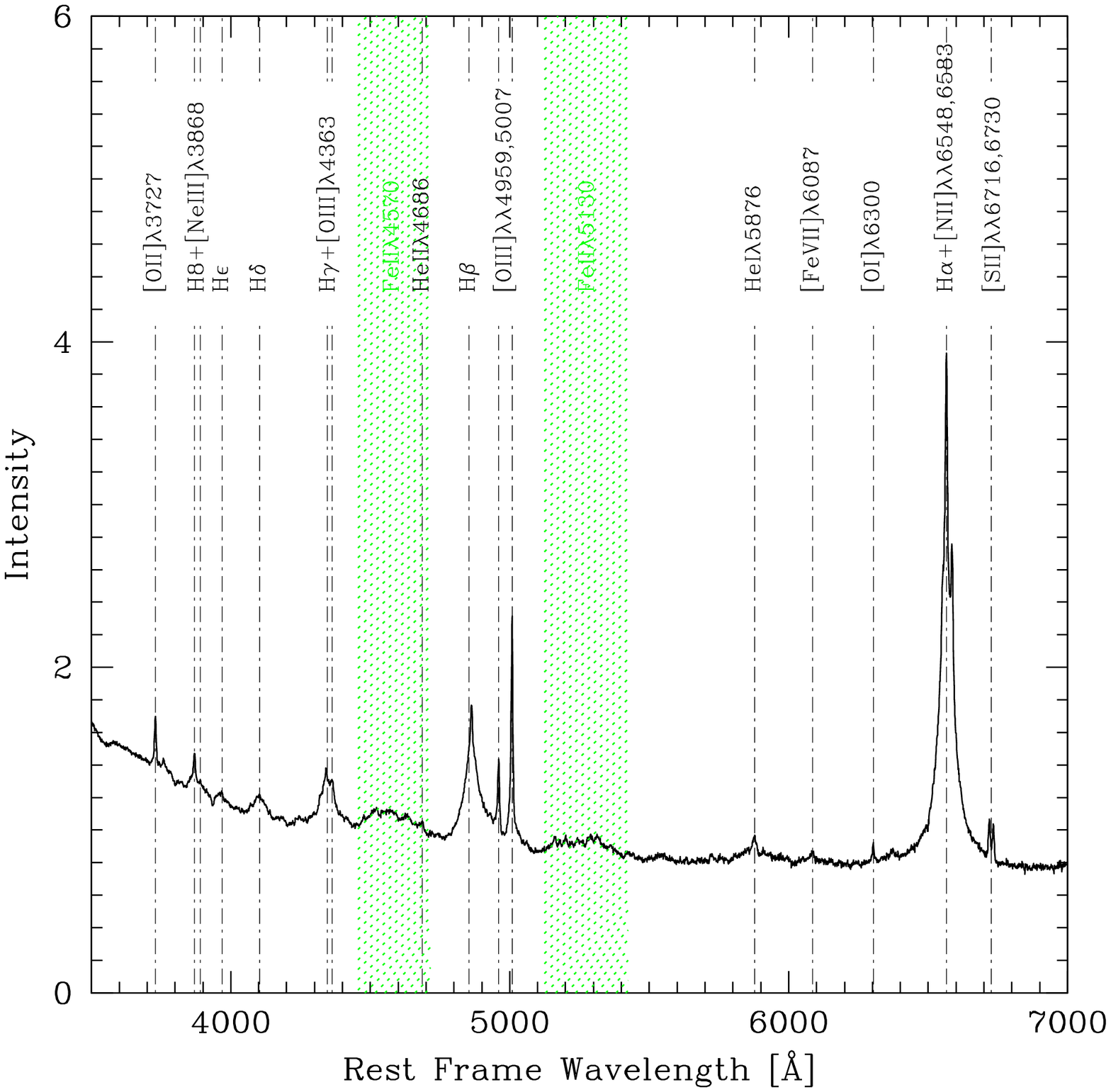} \caption{SDSS median composite
spectrum, in the UV (left panel) and in the optical spectral range (right panel). Major features are identified;
regions with strong emission blends of \feii\ are shaded. For a more exhaustive line list see Ref.
\cite{vandenberketal01}. \label{fig:composite}}
\end{figure}

Fig. \ref{fig:composite} shows a  median composite quasar spectrum, between 800 \AA\ and 7000 \AA, with sources
spanning a redshift range 0.044 $< z <$4.789. The  spectrum  was obtained using a homogeneous data set of over 2200
spectra from the SDSS, at a resolution $\lambda/\Delta\lambda \approx$ 1800. It reaches a peak signal-to-noise ratio
($S/N$) of over 300 per 1 \AA\ resolution element in the rest frame.   See Ref. \cite{vandenberketal01}  for a
comprehensive line list of 80 emission features. Similar composite spectra have been recently constructed from  HST
{\it Faint Object Spectrograph} and FUSE observations \cite{scottetal04,zhengetal97}. The main advantage of composite
spectra is that $S/N$\ is so high that  identification of many faint features that remain invisible in individual
spectra becomes possible \cite{francisetal91}. Most prominent features and overall continuum shape are easily
appreciable; from Fig. \ref{fig:composite} one can realize the extent of \feii\ emission (shaded areas), the
prominence of hydrogen Balmer (especially broad \ha, \hb, \hg), of Lyman lines, and of the \civ\ line, which makes
them, apart from their physical importance, the best studied optical/UV lines of AGNs. Beyond line identification and
some basic information a composite spectrum can yield deceptive results. In principle it is legitimate to median
together all spectra of a given survey if spectral properties scatter randomly with reasonable dispersion around the
median. This is not the case for color selected samples of quasars. It is important to remark that attempts to model a
composite spectrum like the ones from the SDSS and LBQS \cite{francisetal91,vandenberketal01} may be doomed to
failure. Ionization conditions and gas kinematics are not the same for all type-1 AGNs \cite{marzianietal01}. The use
of average line ratios from composite survey spectra has led to an impasse in the attempt at reproducing the
observations through simple photoionization models (see e.g., Ref. \cite{netzer90}, and references therein), an
impasse that still hampers present-day efforts. Recent high quality data \cite{shangetal04} with nearly simultaneous
FUSE, HST and optical observations emphasize the quasar diversity in terms of emission line and continuum
spectrophotometric properties. Optical data for $\approx$ 200 AGNs at $z \simlt$ 0.8 obtained with resolution
$\lambda/\Delta\lambda \sim 1000$\ show the impressive diversity in terms of emission line profiles
\cite{marzianietal03a}.

\subsection{Basis of  Eigenvector Analysis}

Eigenvector techniques are applied whenever many variables appear to be more or less loosely correlated without an
intuitive indication of a dominant correlation or variable. A set of $n$ objects may have $m$\ measured parameters
like flux, FWHM, and equivalent widths of optical and UV emission lines, continuum shape, etc.  We can think that each
set of measurement is a vector in an $m$-dimensional space described by orthogonal vectors $\overrightarrow{v}$, and
define a matrix $M$\ of $n$\ vectors with $m$\ measurements. The {\it Principal Component Analysis} (PCA)  searches
for the best-fitting set of orthogonal axes to replace the original $m$ axes in the space of measured parameters. The
new axis set is sought by maximizing the sum of the squared projections onto each axis i.e., ($M
\overrightarrow{v}$)$^\mathrm{T}(M \overrightarrow{v}$), where $M^{\rm T}$\ denotes the transposed matrix. If the set
of measurements has been previously  centered subtracting the variable average, then $M^\mathrm{T}M$\ can be thought
as a variance/covariance matrix. In spectral PCA  no measurements are performed: the whole spectrum of $n$\ objects is
divided into $m$\ small wavelength bins, and each input variable is the flux in the wavelength bin
\cite{shangetal03,yipetal04}. We seek the maximum of $ \overrightarrow{v} M^\mathrm{T}M \overrightarrow{v}$ imposing
the condition that the norm of $\overrightarrow{v}$ is unity through a Lagrange multiplier $\lambda$,
 and  setting the first derivative to 0 \cite{murtaghheck87}. With our formalism we have $\overrightarrow{v}^{\mathrm
T} M^\mathrm{T}M\overrightarrow{v} - \lambda (\overrightarrow{v}^{\mathrm T}\overrightarrow{v} -1)$, hence
2$M^\mathrm{T}M \overrightarrow{v} - 2 \lambda \overrightarrow{v} = 0$, and then $ M^\mathrm{T}M \overrightarrow{v} =
\lambda \overrightarrow{v}$. This is an eigenvalue equation, which can be solved numerically. More eigenvectors are
sought through a similar procedure, with the additional constraints that the eigenvectors must be orthogonal to each
other. Since the eigenvalues are a measurement of the sum of the squared projections of the original data vectors on
the new normalized vectors, and since we assume to have set as a covariance matrix $M^\mathrm{T}M$, the eigenvalues
are a measurements of the amount of variance carried in each new direction. For example, in a plane we can imagine a
set of almost aligned points; their projections in the original axes may be nearly equal if they are not aligned
preferentially with anyone of the two axes of the two original frame. In a physical context we may think of two
variables that are highly correlated. We can maximize the projections along one axis by simply operating a rotation of
the reference frame. A linear combination of the two original variables (or vectors) will now constitute the one
vector that is needed to account for the variance of the data.  A key aspect of the power of the PCA emerges  from
this simple example: a problem originally treated in two dimensions was inherently one-dimensional i.e., a PCA can
restore a problem with a very large set of variables to its intrinsic dimensionality.

\subsection{The Original Eigenvectors by Boroson \& Green \label{original}}

The first important application of PCA to the interpretation of quasar  spectra was done in the early 1990s  on a
sample of 87 objects of the PG survey at $z < 0.5$\ \cite{borosongreen92}.  T. Boroson \& R. Green measured the most
prominent emission features in the \hb\ spectral region, and found that most of the variance was related to two sets
of correlations, the first being an anti-correlation between the prominence of \oiiiopt\ and optical \feii\ emission
(\feiiopt). The ``eigenvector 1" has been later found in a number of much larger samples
\cite{boroson02,grupe04,kuraszkiewiczetal02,sulenticetal00a,sulenticetal02,yipetal04}, and with spectral parameters
describing phenomenologies as apparently distant as the X-ray continuum continuum shape and the radial velocity shift
of high ionization emission lines. Even if many pieces of the eigenvector analysis had been hinted at by previous
workers \cite{bolleretal96,marzianietal96,wangetal96}, the ``eigenvector 1" provided, for the first time, a powerful
systematic description of quasar spectral diversity. The second eigenvector involved the optical luminosity and the
strength of the high ionization line (HIL) \heii. This second eigenvector had been also found by previous workers, and
is basically the ``Baldwin effect" (\S \ref{be}).

\subsection{The Optical Eigenvector 1 Plane \label{opticale1}}


\begin{figure}[htbp]
\centering
\includegraphics[width=0.45\textwidth]{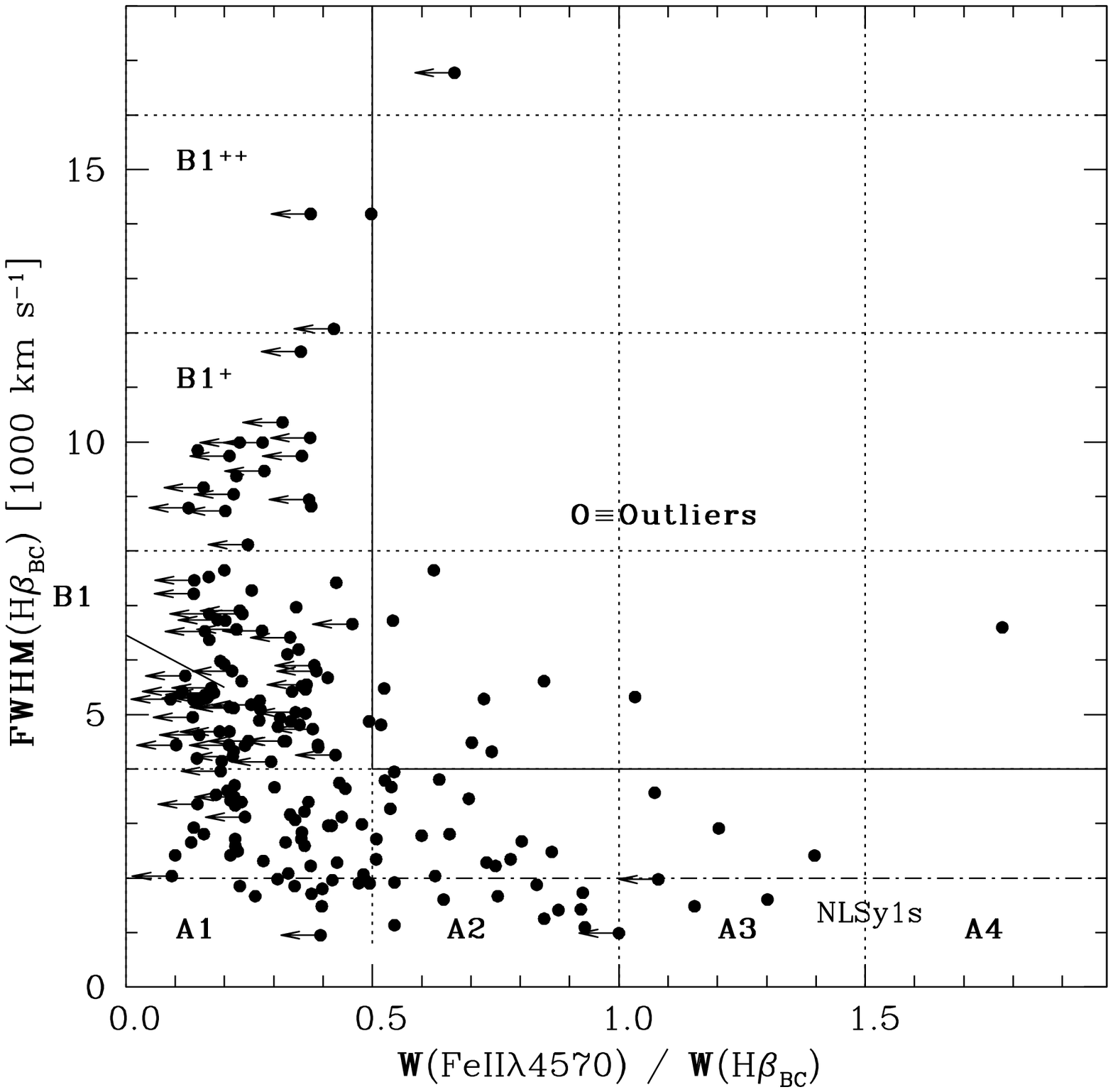}
\caption{The Optical Plane of the
eigenvector 1 of type-1 AGNs; abscissa is the equivalent width
ratio between \feiiq\ and \hbbc, ordinate is FWHM of \hbbc\ in
\kms. The plane has been subdivided in spectral type according to
Ref. \cite{sulenticetal02}. The dot-dashed line identifies Narrow
Line Seyfert 1s, defined by the condition FWHM(\hbbc)$<$ 2000
\kms. In the outlier regions, very few genuine sources (i.e.,
excluding those that are borderline because of errors) are found.
Data are from Marziani et al.
\cite{marzianietal03a}}
\label{fig:e1}
\end{figure}

\begin{center}
\begin{figure}
\hspace{-0.cm}\epsfxsize=7cm\epsfbox{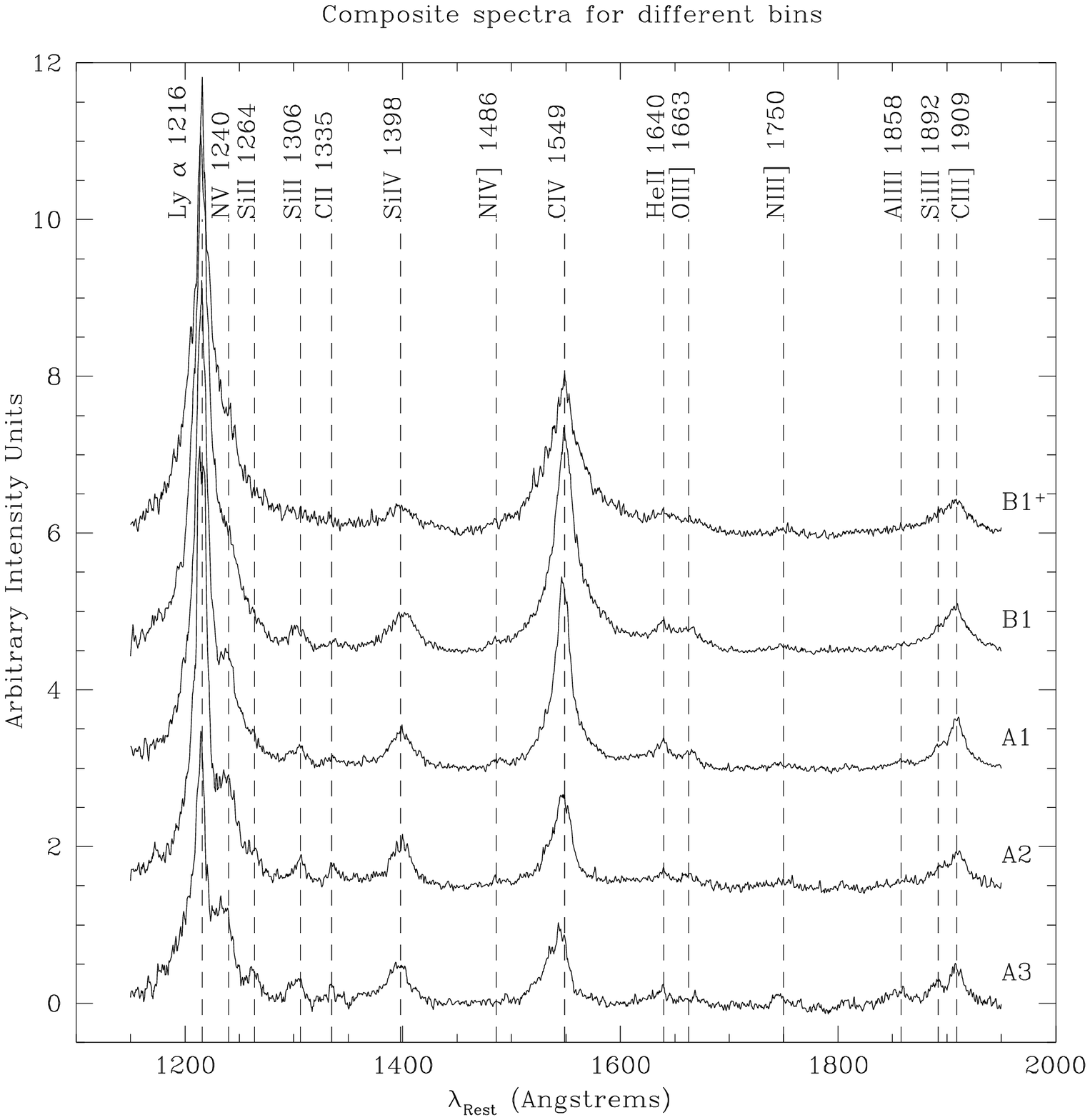}\epsfxsize=7cm\epsfbox{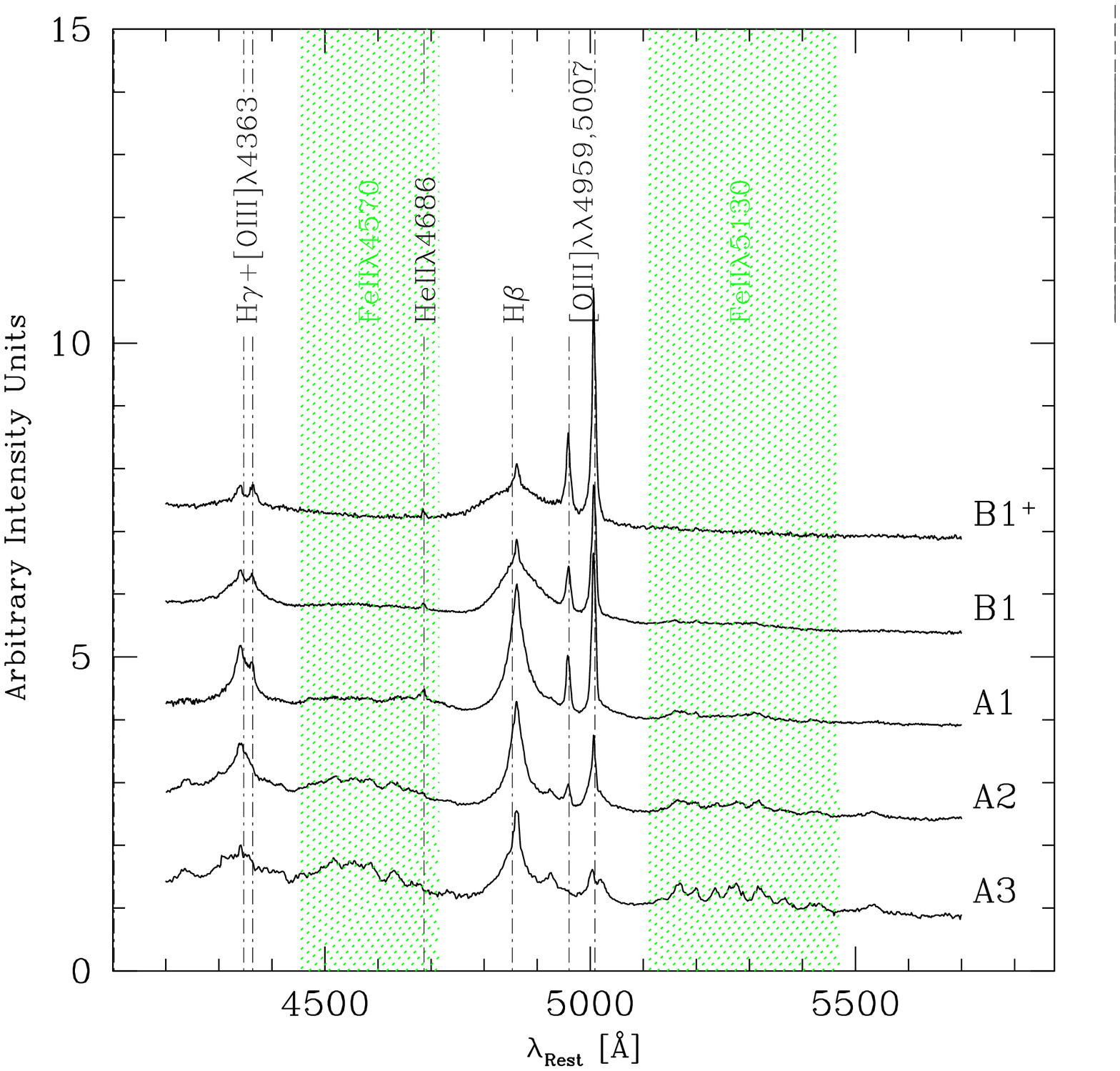}
\caption{Spectral diversity organized along the E1 sequence for UV
(left panel) and optical (right panel) spectra, for the spectral
types defined according to Sulentic et al. \cite{sulenticetal02}.
Prominent blends of \feii\ emission are shaded.
\label{fig:e1spectra}}
\end{figure}
\end{center}

It is remarkable that the spectral diversity of quasars can be reduced (if we exclude any luminosity dependence) to
just two variables. The diversity  and correlation of broad line AGNs  are clearly shown by their distribution in the
so-called ``Eigenvector 1 (E1) optical plane," defined by the FWHM of the H$\beta$\ broad component, FWHM(\hbbc), and
by the equivalent width ratio between the Fe {\sc ii} blended emission at $\lambda$4570 and the H$\beta$\ broad
component, \rfe\ = W(\feiiq)/W(\hbbc) \cite{boroson02,borosongreen92,sulenticetal00a}.  The anti-correlation between
FWHM(\hbbc) and \rfe\ provides perhaps a superior description of E1 than the original one, since it involves
parameters related  to the broad lines only (\oiiiopt\ are narrow lines affected by radio loudness;
\cite{sulenticetal02}). The modest spread of data points drawn from a sample of more than 200 type-1 AGNs  allow the
definition of typical spectral types covering a narrow range in FWHM(\hbbc) and \rfe\ \cite{sulenticetal02}. We set
$\Delta R_{\rm Fe} =0.5$, and, for FWHM(\hbbc)$\simlt$ 4000 \kms, we define spectral type A1, A2, A3 in order of
increasing \rfe\ (see Fig. \ref{fig:e1}; \feii\ emitters are considered strong if \rfe$\simgt 1$; extreme if
\rfe$\simgt$1.5). Sources with FWHM(\hbbc)$\simgt$ 4000 km s$^{-1}$\ very rarely show $R_{\rm Fe} \simgt 0.5$. If
$R_{\rm Fe} \simlt 0.5$, we define a second sequence of increasing FWHM(H$\beta$) (A1,B1,B1$^+ \ldots$ ) with
$\Delta$FWHM(H$\beta$)$\approx$ 4000 \kms. The spectral types  are easy to identify observationally, even from a
visual inspection of the optical spectrum (Fig. \ref{fig:e1spectra}) and most of them may have a direct physical
meaning (\S\ \ref{e1corr}, \ref{e1extremes}, \S \ref{edd}). FWHM(\hbbc) $\approx$ 4000 \kms\ represents a  notable
limit. It separates two groups of AGNs, Population A and B \cite{sulenticetal00a} whose different properties are most
likely related to structural differences in the Broad Line Region (BLR; see \S \ref{blr}).

\subsection{Eigenvector 1 Measurements \label{e1measures}}

\begin{figure}
\hspace{-0.cm}\epsfxsize=6.8cm\epsfbox{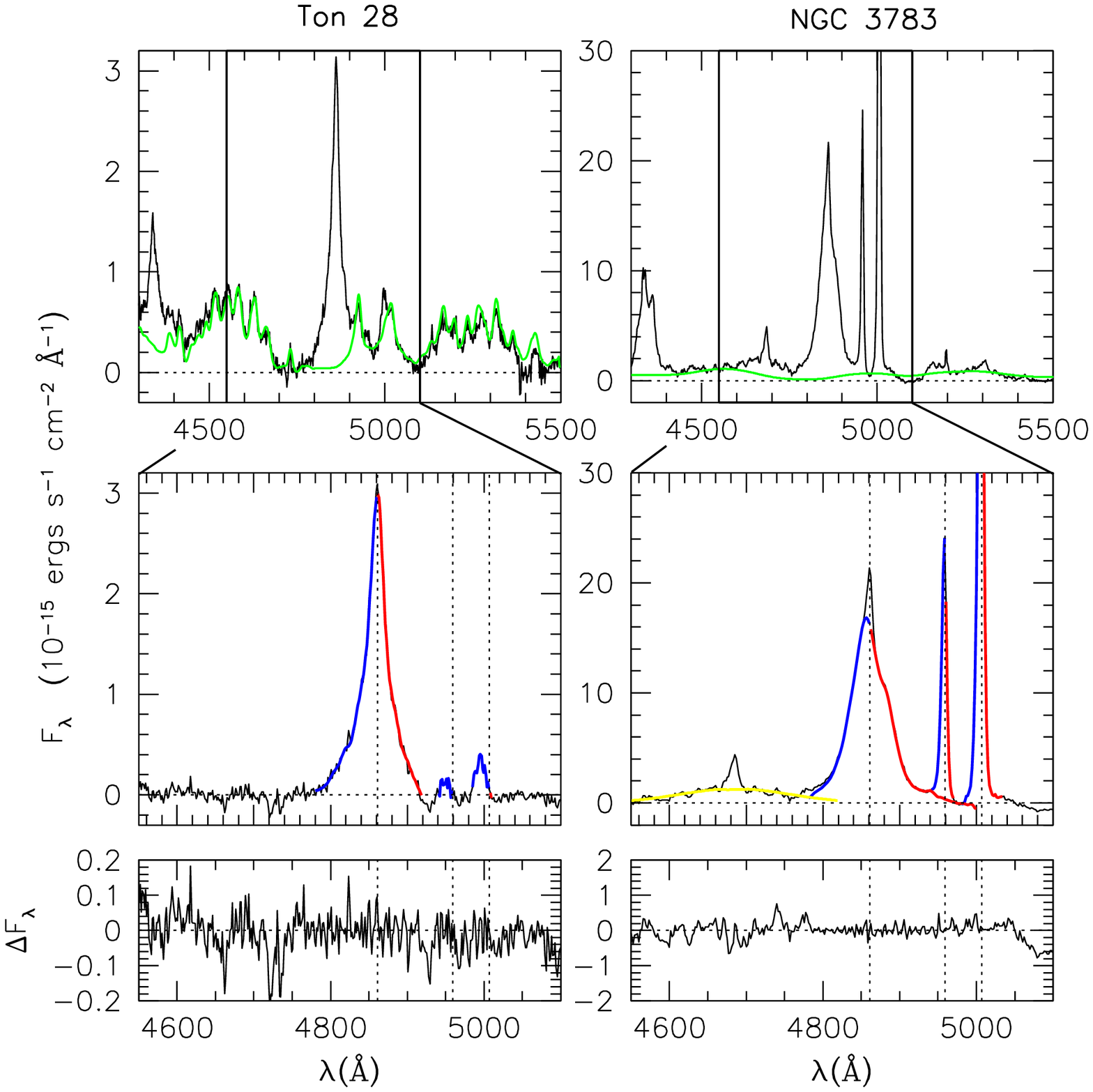} \epsfxsize=6.8cm\epsfbox{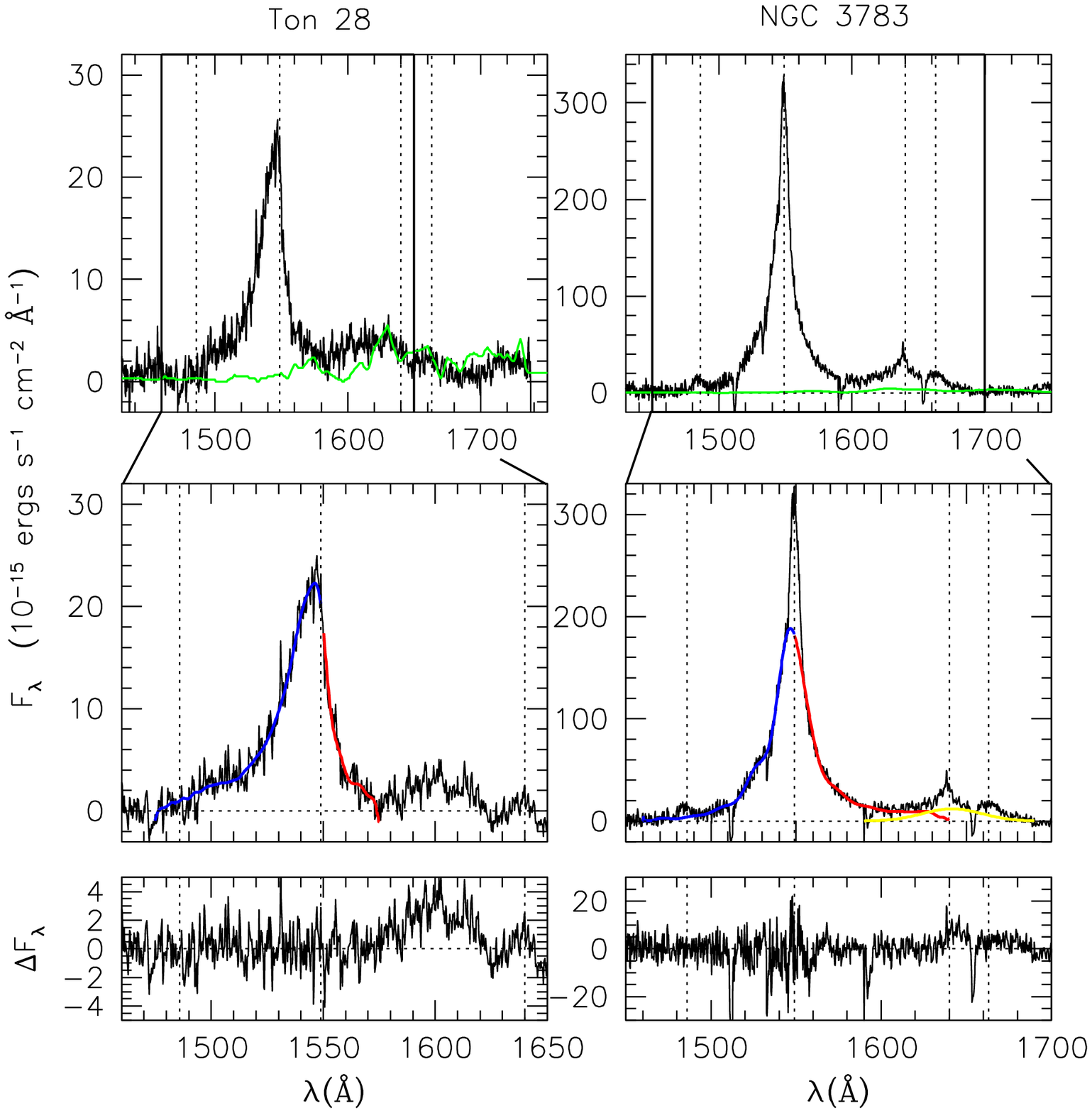}\caption{Line profile analysis for
two sources in widely different positions along the E1 sequence (Ton 28, a NLSy1, and NGC 783), for \hb\ (left panels)
and \civ\ (right panels). The upper panel shows the continuum subtracted spectra, with the \feii\ contribution
highlighted. The middle panel emphasizes the \hbbc\ profile as well as the \heii\ line. Note that the \oiiiopt\ lines
of Ton 28 are significantly blueshifted with respect to the rest frame; Ton 28 is a ``blue outlier".
\label{fig:analysis}}
\end{figure}

Quoting Peterson et al. \cite{petersonetal04}, we can say that {\em accurate line-width measurement depends critically
on avoiding contaminating features, in particular the narrow components of the emission lines}. Fig.
\ref{fig:analysis} illustrates the typical analysis and measurement procedure applied to optical and UV data, in the
\hb\ (left panel) and \civ\ (right panel) spectral region. Two sources -- one representative of Pop.\ A and the other
of Pop.\ B -- are shown. After continuum subtraction, \feii\  can be measured through a template \feii\ spectrum.
Measurement of \feiiq\ (i.e., \feii\ emission integrated over the wavelength range 4434--4684 \AA) parameters like
flux and width  are accomplished by constructing an array of template spectra within reasonable limits of scaling and
broadening factors \cite{borosongreen92,marzianietal96,marzianietal03a}, starting from a template (the I Zw 1 \feii\
spectrum). One then obtains the best scaling and broadening factor by subtracting the templates and by identifying the
one template that minimizes the sum of the least-square residuals in the range 4450--4600 \AA. In principle, \feii\
and continuum  should be fit simultaneously in an automatized way \cite{freemanetal01}, especially if \feii\ is very
strong.  A similar procedure
\cite{dietrichetal02,dietrichetal03,kuraszkiewiczetal04,marzianietal96,vestergaardwilkes01} is applied to the \civ\
spectrum using an \feiiuv\ emission template \cite{marzianietal96,vestergaardwilkes01,willsetal85}. While the
procedure works remarkably well in the optical and in the UV around 1600 \AA, some caution is needed in the UV between
2000 \AA\ and 3000 \AA\ since \feii\ emission around \mgii\ can be strong and different from source to source
\cite{iwamuroetal02}.

We usually  isolate the contribution of \hbnc, \oiiiopt, and of broad\linebreak \heii\ before we apply a high-order
spline fit to produce a model- independent ``description" of \hbbc\ which minimizes the effects of noise (the thick
lines in the side panels of Fig. \ref{fig:analysis}; \cite{marzianietal03a}).  Measurements of FWHM(\hbbc) are one of
the basis of black hole mass estimates, so that it is important to provide a reliable recipe that leads to accurate
results. Subtraction of a narrow-line component \hbnc\ is best done according  to the following criteria
\cite{marzianietal03a}: (1) if a clear inflection is seen, the subtraction is trivial. This is the case of most
sources (see the example of NGC 3783 in Fig. \ref{fig:analysis}); (2) in the case where an  infection is not seen, we
subtracted a Gaussian profile under the condition that FWHM(\hbnc) $\approx$FWHM(\oiiiopt). It is important to stress
that this last condition is not always applicable; (3) for several Pop. A sources (e.g., ``Narrow Line Seyfert 1s"
(NLSy1s) like Ton 28 shown in Fig. \ref{fig:analysis}) with typically Lorentzian profiles and FWHM $\simlt$4000 \kms\
(\S \ref{hbshapes}), the inflection is not observed. Such sources are ``blue outliers" i.e., objects for which the
recessional velocity measured on \oiiiopt\ is lower than that of \hbnc\ by more than 250 \kms\ \cite{zamanovetal02}.
Subtraction of a significant narrow component in many of these sources would imply \hbnc\ stronger than or comparable
to \oiii, which is not consistent with other forbidden line ratios \cite{nagaoetal01}. Typically, \hbnc\ is 1/10 the
strength of \oiii. In the outliers, any \hbnc\ would be appreciably displaced along the \hbbc\ profile. However, since
W (\oiii)$\simlt$ 2.5 \AA, the \hbnc\ becomes too weak to be detected (very high resolution spectra are needed; it is
detected for I Zw 1 \cite{veroncettyetal04}, and references therein). This argues against the subtraction of a strong
\hbnc\ in blue outliers if high-resolution does not resolve a low-ionization \hbnc\ \cite{veroncettyetal04}. If this
recipe is applied, FWHM(\hbbc) measurements with an accuracy of $\pm 10$\% at a 2$\sigma$\ confidence level are
possible for the wide majority of ``good" spectra. We customarily isolate a narrow-component contribution also to
\civ\ before the \civbc\ is fit. Since the presence of significant \civnc\ has been debated, we  discuss the issue in
physical terms in \S \ref{civnc}.

\subsection{Eigenvector 1 Correlates \label{e1corr}}

\subsubsection{\feii}

The emission spectrum of \feii\ in AGNs spans a wide wavelength range from the UV to the IR. It affects not only line
intensity measurements of other ions but the continuum determination as well. In case of the strong \feiiuv, the
emission produces a pseudo-continuum that extends to wavelengths as short as 1000 \AA, and is especially strong in the
near UV between 2000 and 3000 \AA\ (Fig. \ref{fig:composite}; \cite{marzianietal96,verneretal04,willsetal85}).  A
convenient subdivision of \feii\ emission may be as follows: (1) \feiiuv, between 2000 and 3000 \AA\
\cite{iwamuroetal02,verneretal04}; (2) near \feiiuv\ between 3000 and 3500 \AA; (3) \feiiq,  whose values are probably
the most frequently reported \feiiopt\ measurements \cite{borosongreen92,marzianietal03a}. Under the assumption that
the I Zw 1 spectrum is a good representation of optical \feii\ emission, it is \feiiq/\feiiopt $\approx$ 3.3
\cite{marzianietal96}; (4) \feiir\ i.e., the emission blend on the red side of \hb\ (see Figs. \ref{fig:composite} and
\ref{fig:e1spectra}).

\feii\ emission is still rather poorly understood, even if several recent papers have made significant advances after
a decade-long standstill (see Ref. \cite{sulenticetal00a} for a bibliography before late 1999). An  extremely complex
energy level structure of \feii\ makes it very difficult to obtain all the experimental transition probabilities and
therefore calculate line intensities (\cite{baldwinetal04,sigutetal04,verneretal99}, and references therein). Although
$Z \gg Z_\odot$\ does not seem anymore a requirement for explaining very large \feiiuv / \mgii\ ratios
\cite{verneretal04}, there is still no general consensus on the excitation mechanism. There are two lines of thought:
Fe$^+$\ is mainly produced by photoionization; collisional excitation and florescence produce the observed strength
and multiplet ratio. A recent paper \cite{verneretal03} accounts for the observed \feiiuv/\feiiopt\ and \feiiuv /
\mgii\ intensity ratios. Large electron density \ne\ (UV multiplets are favored because of the increase in the upper
level population) and moderate micro-turbolence (broadening $\sim$ 10 \kms) can account for the combination of
important diagnostics such as the \feiiuv / \mgii, and \feiiuv / \feiiopt\ intensity ratio \cite{verneretal04}.
However, models based on photoionization are not void of difficulties: a recent work indicates that photoionized BLR
gas cannot produce both the observed shape and observed equivalent width of the 2200--2800 \AA\ \feiiuv\ bump unless
there is considerable velocity structure corresponding to a microturbulent velocity parameter $v_{\mathrm turb} \sim$
100 \kms, which strongly favor fluorescent processes \cite{baldwinetal04}. A second line of thought
(\cite{collinjoly00,veroncettyetal04}, and references therein) suggest that modelling the spectrum of the BLR requires
a non radiative heating mechanism which increases the temperature in the excited HI region, thus providing the
necessary additional excitation of the \feii\ lines \cite{joly87}. The specificity of such a medium compared with a
photoionized medium is its extremely low degree of ionization. The emission spectrum is made exclusively of low
ionization lines (LILs) like \feii, Balmer lines and \mgii\ for the electron density \ne\ typical of the BLR. The
photoionization models require very high ionizing photon fluxes $\sim 10^{20} - 10^{21} $ \cmq s$^{-1}$\ to explain
very large \feiiuv / \mgii\ ratios, while collisionally heated media  require a shielding mechanism to hide the
emitting regions from the strong AGN continuum. On the basis of our current understanding of the line emitting region
structure (\S \ref{blr}), it is not clear where this shielded medium could be \cite{collinjoly00,baldwinetal04}.

\begin{figure}
\hspace{-0.cm}\epsfxsize=4.7cm\epsfbox{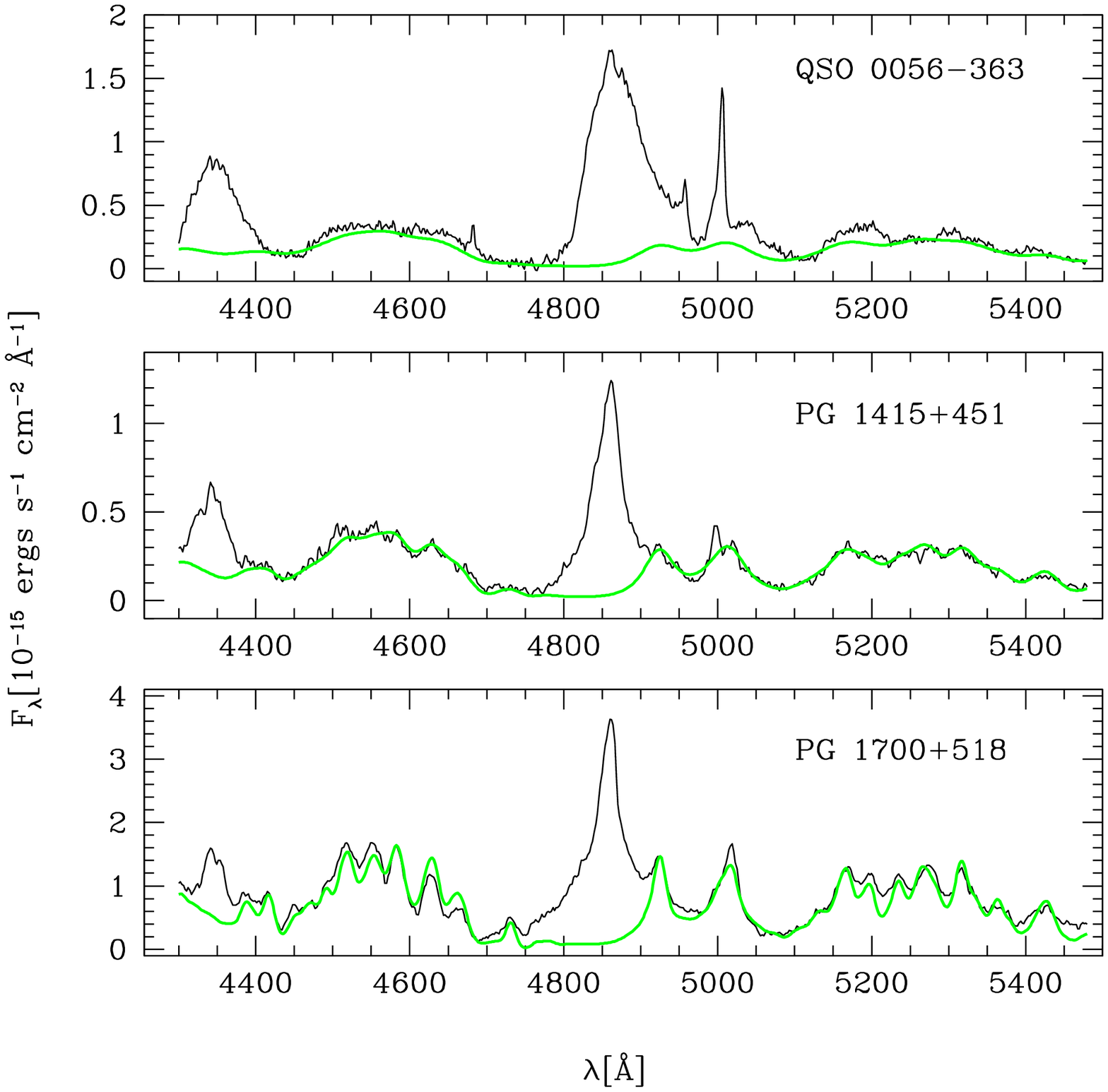}
\epsfxsize=4.7cm\epsfbox{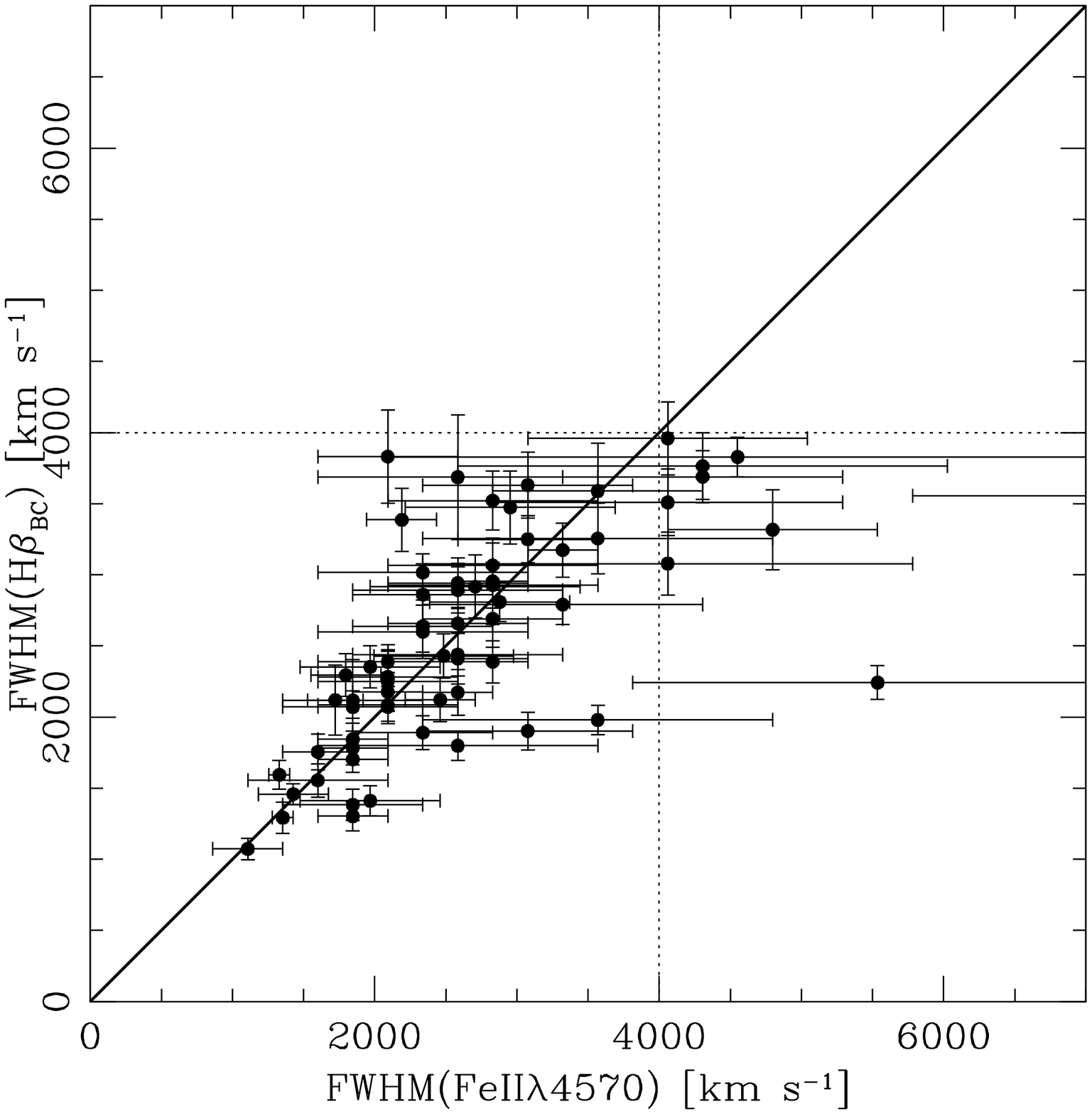}\epsfxsize=4.7cm
\epsfbox{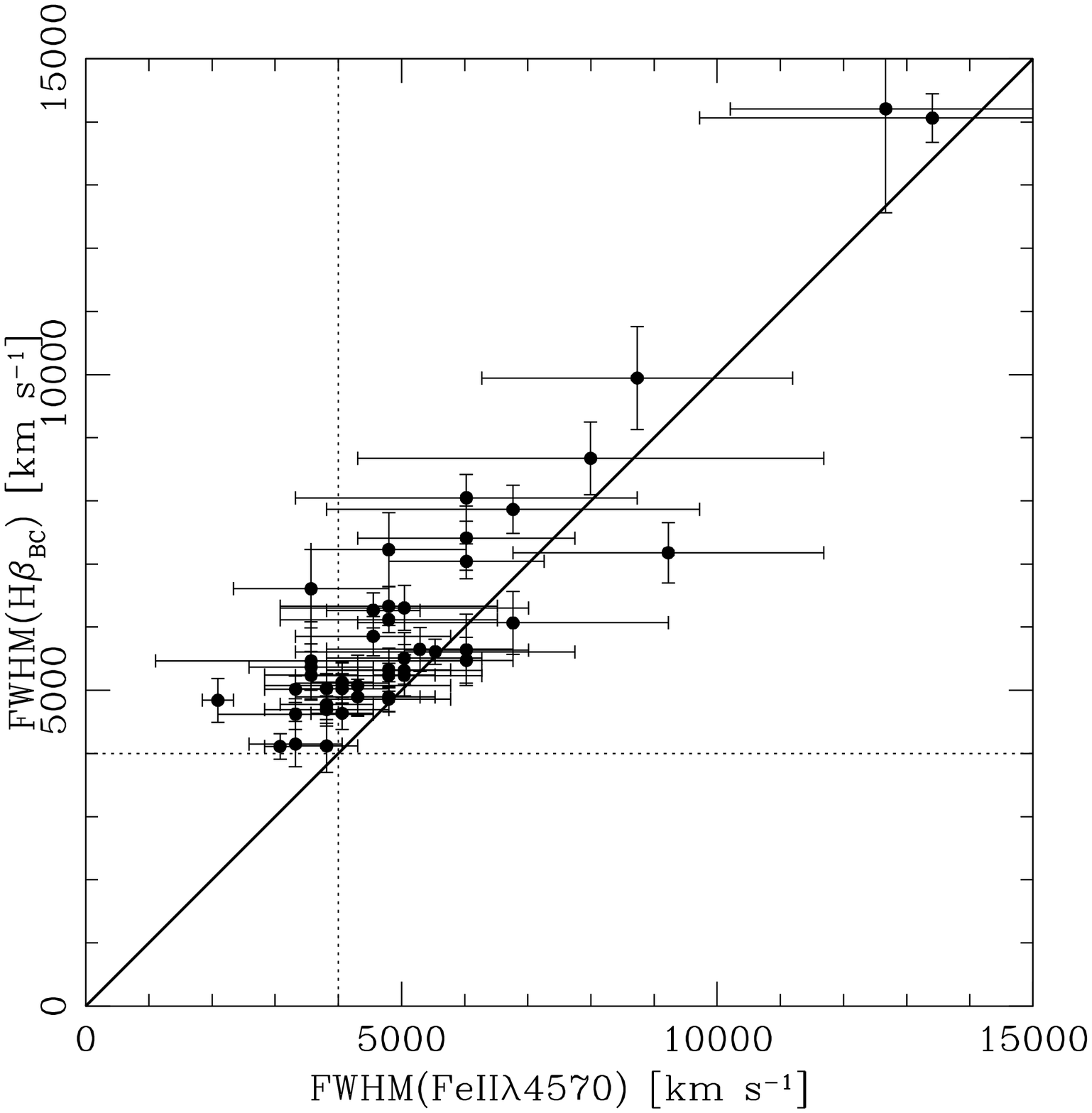}\caption{ The left panel shows three examples
with widely different FWHM(\feiiq); it is shown that the I Zw 1
template, opportunely scaled and broadened, provides a good fit to
all three sources (it provides a satisfactory reproduction to
\feiiopt\ of almost all sources \cite{marzianietal03a}. The middle
and right panel show the correlation between FWHM(\hbbc) and
FWHM(\feiiq), for FWHM(\hbbc)$\le$ 4000 \kms\ and FWHM(\hbbc)$>$
4000 \kms\ respectively. \label{fig:feii} }
\end{figure}

Even if the poor understanding of \feii\ processes  hampers the use of \feii\ emission as a diagnostics, there are
several important constraints related to E1:

\begin{itemize}

\item  \feiiq\ emission is well reproduced, in a sample of 300 AGNs \cite{marzianietal03a,marzianietal03b} by a scaled
and broadened I Zw 1 \feiiopt\ template. All sources did not show  any {\em strong} deviation from the template within
the limits imposed by $S/N$\ and resolution.  Measurements of \feii\ width and strength are intrinsically  difficult.
Non optimal intrinsic width, spectral coverage, and $S/N$\ imply that only an upper limit to the equivalent width can
be measured for faint \feii\ sources (see Fig. \ref{fig:e1});

\item   Fig. \ref{fig:feii} shows that the template method is able to reproduce \feii\ in cases of widely different
width and strength. At zero order, Fig. \ref{fig:feii} confirms that \feii\ and \hbbc\ are approximately of the same
width, suggesting a common origin for these two LILs and confirming  that \rfe\ is a meaningful parameter.  There is
however a difference: while for FWHM(\hbbc)$\simlt$4000 \kms\ we have FWHM(\hbbc)$\approx$FWHM(\feiiq) in a rigorous
statistical sense, a sign test suggests that  FWHM(\hbbc) $\simgt$ FWHM(\feiiq) if FWHM(\hbbc) $\simgt$ 4000 \kms
\cite{marzianietal03c};

\item the \feiiuv / \mgii\ ratio seems to be E1 dependent, in the sense that sources with the narrowest FWHM(\hbbc)
\cite{marzianietal03a} show the largest values of \feiiuv / \mgii\ \cite{iwamuroetal02,kuraszkiewiczetal04}. The
\feiiuv / \mgii\ ratio increase along the E1 sequence may imply higher iron abundance but is also consistent with an
increase in \ne, if density changes from \ne$\simlt$ 10$^{10}$ \cm3\ to \ne $\sim 10^{11}$ \cm3\ \cite{verneretal04}.
Largest \feiiuv / \mgii ($\approx$ 10) ratios require \ne $\sim 10^{11}$ \cm3.  Higher density is supported by several
lines of evidence (\cite{bachevetal04,marzianietal01}; \S \ref{e1othercorr}).

\end{itemize}

\subsubsection{Prominence of \oiiiopt}

The prominence of \oiiiopt\ is one of the most basic correlates along E1, especially if the parameter  actually
measured is the peak height over continuum \cite{boroson02,borosongreen92}. Although not customary, this parameter
yields a better correlation since profiles tend to broaden at low intensity \cite{zamanovetal02}, making W(\oiiiopt) a
worse correlate. Emission of \oiiiopt\ is not straightforward to interpret. If a sample covers low-$z$\ sources and
spans a large $z$\ range, \oiiiopt\ measures are prone to aperture effects  since the \oiiiopt\ emission morphology
can be  anisotropic \cite{hesetal93}. Resolved ``narrow line regions" (NLRs i.e., where \oiiiopt\ is produced) reveal
that line emission is not only anisotropic but also strongly influenced by radio jets, in nearby RQ Seyfert nuclei and
in BLRG alike \cite{bicknell02,capettietal96,falckeetal98,whittle92}. Not surprisingly, the EW of \oiiiopt\ is in RL
than in RQ sources of the same optical spectral type \cite{laor00,boroson02,marzianietal03b}. In addition, \oiiiopt\
profiles can be strongly asymmetric with a predominance of blue-ward asymmetries (\cite{whittle92}, and references
therein). In the E1 context we find that large W(\oiiiopt) sources can have a blue-ward asymmetric wing; low
W(\oiiiopt) sources (located in the lower left part of the optical E1 diagram; \cite{zamanovetal02})  have emission
entirely ascribable to a semi-broad, blueshifted feature that may resemble the blueward wing observed in stronger
\oiiiopt\ emitters.

\subsection{Eigenvector 1 Extremes \label{e1extremes}}

\paragraph{Narrow Line Seyfert 1s} NLSy1s have  been recognized  as  a distinct type of Seyfert nuclei since 1985 \cite{osterbrockpogge85}.  The defining criterion is that the width of the Balmer
lines be  less than 2000 \kms.  The \oiii/\hb\ ratio (but \hb\ is not necessarily \hbnc!) is  often smaller than 3
\cite{constantinshields03,stepanianetal03}. NLSy1s usually (but not always) show rather strong  \feiiopt\ emission.
Their soft X-ray spectra are very steep \cite{vaughanetal99,wangetal96} and variable (e.g.,
\cite{grupeetal98,turneretal99}). The original definition is not void of difficulty, since: (1) a sharp discontinuity
in many properties of type-1 sources is seen at FWHM(\hbbc)$\approx$ 4000 \kms, and not at FWHM(\hbbc)$\approx$ 2000
\kms\ (\S \ref {e1corr}). In other words, in the range 2000 \kms\ $\simlt$ FWHM(\hbbc) $\simlt$ 4000 \kms\ sources
show properties consistent with the ones of NLSy1, although less extreme. (2) Sources with \rfe $\simlt$ 0.5 (spectral
type A1; \cite{sulenticetal02}) may not be true NLSy1s since they often show prominent \oiiiopt, and their \hbbc\
profiles can be different from the typical NLSy1 and bin A2  and A3 profiles; several sources are core-dominated RL
sources and some belong to the second extremal type mentioned in \S \ref{samples}. With these caveats in mind, NLSy1s
seem to be drivers of the E1 correlations since they show and minimum FWHM(\hbbc) and may have extreme \rfe\ in the
``main sequence" of the optical E1 plane \cite{sulenticetal00a}. In the context of E1 they are interpreted as young or
rejuvenated gas-rich objects with low black  holes mass and high accreting rate
\cite{constantinshields03,grupe04,krongoldetal01,mathur00,petersonetal00}.


\begin{figure}[htbp]
\centering
\includegraphics[width=0.5\textwidth]{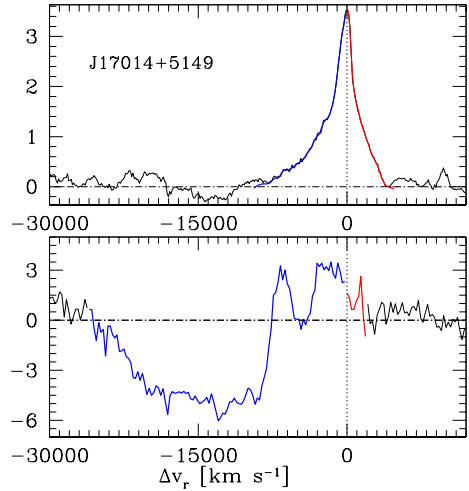}
\caption{Inter-profile \hb\ and \civ\
comparison for the prototypical BAL QSO PG 1700+518. The profiles
are shown here after continuum and \feii\ subtraction. Note that
the \civ\ emission component is almost fully blueshifted with
respect to the quasar rest frame. }
\label{fig:bal}
\end{figure}

\paragraph{BAL QSOs} High-$z$ and
low-$z$ BAL QSOs share some of the properties of the extreme Pop.\ A sources in the E1 optical diagram: weaker \civbc\
in emission \cite{hartigbaldwin86,turnshek84}; stronger \feiiuv\ and \aliii\ \cite{wampler86}, as well as stronger
\nv. BAL QSOs occur with a frequency of $\approx$ 15\% with respect to all quasars
 \cite{reichardetal03a}; the fraction of radio-loud
BAL quasars may be  consistent with the fraction of radio-loud ordinary quasars \cite{toleaetal02}.  The  ``Balnicity"
Index \cite{weymannetal91} distribution  rises with decreasing index values. This suggests that the fraction of
quasars with intrinsic outflows may be significantly underestimated \cite{reichardetal03a}. Several high-$z$\  BAL
QSOs have extremely strong \feii\ and very weak \oiiiopt, extending the inverse relationship found for low-$z$\ QSOs
and typical of Pop. A \cite{yuanwills03}. An important information that is still missing for most high-$z$ BAL sources
is a reliable estimate of the systemic redshift \cite{koristaetal93}. Without this information, the interpretation of
the \civ\ absorption/emission profile is rather ambiguous. Identification of low-$z$ BAL sources had to await
space-based UV observatories \cite{borosonmeyers92,sulenticetal05,turnsheketal97}, and only a handful  of sources is
known to-date. Classical BAL sources with Balnicity Index $\gg$ 0 \kms\ at low and high $z$\ for which the rest frame
can be reasonably set from narrow lines show an almost fully blueshifted emission component of \civ\
 in addition to a fully blueshifted absorption (\cite{sulenticetal05}; Fig.
\ref{fig:bal}). Most low-$z$\ BAL AGNs with terminal wind
velocities in the range $\approx$ 20000-30000 \kms\ are either
outliers or are located along the upper envelope of the E1 ``main
sequence'' for Pop. A sources. Most revealing are the cases of Mkn
231 and IRAS 07598+6508, which are basically the only 2 outliers
in a sample of $\sim$300 low-$z$\ AGNs. The sources show
FWHM(\hbbc) significantly larger than FWHM(\feiiopt). The
difference is due to a strong blue-ward asymmetry visible in the
\hbbc\ profile. The \hbbc\ profile can be interpreted as due to a
narrower component with FWHM $\approx$ FWHM(\feiiopt), and to a
completely blueshifted feature, most probably associated to an
high ionization outflow emitting most of HILs like \civ. If this
effect is taken into account, and FWHM(\feiiopt) instead of
FWHM(\hbbc) is used, Mkn 231 and IRAS 07598+6508 move into the
``main sequence'' optical E1 diagram to become extreme (in terms
of \rfe) Pop.\ A sources.

\begin{figure}[htbp]
\centering
\includegraphics[width=0.45\textwidth]{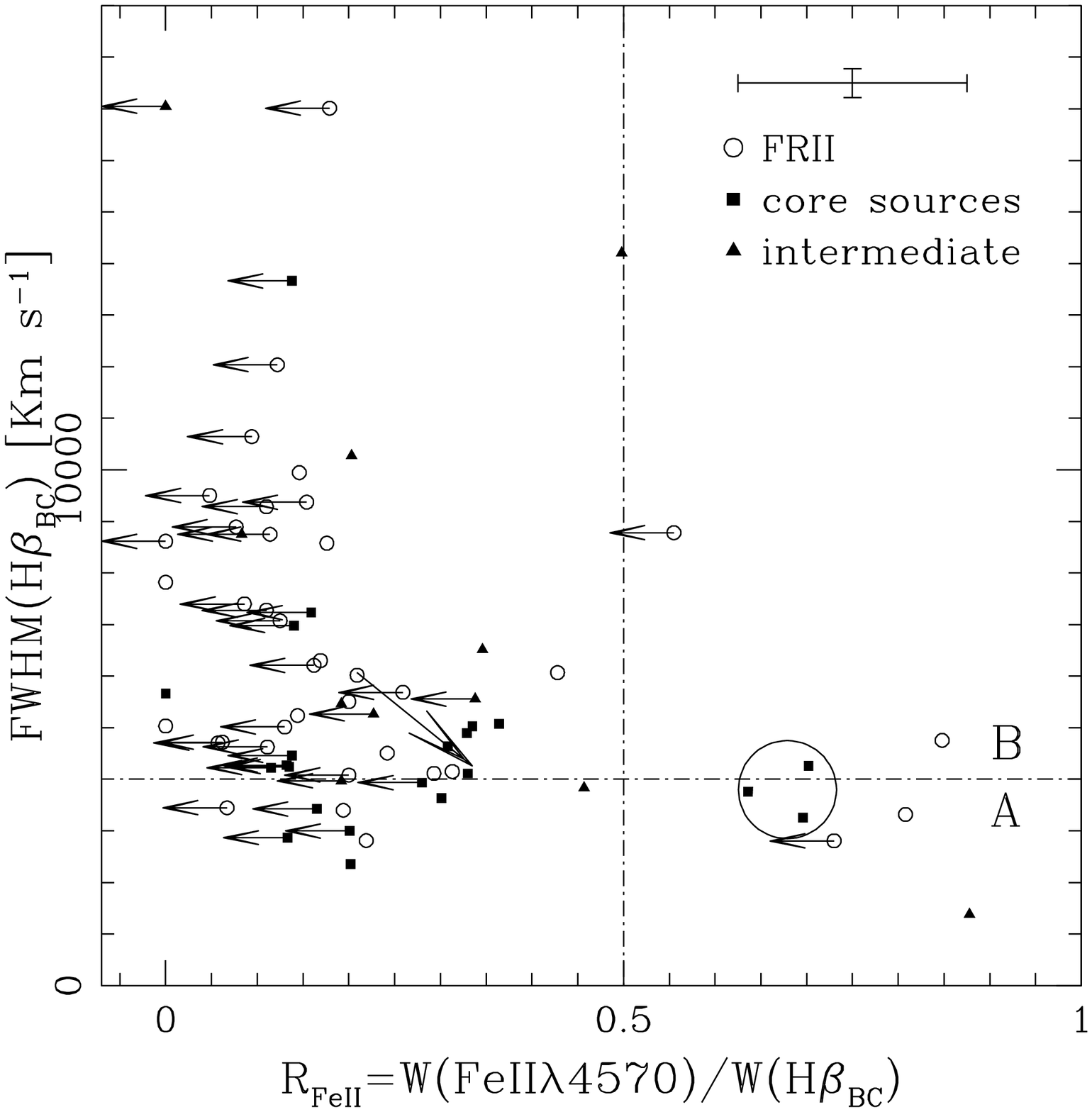}
\caption{Distribution of RL sources in the
optical plane of the eigenvector 1 of type-1 AGNs; abscissa is the
equivalent width ratio between \feiiq\ and \hbbc, ordinate is FWHM
of \hbbc\ in \kms.  The big arrow indicates displacement between
the median \rfe\ and FWHM(\hbbc) values from lobed and pure core
RL sources.  The circle identifies CD sources that genuinely have
\rfe\ $\simgt$0.5 \cite{sulenticetal03}; other sources at \rfe
$\simgt$ 0.5 are strongly affected by measurement errors.}
\label{fig:e1radio}
\end{figure}

\paragraph{Radio Quiet \& Radio Loud Quasars}
FWHM(\hbbc)  in RL sources are at least twice as large as the RQ
majority \cite{sulenticetal00a}. The average broad \feiiq\
emission line strength is also about half that for RQ sources
\cite{borosongreen92,joly91,marzianietal96}. Fig.
\ref{fig:e1radio} shows that the Fanaroff-Riley II (i.e.,
lobe-dominated, LD) ``parent population'' of relatively un-boosted
RL sources (median radio/optical flux ratio \rk \ $\sim$ 500)
shows the most restricted occupation.  The Doppler boosted
core-dominated (CD) RL sources (median \rk\ $\sim$1000) lie
towards smaller FWHM(\hbbc) and stronger \feii\ in E1 as expected
if the lines arise in an accretion disk.

\subsection{Additional E1 Correlates \label{e1othercorr}}

\paragraph{\hbbc\
Profile Shape \label{hbshapes}}  Median spectra of the \hbbc\ profile were extracted  for each spectral type (Fig.
\ref{fig:e1}) after \feiiopt\ and continuum subtraction in a sample of $\sim$ 200 AGNs
\cite{marzianietal03a,sulenticetal02}. The median \hbbc\ profiles in spectral types A are almost symmetric, with a
slight blueward asymmetry in bin A2 (see Fig. \ref{fig:e1} for the definition of the spectral types). The blue
asymmetry becomes visually apparent in bin A3 which contains sources that are extreme in many ways. The median
profiles of Pop. B sources are very redward asymmetric, with the strongest asymmetry in the (few) bin B1 sources. The
\hbbc\ profile in NLSy1 and Pop. A sources is well fit by a Lorentz function \cite{veroncettyetal01}.  The best fit to
\hbbc\ bin B1 and B1$^+$ profiles is achieved through the sum of two Gaussians: (1) an unshifted Gaussian core
(FWHM(\hbbc) $\sim 5000$ \kms) and (2) a broader redshifted Gaussian base with FWHM$\sim$ 10000 \kms\
\cite{brotherton96}. We remark that a Double Gaussian fit to \hbbc\ in Pop.\ B sources  is a formal result.  The two
components have a physical justification if, for the generality of AGNs, the broader one can be ascribed to gas that
lies closest to the continuum source in an optically thin (to the \hi\ ionizing continuum) region with large covering
factor $ f_{\mathrm c} \rightarrow$ 1 \cite{brotherton96,brothertonetal94b,marzianisulentic93,shieldsetal95}.


\paragraph{\oiiiopt\ blueshifts}
The narrow lines of \oiiiopt\ usually provide a reliable measure of the systemic, or rest-frame, quasar velocity.
However, several observations indicate that the  NLSy1 galaxy prototype I Zw 1 shows an \oiiiopt\ blueshift of
$\approx$ --500 \kms\ relative to other rest-frame measures \cite{borosonoke87,marzianietal96}, with a second
component at $\approx$ --1500 \kms\ (see references in Ref. \cite{veroncettyetal04}). I Zw 1 is not alone. It is
possible to identify sources with \oiiiopt\ $\Delta v_{\rm r}\simgt -300$ \kms\ with respect to \hbnc. Such large
shifts are unlikely to be due to measurement errors \cite{zamanovetal02}. The two  known largest blueshift after I Zw
1 reach $\Delta v_{\rm r} \sim -1000$ \kms\  \cite{aokietal04}. Since these sources are apparently rather rare (but
not unique! \cite{grupeetal99,grupeetal01,marzianietal03b}), and since they seem to lie out of a continuous shift
distribution, they have been dubbed ``blue outliers".  The distribution of blue outliers in the E1 optical plane  is
different from that of the general AGN population \cite{zamanovetal02}. The blue outliers occupy the lower right part
of the diagram and are exclusively Pop. A/NLSy1 nuclei \cite{zamanovmarziani02}. Low W(\oiiiopt) may show only a
blue-shifted component as if only the asymmetric part of the profile observed in stronger sources is being emitted. It
is relatively straightforward to interpret the large blueshift and the profile of the \oiiiopt\ lines as the result of
an outflow \cite{aokietal04,zamanovetal02}, and the low W(\oiiiopt) as a signature of a compact NLR
\cite{veroncettyetal04,zamanovetal02}.

\paragraph{High Ionization Lines} The peak shift of the prototypical HIL \civ\  is an important E1
correlate in the sense that large shifts are shown only for FWHM(\hbbc) $\simlt$ 4000 \kms\
\cite{bachevetal04,marzianietal96,sulenticetal00b} at low $z$. W(\civ) decreases along the E1 sequence from sources
with large FWHM(\hbbc) and low \rfe\ (typical W(\civbc)$\sim$ 100$\div$200 \AA) to the most extreme NLSy1s (typical
W(\civbc)$\simlt$50 \AA; \cite{bachevetal04,marzianietal96,rodriguezpascualetal97,sulenticetal00b}).


\paragraph{Ionization Level/Electron Density}
A decrease in equivalent width of \civ\ and \hbbc\ along with an
increase in prominence of \feiiq\ in the E1 sequence toward NLSy1
sources suggest a systematic decrease in ionization level from RQ
Pop. B to Pop.\ A \cite{marzianietal01}. The behavior of W(\civ)
and W(\hbbc), and of the diagnostic ratios \siiii / \ciii\
 and \aliii / \ciii\ can be understood in terms of a
decrease in ionization parameter $U$\ (defined as the ratio
between the ionizing photon flux and the electron density \ne) and
an increase in \ne\ from $\log U \sim -1$, $\log$ \ne $\sim$ 9.5,
to $\log U \sim -2 \div -2.5$, $\log$ \ne $\sim 10.5 \div 11$\
\cite{bachevetal04,koristaetal97,marzianietal01}. Notably, the
diagnostic ratios indicate that $\log$ \ne\ $\sim$ 10.5 $\div$ 11
towards the NLSy1 domain (see also Ref. \cite{veroncettyetal04}
\linebreak for I Zw 1).

\paragraph{X-ray Continuum}  The soft (energy
$\simlt$ 2-3 KeV) X-ray excess \cite{comastrietal92,wilkeselvis87} is perhaps the most distinctive X-ray feature that
differentiate sources along E1 \cite{wangetal96,sulenticetal00a}. The nature of the soft X-ray excess is still
debated, although direct thermal emission from the accretion disk seems to be ruled out  on the basis of the very high
temperature derived for this optically thick component \cite{piconcellietal04}. Even the newest X-ray data cannot
constraint the current models \cite{brinkmannetal04,gierlinskidone04}. In the context of E1 several recent studies
(including  XMM-Newton observations; \cite{piconcellietal04,porquetetal04}) reveal systematic differences in the X-ray
properties of RL and RQ quasars. XMM observations \cite{piconcellietal04} confirm a previously reported
anti-correlation between soft and hard photon indices and FWHM(\hbbc)   \cite{reevesturner00,wangetal96}. The soft
X-ray spectral index correlates with the hard-X ray spectral shape, in the sense that NLSy1 and Pop. A sources have a
steeper spectral shape \cite{leighly99}.

\paragraph{UV Continuum} Inferences on UV/FUV continuum shape are complicated by the lack
of data at low $z$\ before the advent of HST and FUSE, as well as by the possibility of strong extinction. A
breakthrough occurred with the discovery in the spectrum of 3C273 of an optical-UV feature called the ``Big Blue
Bump", corresponding to an important fraction of the bolometric luminosity. Some soft X-ray selected AGNs are NLSy1
galaxies with an enhanced big blue bump emission component relative to the underlying continuum making the optical
continuum blue and the soft X-ray spectrum steep \cite{grupeetal98}. On the other hand, several recent works indicate
that NLSy1s have a redder continuum than the rest of type-1 AGNs
\cite{crenshawetal02,constantinshields03,warneretal04}. After correcting with  an extinction curve obtained from
$\sim$1000 AGNs, it seems that radio-quiet AGNs have very similar intrinsic UV-to-optical continuum shape over several
orders of magnitude in luminosity, and that that radio-loud and radio-quiet AGNs probably share the same underlying
continuum shape \cite{gaskelletal04}. A recent investigation of ultraviolet-to-optical Spectral Energy Distribution
(SED) of 17 AGNs used quasi-simultaneous spectrophotometry spanning 900-9000 \AA\ (rest frame) with FUSE, HST and the
2.1-meter telescope at KPNO to study the big blue bump. Most objects exhibit a spectral break around 1100 \AA, with a
slope slightly steeper for RL sources \cite{elvisetal94,shangetal04,vandenberketal01,zhengetal97}, but otherwise there
is no obvious trend which may be related to the source location in the E1 sequence (apart from extinction).

\paragraph{FIR excess} Several quasars show an SED with an extremely strong FIR excess
\cite{bolleretal02,braitoetal04,rodriguezardilaviegas03,sulenticetal05}. The strong excess can be due, at least in
part, to the contribution of circumnuclear starburst, as in the case of Mkn 231 \cite{daviesetal04}, or to reprocessed
radiation from a dusty torus encircling the accretion disk and the BLR. NLSy1s and extra-strong \feii\ sources are
known to be found often among sources  detected by the {\it Infra-Red Astronomical Satellite}
\cite{liparietal93,moranetal96}, although it is as yet unclear how frequent is a FIR bump among NLSy1s
\cite{stepanianetal03}.

\paragraph{Bolometric Correction \label{bolcorr}} It is not easy to obtain \lbol\ for individual quasars because
they emit significant power over a large part of the electromagnetic spectrum, most notably in the FUV and in the FIR.
Bolometric correction factors are found to be \lbol $ \approx 9 \div 13 \lambda L_\lambda $ \ergss\ for $\lambda
\approx$ 5400 \AA, with possible deviations for single sources of amplitude $\pm$50\%\
\cite{elvisetal94,shangetal04,woourry02a}. A bolometric relationship \lbol $ = 9 \div 10 \lambda L_\lambda $ \ergss\
has been often adopted \cite{collinetal02,kaspietal00,marzianietal03b,wandeletal99,woourry02a}. Recent work based on
good coverage of the UV, FUV and X but still poor sampling of the IR SED confirms that the assumption  $L = 13 \lambda
L_\lambda $ \ergss\ at 5400 \AA\ is a fairly good approximation  \cite{shangetal04}. Again, scatter is pretty large
($\pm$50\%). To test any  dependence of the bolometric correction along E1, we analyzed 44 sources that share accurate
\hbbc\ measurements \cite{marzianietal03a}  and SEDs constructed from archival data \cite{woourry02a}. Comparison of
\lbol\ values computed from the SED and from the bolometric correction showed no significant systematic difference:
the average $ {\Delta \rm \log L_{\rm bol}}$ is $\approx -0.06$\ for all 44 sources, with a standard deviation $\sigma
\approx 0.24$. Similar results hold if RQ Pop.\ A sources and RL AGNs are considered separately. It is worth noting
that the scatter is due to a minority of badly behaving data points ($\approx$20\%). If they are removed, the standard
deviation becomes $\sigma \approx 0.1$\ in all cases, with systematic differences always $\simlt$0.05. Although big
blue bump studies show that there may be a gross uniformity as far as the FUV is concerned, the difficulty to assess
prominence and inherence of any FIR bump as well as the still uncertain shape in the EUV warrant caution before claims
of accuracy better than $\pm$50 \%\ can be believed for the bolometric correction of individual objects. It is also
important to stress that sources which are RL CD could add an appreciable scatter to an average SED, since their
continuum may be strongly affected by relativistic beaming. RL CD sources should be considered on an individual basis
and not included in averages.

\paragraph{X-ray Variability} A common phenomenon  observed in  AGNs is
variability (for a review see e.g., Ref. \cite{ulrichetal97}). The most vigorous variability observed is that of the
X-rays \cite{mushotzkietal93}. Giant-amplitude X-ray variability by more than a factor of 15  has been found in NLSy1
galaxies. The most extreme one is IRAS13224-3809 which raised by a factor of about 57 in just two days
\cite{bolleretal97}. This giant variability is persistent \cite{galloetal04,youngetal99}. Examination of X-ray
properties across the Seyfert population reveals an anti-correlation  between variability amplitude  and FWHM(\hbbc)
\cite{grupeetal01,turneretal99}.

\paragraph{Optical Variability} Optical variability has been established as an identifying property of type-1 AGNs  for more
than three  decades. Typically these objects show continuum variations by 1-2 magnitudes  with timescales ranging from
days to years \cite{picaetal88,webbetal88}. Broad emission lines have also been found to vary. Emission line
variations  lag the continuum variations with delays ranging from a  few days  to months
\cite{crenshawetal96,petersonetal89}. The conclusion of an optical photometric study   on six NLSy1s is
 that, as a class, there is no evidence that NLSy1s behave any differently than broader-lined Seyfert 1s \cite{klimeketal04}. However,   an earlier study   \cite{webbetal00}
conducted on a large sample (23 AGNs)  found some evidence that  AGNs with the strongest \feii\ emission lines were
less likely to show strong optical variability. In the E1 context, an unpublished analysis of a survey of PG quasars
\cite{giveonetal99} suggests that strong optical variability $\Delta m \sim 0.5$ is not present among sources with
FWHM(\hbbc)$\simlt$4000 \kms, and \rfe $\simgt$0.5; it seems to be frequent for FWHM(\hbbc) $\simgt$ 4000 \kms, and,
interestingly enough, for sources with FWHM(\hbbc)$\simlt$4000 \kms\ and \rfe$\simlt$0.5.

\paragraph{} Summing up, along the E1 sequence we see extreme X-ray variability in correspondence of the high \rfe, low
FWHM(\hbbc) end; optical variability, on the other hand, seems to follow the opposite trend: to be modest or even
undetected in NLSy1s sources, and to be of relatively large amplitude for some sources with broader lines.

\begin{center}
\begin{tabular}{lcc}\\
\multicolumn{3}{c}{Table 1: Main Trends Along the E1 Sequence}\\
\\
\hline\hline
& Large \rfe  & Low \rfe \\
& Small FWHM(\hbbc)& Large FWHM(\hbbc)\\ \hline\\
W(\hbbc) & low & large\\
\hbbc\ shape & Lorentzian & redward asymmetic; \\
& & double Gaussian\\
\civ\ shift  & large blueshift & no shift\\
W(\civ) & low & high\\
\civnc\   & no & may be prominent\\
W(\oiiiopt) & lower & larger \\
\oiiiopt\ shift & may show blueshift & agrees  with \hbnc\\
width \feii\ & equal to \hbbc & may be less than \hbbc\\
optical variability & possible & higher amplitude \\
X ray variability & can be extreme & possible\\
Soft X photon index & can be large ($\simgt 2-4$) & $\simlt 2$\\
Radio & predominantly RQ & RQ and RL \\
BALs   &  extreme BALs &  less extreme BALs \\
ionization & lower & higher \\
LIL em. region \ne & higher & lower \\
 & &  \\ \hline\hline
\end{tabular}
\end{center}

\section{Inferences on the Broad Line Region Structure \label{blr}}

Two observational strategies have been proved to be very powerful to investigate the structure of the broad line
region: reverberation mapping \cite{horneetal04,peterson93} and statistical inter-profile comparisons, especially
between HILs and LILs \cite{brothertonetal94a,corbinboroson96,marzianietal96}. Two dimensional reverberation mapping
(i.e., mapping done considering narrow radial velocity intervals in the line profile) requires special observational
capabilities, and attempts to perform such measurements have yet to provide convincing results on BLR structure
\cite{horneetal04}. The correlations of E1 already allow to organize  data in a  fashion that provides also
appropriate input to reverberation mapping studies. Table 1 summarizes the main trends along the E1 sequence from
large FWHM(\hbbc) and low \rfe, to the narrowest FWHM(\hbbc) and largest \rfe.

\subsection{A Different BLR Structure Separates Two AGN Populations}

\begin{figure}[htbp]
\centering
\includegraphics[width=0.5\textwidth]{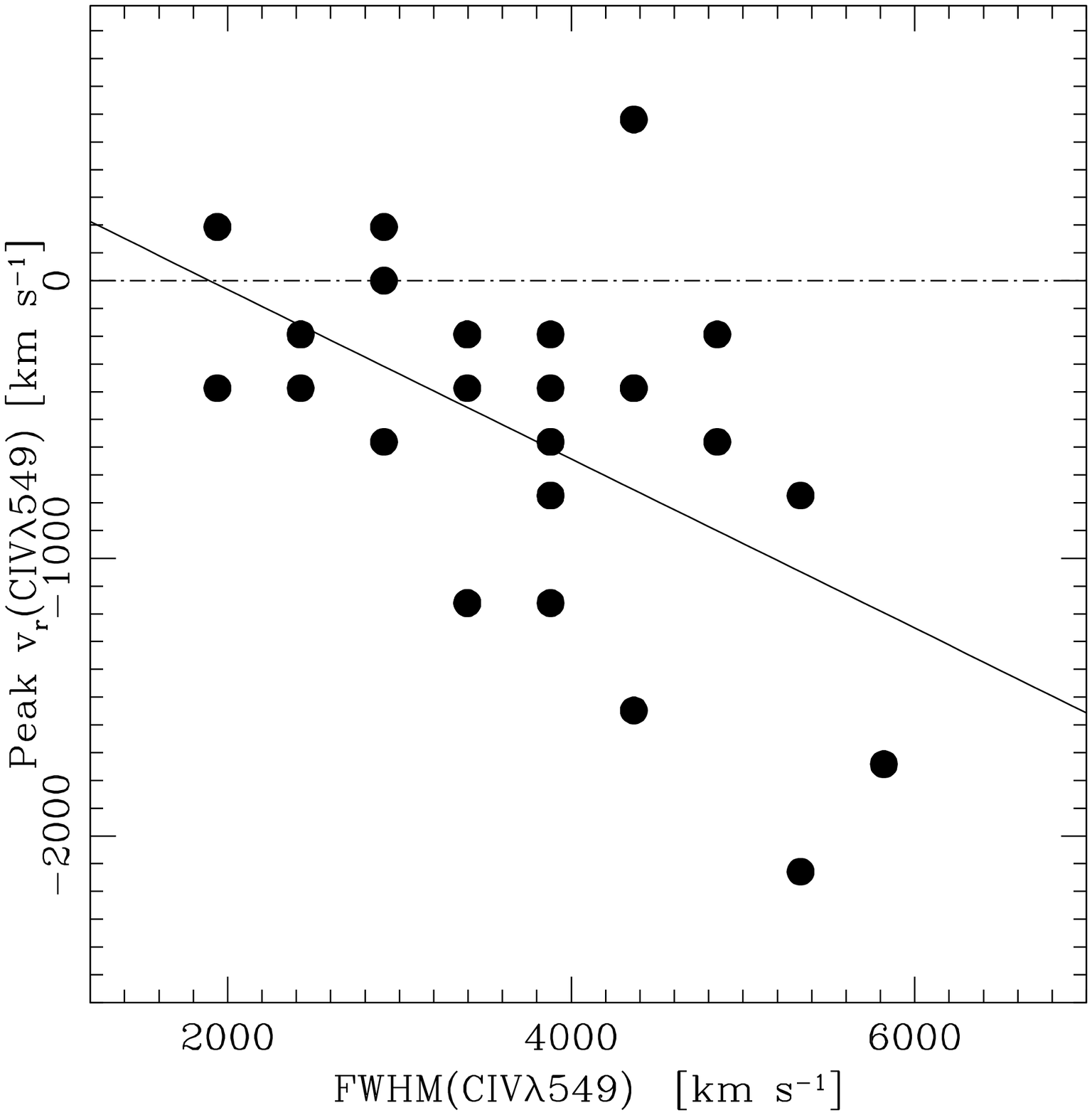}
\caption{Correlation between \civbc\ shift and FWHM for radio quiet, Population A (FWHM(\hbbc $\simlt$4000 \kms)
sources. A formal least-square fit is shown.} \label{fig:civshift}
\end{figure}

An obvious question is  whether there is a continuity in
properties all across the sequence in the optical E1 plane.  This
does not seem to be the case. Three results stand out:

\begin{enumerate}
\item a discontinuity cannot occur at FWHM(\hbbc)$\approx$ 2000 \kms, the formal limit of NLSy1s, since  sources show
continuity in properties at least up to FWHM(\hbbc) $\approx$ 4000 \kms\ \cite{sulenticetal00a}. For example,
ultra-soft sources are found at FWHM(\hbbc)$\approx$ 3000 \kms\ \cite{breeveldetal01};

\item  at FWHM(\hbbc)$\approx$ 4000 \kms\ we observe a rather abrupt change in \hbbc\ profile shape;

\item large \civ\ and \oiiiopt\ blueshifts occur for FWHM(\hbbc)$\simlt$ 4000 \kms. Fig. \ref{fig:civshift} shows a
correlation between FWHM(\civbc) and \civ\ peak shift that is valid for RQ Pop.\ A sources only; the same variables
for Pop.\ B sources would show a scatter plot.
\end{enumerate}

These  results have strong structural implications for the BLR. The simplest model able to explain \hbbc\ and \civ\
properties for sources with FWHM(\hbbc) $\simlt$ 4000 \kms\ is a disk and wind system. An optically thick,
geometrically thin disk emits most of \hbbc\ and all of \feii;  optically thin  gas with an outflow radial velocity
component $v_{\rm out}$\ emits \civ\ and other HILs (\cite{bottorffetal97,murraychiang97,progaetal00}; due to the
absence of intervening absorption we may expect that the material is highly ionized). The optically thick disk
obscuring the receding part of the flow and leaves a net blueshift in the \civbc\ line profile
\cite{bachevetal04,dultzinhacyanetal00,marzianietal96}. Other lines, like \siiii\ and \ciii\ are most probably emitted
at the base of the flow in the LIL emitting regions as also suggested by some reverberation mapping results
\cite{peterson93}. This model introduces a dependence on the aspect angle $\theta$\ that can be strongly different for
HIL and LILs.

Observations first supported this scheme in the context of a RQ/RL comparison
\cite{dultzinhacyanetal00,marzianietal96}. It seems now that a more meaningful comparison is between sources
with\linebreak FWHM(\hbbc) $\simlt$ 4000 \kms\ (note that this boundary may be somewhat luminosity-dependent, \S
\ref{beevol}), and the rest of type-1 AGNs \cite{sulenticetal00a} because sources with\linebreak FWHM(\hbbc)$\simgt$
4000 \kms\ can be both RL and RQ, and their broad line spectra can be indistinguishable. The evidence in favor of this
model is not only statistical but also based on detailed inter-profile comparisons and monitoring of individual
sources. For example, NLSy1s IRAS 13224--3809 and 1H 0707--495 are characterized by very blue continua; broad,
strongly blueshifted HIL  \civ\ and \nv; narrow, symmetric intermediate-ionization  lines (including \ciii, \siiii,
and \aliii) and LILs like \mgii\ centered at the rest wavelength \cite{leighlymoore04}. In NGC 4051,  the \heii\ line
is almost five times broader than \hb\ and is strongly blueward asymmetric. Variability and single-epoch data are
consistent with the Balmer lines arising in a low-inclination (nearly face-on) disk-like configuration, and the
high-ionization lines arising in an outflowing wind, of which the near side is preferentially observed
\cite{petersonetal00}. Akn 564 is the most extensively monitored NLSy1 galaxy in the UV \cite{collieretal01}. Absence
of response in the canonical HIL \civ\ line is consistent with  matter-bounded emission.

\subsubsection{Extreme BAL QSOs as Extreme Pop.\ A Sources \label{bal}}

\begin{figure}[htbp]
\centering
\includegraphics[width=0.5\textwidth]{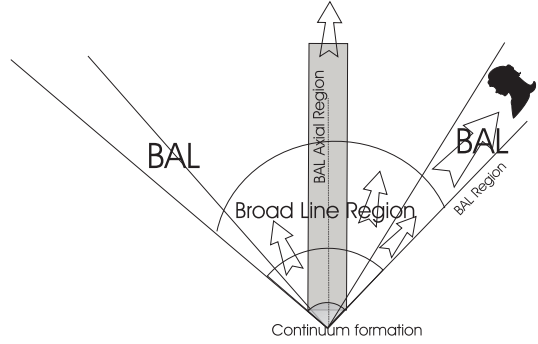}
\caption{Schematic view of the central engine of extreme
Population A sources. The region co-axial with the disk is
postulated because of the presence of a secondary, broad
absorption in most classical, low-$z$\ BAL QSOs. It may be present
only in sources with significant black hole spin.}
\label{fig:model}
\end{figure}

Assembling the previous results with the main constraint set by
much earlier work (see, e. g., Ref. \cite{elvis00}) can lead to a
simple geometrical and kinematical model which encompasses Pop.\ A
sources as well as BAL quasars (see Fig. \ref{fig:model}).  The
full blueshift of the emission component of \civ\ (as in Fig.
\ref{fig:bal}) requires that the opening angle of the flow is $
\Theta \simlt 100^\circ$. Closer to the boundaries of the outflow
and beyond the BLR proper extends the BAL region in a conical
corona of divergence angle $\simlt 10^\circ$.  Whenever a
secondary absorption is present (Fig. \ref{fig:bal}; see also Ref.
\cite{koristaetal93} for an atlas of BAL profiles), we may observe
an axial region covering all of the continuum-emitting region and
part of the BLR as well \cite{sulenticetal05}. The depth of the
absorption implies $f_{\rm c} \simgt 0.5$. It is important to
stress that a shell of absorbing material which could provide an
adequate $f_{\rm c}$\ is not viable in the geometrical context of
our model because of the absorption relatively narrow width. Bent
flow lines seen at large viewing angle or along the flow could
give rise to NAL and BAL respectively \cite{elvis00}. They would
require an opening angle $\sim 180^\circ$, which does not seem
supported by the \civ\ profile in our sources.   While the
cylindrical sheet may appear rather {\em ad hoc}, it may have a
straightforward physical explanation if it is associated to the
axial flows which are likely present if the black hole spin is
significant (\S \ref{spin}).

\subsubsection{On the Structure of Population B Sources}

The observational structural constraints are by far more ambiguous
for Pop.\ B sources where we do not find a kinematical decoupling
between HILs and LILs \cite{dultzinhacyanetal00}. This is not to
say that a disk and wind system is not applicable, but that the
evidence in favor of a wind is less compelling, probably because
the ratio between outflow velocity and rotational velocity $v_{\rm
out}/v_{\rm rot}$\ is much lower than in Pop.\ A sources. It is
possible to fit the HIL profiles of Pop.\ B source NGC 5548 with a
wind model, with parameters consistent with reverberation mapping
results \cite{chiangmurray96}. Modest outflows ($\sim$ 1000 \kms)
are seen also in Pop.\ B sources like NGC 3783
\cite{krongoldetal03}. A mild stratification can explain why
higher ionization lines show broader profiles
\cite{marzianietal03c}: the FWHM of \heii, \hbbc, and \feii\
decreases in this order. A ``split" BLR  explains the same trend
in Pop. A \cite{bachevetal04} and why  the \heii, \hei, and \hb\
lines respond to continuum change with increasing delay in Mkn
110\ \cite{kollatschnyetal01,kollatschny03b}.

\subsection{The BLR and the Accretion Disk/Wind Paradigm}

Emission lines originating from a geometrically thin, optically
thick disk \cite{collinsouffrinetal88} will show extremely small
FWHM when observed face-on. Observed properties of CD RL sources
suggest that we need a considerable velocity dispersion in the
vertical direction to account for observed LIL widths (FWHM(\hbbc)
$\approx 3000$ \kms).  A candidate for the line emitting region
involves the outer, self-gravitating part of  the disk
\cite{collinhure01,bianzhao02}. At some distance \rs\ the disk is
expected to become gravitationally unstable and to dissolve into
individual self-gravitating clouds or rings.  For the  face-on
case, assuming orbital motion with Keplerian angular velocity
($\Omega_{\rm K}$), one can write:

$${\rm FWHM(H}\beta_{\rm BC}) \simeq \Delta v \simeq \nu \Omega_{\rm K} R_{\rm SG}, $$
where $\Delta v$ is the vertical velocity dispersion, assumed to
be proportional to the Keplerian velocity by a factor of $\nu$. A
reasonable guess for $\nu$\ is about 0.1--0.2. One can easily show
that FWHM of $\approx$ 1000 \kms\ implies an \rs $\sim$ 5000 \rg,
and 3000 \kms\ implies \rs $\sim$500 \rg. The Toomre stability
criterion applied to a standard Shakura--Sunyaev disk
\cite{shakurasunyaev73}  yields results dependent on the
assumption of the dominant source of opacity, making the \md\ and
\mbh\ dependence of \rs\ highly uncertain
\cite{bianzhao02,collinhure01}. It is however reasonable to
conclude that \rs\  can be smaller in Pop.\ B sources by a factor
$\sim 10$, and that this may leave a very ``small" emitting
surface for any standard optically-thick geometrically-thin disk.
Part of the line profile may be produced in the fragmented disk if
it is illuminated by a geometrically thick, hot inner region ({\it
Advection Dominated Accretion Flow}, ADAF, or evaporated disk
\cite{czernyetal04}).

The clumpy structure expected for $r \simgt $ \rs\ may be
resolvable through high-resolution spectroscopy. We note also that
double-peaked profiles are expected to be resolved if the outer
BLR radius is \rb\ $\simlt$ 10$^5$ \rg. However, double-peaked
line profiles from a disk can easily be turned into single-peaked
profiles by the presence of a disk wind \cite{murraychiang97}.
Although $v_{\rm out} \ll v_{\rm rot}$, the outflow velocity
gradient is as large as the rotational velocity gradient. Since
photons can escape much more easily along lines of sight with a
small projected velocity, the resulting line profiles are single
peaked with broad wings even though the emission comes from gas
that is essentially on circular orbits \cite{murraychiang97}.

%

\section{Black Hole Mass Determination \label{mass}}

\subsection{The Virial Assumption}

One can estimate the mass of the  supermassive black hole using
FWHM(\hbbc) and a reverberation BLR ``radius" \rb\
\cite{kaspietal00} along with the assumption of virialized motions
\cite{krolik01}.  The virial mass is
$$M_{\rm bh } = f \frac{r_{\rm BLR} v^2 }{ G},$$
where $G$\ is the gravitational constant. If $v =  $\ FWHM of a
suitable line, $f \approx \sqrt{3}/2$ if the orbits of the BLR gas
elements are randomly oriented.

The first question is whether the virial assumption is consistent
with the data. If the virial assumption is valid, the BLR dynamics
should be dominated by the gravity of a central point mass. In
this case  the characteristic line broadening should correlate
with the time lag for different lines. The virial relationship is
only marginally consistent with the best time-delay data
\cite{petersonetal04,krolik01}. However, emission line profiles of
Pop.\ A sources are relatively symmetric and smooth
\cite{marzianietal03a}, and the optically thick part respond in a
roughly symmetric fashion to continuum changes in 3C 390.3
\cite{shapovalovaetal01} and NGC 5548 \cite{shapovalovaetal04}. In
the following, we will discuss  the appropriateness of the virial
assumption to derive \mbh\ considering the correlations and trends
along the E1 sequence of \S\ \ref{e1}, as well as the structural
constrains on the BLR derived in \S\ \ref{blr}.

\subsection{Photoionization Mass}

The first estimates of \mbh\ were based on the rough similarity of
AGN spectra, and on the consequent assumption of constant
ionization parameter $U$\ or of constant product $U$\ne\
(\cite{wandeletal99} and references therein). A cumbersome
evaluation of the ionizing luminosity is still needed with these
methods. Even if these estimates are very rough, and the
assumption of constant $U$\ is debatable (but see Ref.
\cite{mclurejarvis02}), these studies provide a consistency check
for photoionization models and reverberation mapping results
\cite{wandeletal99}. Further refinements may be possible when the
behavior of $U$\ along the E1 sequence will be better understood
and if \feii\ can be used to constrain ionizing photon flux and
\ne\ \cite{marzianietal01,verneretal04}.

\subsection{Mass Determination  through Reverberation Mapping}

The distance of the BLR \rb\  can be deduced from reverberation data, most notably from \hbbc\ data. The  cross
correlation function between the continuum and the emission line light curve  measures a time lag $t_{\rm L}$\ between
continuum and line variations. The time lag, due to the light travel time across the broad line emitting region,
yields an estimate of  the \rb $\approx c t_{\rm L}$. It is important to stress that the derivation of \rb\ follows
from several assumptions: (1) the continuum emitting region is much smaller than the line emitting region; (2)
observable and \hi\ ionizing continuum are related. This assumption seems to hold well since the monochromatic
luminosity at UV selected wavelengths strongly correlates with \rb\ \cite{mclurejarvis02,vestergaard02}; (3) the light
travel time across the BLR is a most important parameter, in the sense that it is shorter than the dynamical time (so
that the BLR structure is not changing over the light travel time); (4) no dynamical effects of radiation are
considered. Some peculiar structures in the line profiles may however derive  from radiation pressure
\cite{marzianietal93}; (5) the line response is linear. The responsivity of the Balmer lines are generally
anticorrelated with the incident photon flux. Thus, the responsivity vary with distance within the BLR for a fixed
continuum luminosity and changes with time as the continuum source varies \cite{koristagoad04}. More technical
problems involve unevenly sampled data, errors in flux calibration, normalization of spectra (different setup,
aperture effects), and dilution by stellar continuum \cite{peterson93,horneetal04}. The coupled effects of a broad
radial emissivity distribution, an unknown angular radiation pattern of line emission, and suboptimal sampling in the
reverberation experiment can cause  systematic errors as large as a factor of 3 or more in either direction
\cite{krolik01}.

The line FWHM observed at a single epoch is not necessarily the best estimator of the gas velocity dispersion. The
highest precision measure of the virial product is obtained by using the cross-correlation function centroid (as
opposed to the cross-correlation function peak) for the time delay and the radial velocity dispersion of the variable
part of the line \cite{petersonetal98,petersonetal04}. The velocity dispersion for the variable part of the spectrum
(which is  due to portion of the BLR that is truly optically-thick) has now become available for 35 sources
\cite{petersonetal04}. With those data Peterson et al. \cite{petersonetal04} found that the random component in the
error of reverberation-based \mbh\ measurements is  typically around 30\%.

\subsubsection{Extending Reverberation Mapping Results through the
BLR Size-Luminosity Relationship \label{kaspi}}

It is now common to estimate \mbh\ by assuming that the BLR distance from the central continuum source is $\rm r_{\rm
BLR} \propto (L_{5100})^{\alpha}$ and $\alpha =$ 0.6 -- 0.7, as derived from the reverberation data by Kaspi et al.
(\cite{kaspietal00}; \S \ref{kaspi}). We therefore can write the black hole virial mass as follows: \mbh $\propto$
FWHM(\hbbc)$^2 (\lambda L_{5100})^\alpha$, where L$_{5100}$\ is the specific luminosity at 5100 \AA\ in units of
\ergss \AA$^{-1}$\ \cite{marzianietal03b}.

Any deviation from $\alpha$ = 0.6 $\div$ 0.7 has quantitative effects. The mass of luminous quasars at $z \simgt 0.4$
has been computed by extrapolating this relationship, so that it is important to stress that the relationship is based
exclusively on quasars of $z \simlt 0.4$\ and that the high (and low) luminosity ranges of the correlation are poorly
sampled. If we consider sources in the luminosity range $43.4 \simlt \log \rm L/L_\odot \simlt 45$\ (i.e. where the
sources in Ref. \cite{kaspietal00} show uniform luminosity sampling), the slope of the best fit is $\alpha$ = 0.8, and
could easily be as high as $\alpha$ = 1 without increasing significantly the fit standard deviation.  This case may be
appropriate even for the PG quasar luminosity range \cite{marzianietal01}. In dwarf active galaxies, $\alpha \approx$
0.5 seems appropriate i.e., dwarf active galaxies show larger BLRs than the values predicted by the \rb –- $\nu
L_\nu$\ relation for more luminous  AGNs \cite{wangzhang03}. One must remain open to the possibility that $0.5 \simlt
\alpha \simlt 1$, and that $\alpha$\ might even be  a function of $L$. Changing $\alpha$\ implies an $L$-dependent
change in mass estimates: the slope of the luminosity-to-mass relationship  is affected as well as the location of
points in the \lm\ vs. \mbh\ diagram in Fig. \ref{fig:ml}. If a restriction is made to the most likely $\alpha$ range,
$\alpha$ = 0.6 $\div$ 0.7, the effect on \mbh\ estimates is $\approx$ 0.2 dex \cite{mcluredunlop04}.

Summing up all sources of uncertainty, individual \mbh\ obtained
from single-epoch \hbbc\ observations and from the \rb --
luminosity correlation seems to be affected by errors as large as
a factor $2-3$\ at 1 $\sigma$ confidence level
\cite{vestergaard02,petersonetal04}.

\subsection{\mbh\ Estimation  through an Optically Thin VBLR
\label{thin}}

\begin{figure}[htbp]
\centering
\includegraphics[width=0.5\textwidth]{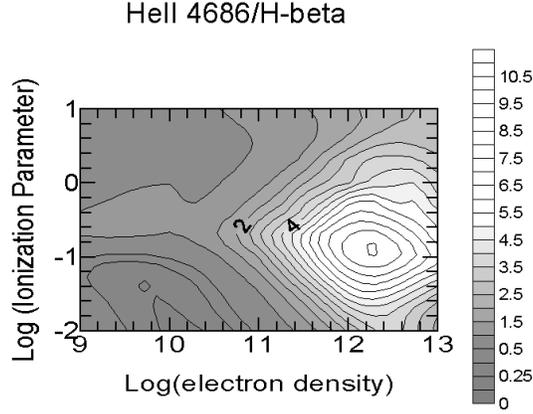}
\caption{Behavior of the intensity ratio
\heii/\hbbc\ as a function of the ionization parameter $U$ and
electron density \ne.}
\label{fig:heii}
\end{figure}

Reverberation mapping-based \mbh\ determinations are probably affected by the\linebreak non-negligible size of the BLR
that is optically thick to the Lyman continuum, so that the derived \rb\ is not a very well defined quantity. The
presence of high-velocity, optically thin line emission is likely rather common in AGNs
(\cite{shieldsetal95,sulenticetal00c} and references therein). Typical supporting evidence includes variability in the
\hbbc\ line core, coupled with the absence of variability in the line wings, or strong response in \heii\ without
change in \hbbc. Observations of  quasar PG 1416--129 revealed a large decline in its continuum luminosity over the
past 10 years \cite{sulenticetal00c}. In response to the continuum change, the ``classical'' broad component of \hb\
almost completely disappeared (the flux decreased by a factor $\approx$ 10). A redshifted  very broad component
\hbvbc\ persisted after the demise of the broad component \cite{sulenticetal00c}. In an optically thick medium the
intensity of a recombination line is governed by the luminosity of the ionizing continuum. If the medium is optically
thin the intensity of the same recombination line is governed by the volume and density of the  emitting gas and is
not directly related to the luminosity of the ionizing continuum. The \hbvbc\ luminosity can be written as:

$$L(H\beta_{\rm VBC}) = 4 \pi r_{\rm VBLR}^2 n{\rm _e}^2 h \nu_{\rm l} \alpha_{\rm
l} \Delta r, $$

where $\alpha_{\rm l}$\ and $\nu_{\rm l}$\ are the effective
recombination coefficient and the line frequency for \hb, $r_{\rm
VBLR}$\ is the distance from the central continuum source of a
shell of density \ne\ and $f_{\rm c} \approx$ 1. The unknown
$r_{\rm VBLR}$ can be  computed given the luminosity of \hbvbc\
after having inferred  \ne\ from the {\em profile} ratio in the
line wings of \heii\ and \hb\
(\cite{marzianisulentic93,koristaetal97}; see Fig. \ref{fig:heii}
for the expected dependence of the \heii/\hb\ intensity ratio).
$r_{\rm VBLR}$\ is a well-defined quantity because $\Delta r \ll
r$ if the optical thickness to the Lyman continuum is less than
unity at $n_e \gg 10^{10}$ cm$^{-3}$.  In principle, if we could
be sure that the wings of the Balmer lines are due to optically
thin gas, the determination  of $r_{\rm VBLR}$\ would be possible
even from a single line profile observation of \hbbc, if the
density and the covering factor are known. It id tempting consider
the FWZI and the luminosity of the non-variable part of the line
profile of \hbbc\ from reverberation mapping spectra. This
approach could be attempted in the near future since excellent
data are now becoming available from reverberation campaigns
\cite{petersonetal04}.

\subsubsection{Not Only Gravitational Redshift \label{red}}

The amplitude of the redward \hbbc\ asymmetry observed in Pop.\ A sources seems to be mass dependent
\cite{marzianietal03b}. A factor $\approx$6 increase in the redward displacement of the \hbbc\ centroid at 1/4 peak
intensity  [$c(1/4)$]  means that the \hbbc\ asymmetry may be due to gravitational and transverse redshift
\cite{corbin97}.   A non-Doppler shift is due to a purely gravitational term $\approx G$\mbh/\rb. A second term is due
to Doppler transverse shift, which is $\approx$ 1/$\gamma$, where $\gamma$ is the Lorentz factor. If gas motions are
virial at \rb ($\gg$ \rg), the two terms yield a shift

$$\Delta  z \approx \frac{3}{2} \frac{GM}{c^2 r_{\rm BLR}}.$$

The Pop.\ B sources considered by a recent study \cite{marzianietal03b} have been subdivided in narrow ranges of \mbh\
and \lm. In the range $3.5 \simlt \log $ \lbol/\mbh $\simlt 3.9$, the resulting  $ r_{\rm VBLR}$\ values are $\approx
0.005$\ pc and 0.01 pc for $\log M = 8$ and $\log M = 9$ respectively, if we take the $c(1/4)$\ value as a
conservative estimate of the redshift. If we also model the VBLR  as a gas shell ($ f_{\rm c} \approx$ 1) with optical
depth to the Lyman continuum $\tau \simlt 1$,   {\tt CLOUDY} \cite{ferland00} simulations show  the shell emission
falls far short in explaining the VBLR luminosity \cite{marzianietal03b}. The difference between the expected and
observed VBLR luminosity is largely a consequence of the small shell radius required to explain the large $\Delta v_r
$\ in the c(1/4).  We conclude that, even if c(1/4) is mass dependent, the $c(1/4)$\ shift amplitude cannot be
explained by gravitational and transverse redshift {\em alone} \cite{marzianietal96}.


\subsection{\mbh\ and Host Mass}

It seems that luminous (--24 $>M_{\rm V} >$  --28) quasars (both RL and RQ) are hosted in galaxies which are
spheroidal or, at least, possess large bulges \cite{floydetal04}. A correlation of nuclear black hole \mbh\ with
stellar bulge velocity dispersion $\sigma_\star$\ is now well established in nearby galaxies
\cite{ferraresemerritt00,onkenetal04}. Supermassive black holes in galactic nuclei are though to be closely related,
even in an evolutionary sense, to the bulge of the host galaxy. Reverberation-based \mbh\ estimates can be calibrated
using the the correlation between \mbh\   and $\sigma_\star$, even if indirectly (i.e., one cannot use the same
objects unless the luminosity drops so much that $\sigma_\star$\ becomes measurable; \cite{nelsonetal04}). Seyfert
galaxies (if we exclude NLSy1s, \cite{mathuretal01}) follow the same \mbh\ relation as nonactive galaxies, indicating
that reverberation mapping measurements  are consistent with those obtained using other methods \cite{onkenetal04}.
Results based on the reverberating part of \hbbc\ \cite{petersonetal04} suggest that the systematic uncertainty in the
\mbh\ is  less than a factor of 3 \cite{onkenetal04}. The relationship between \mbh\ and $\sigma_\star$\ should be
taken with care at low \mbh. NLSy1s seem to be often host  in dwarfish galaxies \cite{krongoldetal01}. Analysis of PG
quasar observations suggests a nonlinear relation between  \mbh\ and  bulge mass (\mbh $\approx M_{\rm bulge}^{1.53
\pm 0.14}$) although a linear relation cannot be ruled out \cite{laor01}. The mean \mbh/$M_{\rm bulge}$\ ratio may
 drop from ~0.5\% in bright ($M_{\rm V} \sim -22$) ellipticals to ~0.05\% in
low-luminosity ($M_{\rm V} \sim -18$) bulges (see also Ref. \cite{mathuretal01}).

Even more uncertain is the relationship between \mbh\ and bulge absolute magnitude.  Seyfert galaxies are offset from
nonactive galaxies but  the deviation can be entirely understood as a difference in bulge luminosity, and not in \mbh;
Seyfert galaxy hosts are brighter than normal galaxies for a given value of their velocity dispersion, perhaps as a
result of younger stellar populations \cite{nelsonetal04}. We indeed observe post-starburst quasars i.e., type-1 AGNs
that also display the strong Balmer jumps and high-order Balmer absorption lines characteristic of a very massive
stellar population with ages $\sim$ 100 Myr, even if they are a few percent of the quasar population
\cite{pauletal05}.

\subsubsection{Mass Determination from \oiiiopt}

In high-luminosity quasars, the relationship between \mbh\ and $\sigma_\star$\ can be studied comparing  \mbh\ derived
from the width of the broad \hbbc\ line and \mbh\ from the width of the narrow \oiiiopt\ lines used as  a proxy to
measure $\sigma_\star$. RQ AGNs seem to conform to the established \mbh-$\sigma_\star$ relationship up to values of
\mbh $\sim 10^{10}$  \msol, with no discernible change in the relationship out to  $z \approx 3$\
\cite{shieldsetal03}. There are two major difficulties here. Even if an \mbh\ –- FWHM(\oiii) correlation is present
\cite{nelson00}, scatter is large. FWHM(\oiii) measures may not always provide a way to estimate \mbh. \mbh\ values
from \oiiiopt\ can be considerably higher than values calculated using FWHM(\hbbc) \cite{marzianietal03b}. This is not
surprising because NLSy1-type AGNs apparently do not follow the same relationships as other type-1 AGNs
\cite{mathuretal01}. It is important to stress that low-W(\oiiiopt) sources can have FWHM(\oiiiopt) $\simgt$
FWHM(\hb), invalidating the virial assumption. Blueshifted \oiiiopt\ emission arises in outflowing gas possibly
associated to a disk wind \cite{zamanovetal02}. The NLR in blue outliers may be very compact and its velocity field is
not likely to be dynamically related to the host galaxy stellar bulge. This points to a limiting W(\oiii) below which
FWHM(\oiii) ceases to be a useful mass estimator. Only large W(\oiii) RQ sources in Pop.\ B may have very extended NLR
whose motions is dominated by gravity due to bulge stars.

\subsection{Masses of ``Special" Sources}

\subsubsection{``Double Peakers"}

A small fraction of AGNs exhibits exceptionally broad, double peaked LILs. The \habc\ emission line profile is
strikingly peculiar (see Ref. \cite{sulenticetal00a} for a few typical examples, or recent surveys such as the ones of
Refs. \cite{eracleoushalpern03,stratevaetal03}). AGNs with double peaked LILs remain  relatively rare specimens, even
if the SDSS has allowed to identify $\sim 100$\  sources: $\approx$ 4\%\ (with many dubious cases) of all RQ and RL
SDSS sources \cite{stratevaetal03}. Although predominantly RQ, they are more likely to be found in RL sources, and
account for about 20\%\ of RL AGNs \cite{stratevaetal03,eracleoushalpern03}. Their relative rarity and their
exceptionally broad lines prompted workers to search for a specificity either in terms of peculiar views or physics
soon after the discovery of the prototypical source Arp 102B \cite{sulenticetal90}.

Prototype ``double-peakers''  Arp 102B, 3C 390.3, and 3C 332 have been now monitored form more than 20 years
\cite{lewisetal04}. A common property of double-peaked  lines  is slow, systematic variability of the profile shape on
a timescale of years i.e., the timescale of dynamical changes in  accretion disks
\cite{eracleoushalpern03,newmanetal97}. Lower limits on the plausible orbital periods from the absence of peak radial
velocity changes would require supermassive binary black holes \cite{gaskell96} with total masses in excess of
10$^{10}$ \msol. Such large \mbh\ values are difficult to reconcile with a maximum expected limit \mbh $\sim 3 \cdot
10^9$ \msol\ (\S \ref{big}; \cite{eracleousetal97}), and with the known \mbh\ -- host bulge mass relationship. Other
recent results from a five year monitoring  of  \hb\  further support the dismissal of the binary black hole
hypothesis for 3C390.3 on the basis of the masses required \cite{shapovalovaetal01}. Hot spots, spiral waves and
elliptical accretion disks have been recently favored \cite{lewisetal04,storchibergmannetal03}. If the hot spot lies
within the accretion disk, it is possible to estimate \mbh\ and also the BLR physical size. The determination of \rb\
in physical units removes the degeneracy due to the rotational velocity field (i.e., the velocity scales as (\mbh /
$r$)$^{-0.5}$; disk model profiles yield a distance normalized to \rg). \mbh\ $\approx ( 2.2 \pm 0.7 ) \cdot 10^8$
\msol\ was inferred from the period and the distance of the hot spot in Arp 102B ($4.8 \cdot 10^{-3}$ pc;
\cite{newmanetal97}). It interesting to note that, within the framework of the model,  \mbh/$r \approx 3 \cdot
10^{25}$ g cm$^{-1}$\ is still below the formal requirement for a black hole but  the implied density $\rho \simgt
10^{14}$ \msol pc$^{-3}$\ is a stringent limit. The \mbh\ in double peakers can be also estimated using the empirical
relationship between \rb\ and optical continuum luminosity \cite{kaspietal00}.  \mbh\ and \md\ computations have been
carried out  for 135 objects from the SDSS \cite{wuliu04} and from a survey of RL broad emission line AGNs
\cite{eracleoushalpern03}. \mbh\ values range from 3 $\cdot 10^7$ \msol\ to 5 $\cdot 10^9$ \msol, and \md\ is between
0.001 and 0.1. Double peakers are found up to $L \sim 10^{46}$ \ergss.

\subsubsection{Black Hole Binaries}

Several observational  properties of extragalactic radio  jets, such  as   bending,  misaligment,  and wiggling (often
associated with  knots  superluminally moving along  different-scale curved trajectory)  have  been interpreted in
terms  of  helical structures of the jets.  This structure is likely caused by the precession of the jet in a binary
black hole system \cite{camenzindandcrockenberger92,caproniabraham04,conwayetal95,ostoreroetal04,villataetal99}. In
the  case of OJ 287 the bending of the VLBI jet was reported in Ref. \cite{vicenteetal95}. A very small change in the
orientation of the jet  is needed  in  order to change the Doppler boosting dramatically, thereby producing long-term
periodic brightness modulations. OJ287 was first recognized as a candidate binary black hole system by noticing the
regularly spaced outburst pattern in its historical optical light curve \cite{sillanpaaetal88}.  Photographic
information on the brightness of this blazar (thought to be  a variable  star) extends  for about  100 years. Even
though the observations were scanty in the beginning, there  still was a convincing pattern which led to  the
prediction  of the next major  outburst in OJ287 in the fall  of 1994 .  For the first time a  predicted cyclic
phenomenon was indeed observed in an extragalactic object \cite{sillanpaaetal96}. Sillanp\aa\aa\ et al.
\cite{sillanpaaetal88} modelled the periodic outbursts using a binary system consisting of a pair of supermassive
black holes with an orbital period of 8.9 yr (in the rest frame of OJ 287). The light variations in this model are
related to tidally induced mass flows  from the accretion disks into the black holes \cite{kidgeretal92}. A variant of
this model  explains and predicts other features of the observed light curve \cite{lehtovaltonen96}. The variant
allows for relativistic precession in the binary black hole system, and associates the major flares with the times at
which the secondary black hole crosses the plane of the disk of the primary. The planes of the disks are perpendicular
to each other. The monitoring campaign organized to observe OJ 287 in 1994 gave such a detailed light curve that a
unique determination  of the orbital parameters was possible. The model gave masses of 10$^8$ \msol\ and 1.7 $\cdot$
10$^9$ \msol\ for the secondary and primary black hole respectively, and also predicted an eclipse of the secondary
black hole disk by the disk of the primary in 1989. The eclipse was indeed observed
\cite{valtonenetal99,pursimoetal00} in the optical but not in the radio, as expected.  The observation of the next
predicted flares for 2006 or  2007 (the date depends on the model) will allow  to refine models and make a direct
measurement of the orbital energy loss in the system.

\subsection{At Low and High $z$: \mbh\ Determination from \civ}

\begin{figure}
\hspace{-0.5cm} \epsfxsize=5cm \epsfysize=5cm \epsfbox{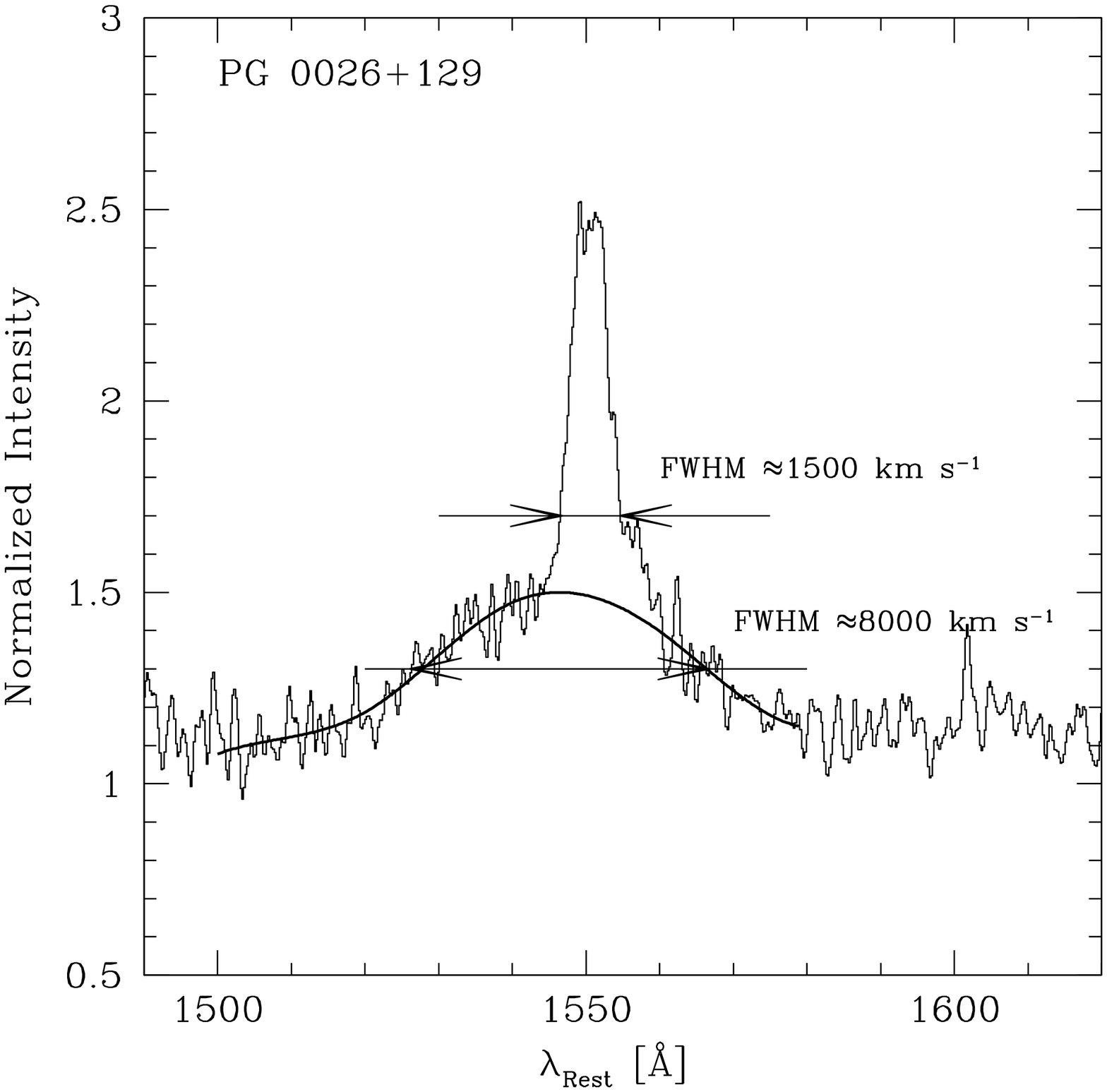}
\epsfxsize=5cm \epsfysize=5cm \epsfbox{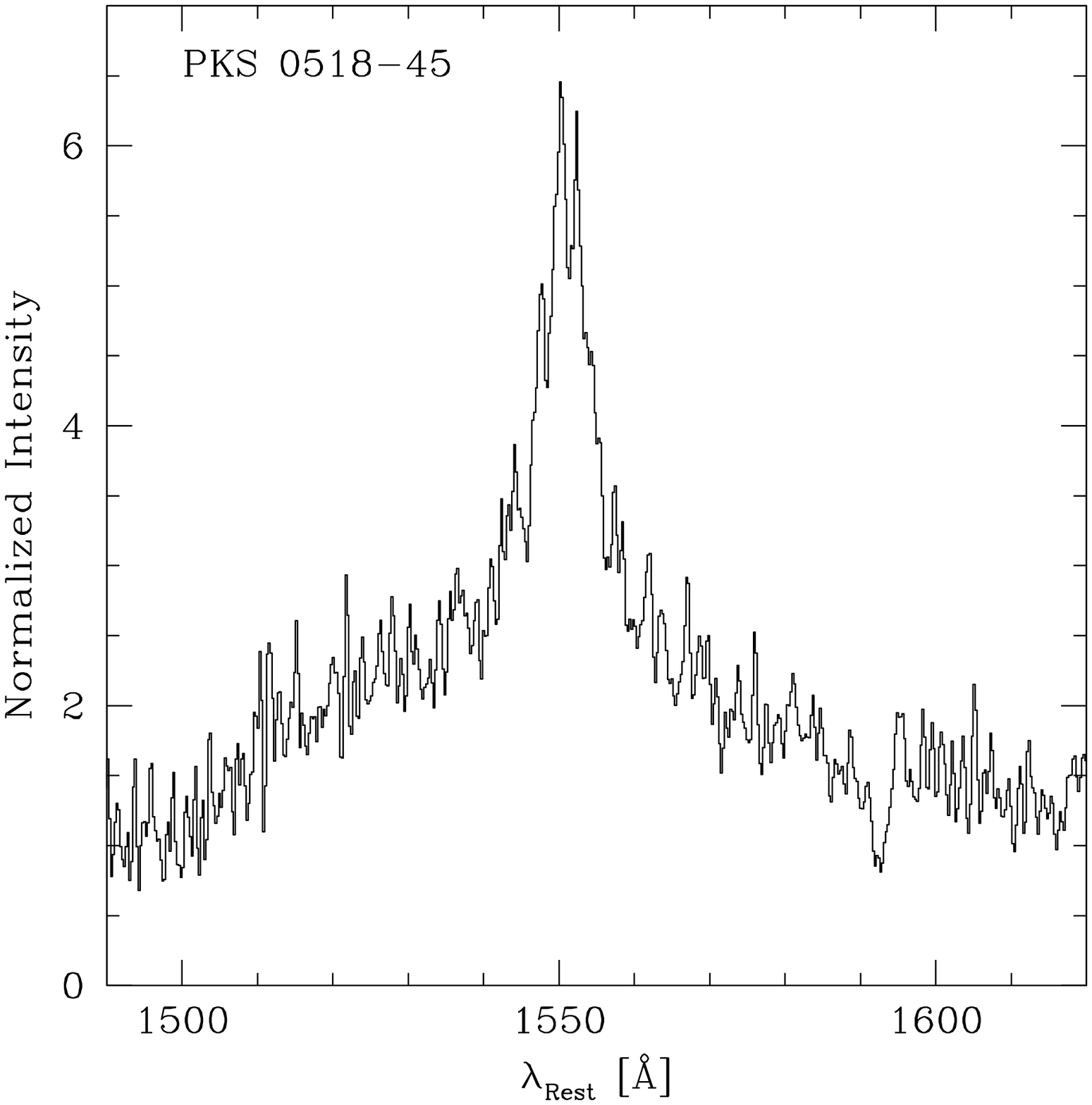}\epsfxsize=5cm
\epsfysize=5cm \epsfbox{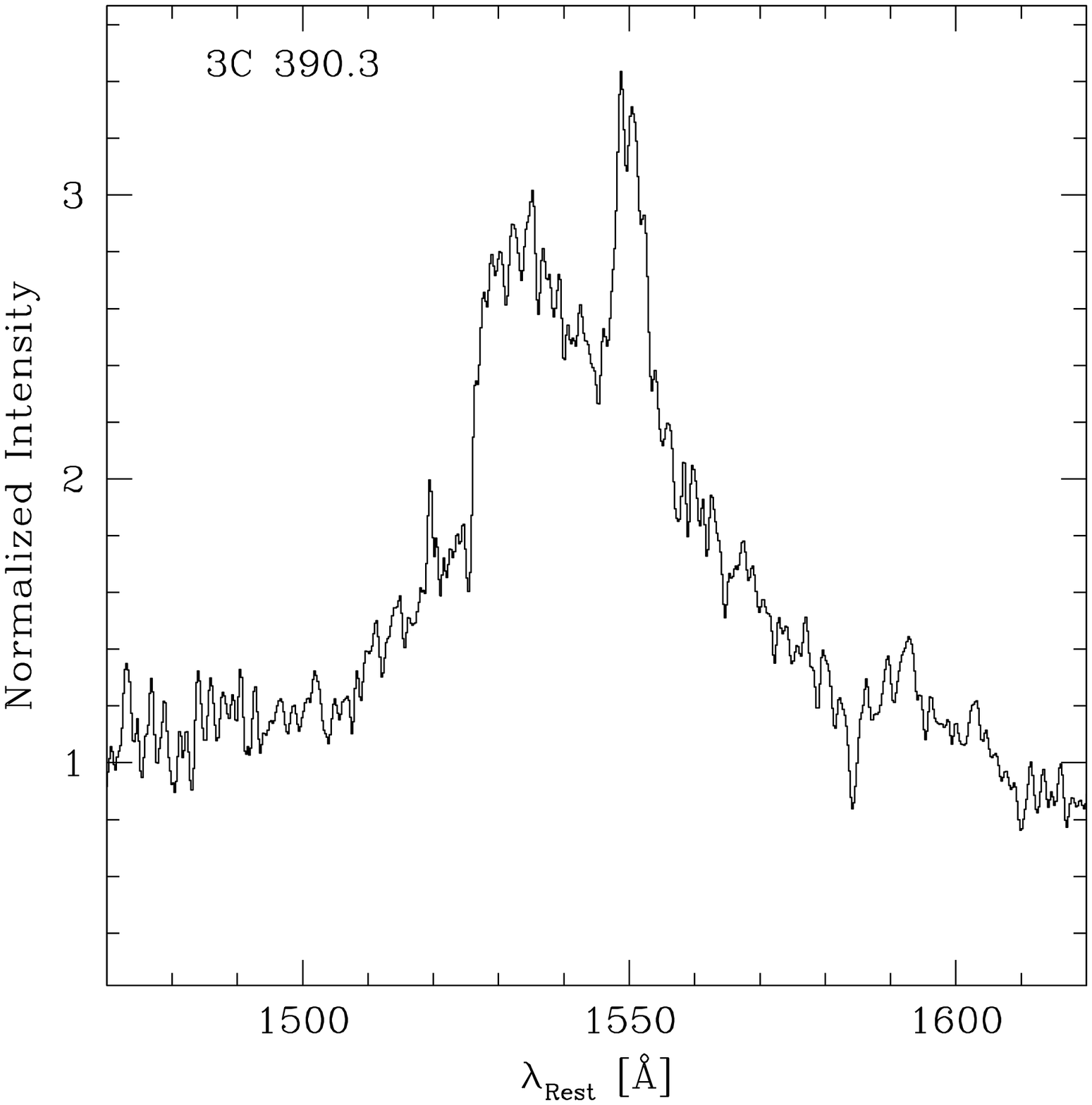} \caption{Examples of sources
with unambiguous \civnc\ emission. The left panel shows two
FWHM(\civ) measurements: one for the whole \civ\ profile, and the
second one for the \civbc\ (thick solid line) only. Neglecting
\civnc\ results in large FWHM(\civ), and hence \mbh\ measurement
errors. The middle and the right panel show two radio loud sources
with very high W(\oiiiopt) and widely different W(\civnc).
\label{fig:civnc}}
\end{figure}

\subsubsection{Existence of a \civnc\ Contribution \label{civnc}}

The very existence of significant \civ\ narrow emission (\civnc) has been a contentious issue. There is plenty of
evidence that at least some AGNs with strong narrow lines have a strong \civnc. Perhaps two of the most striking
examples are PG 0026+12 (RQ; a borderline NLSy1 with strong NLR emission) and Pictor A (RL; Fig. \ref{fig:civnc}).
These are two cases in which an obvious inflection between broad and narrow component is seen, and in which a \civnc\
can be easily isolated with FWHM$\simgt$1000 \kms. The visual impression (already suggested in Ref.
\cite{marzianietal96}) of a connection between \oiiiopt\ and  the \civ\  narrow core is confirmed by a loose
correlation between W(\oiiiopt) and W(\civnc) which we found considering {\em only} the cases of  least ambiguous
inflection. The data dispersion is probably intrinsic: at the two \civnc\ extremes of large \oiiiopt\ we find Pictor A
(W(\civnc) $\approx$ 30 $\pm 6$ \AA) and 3C 390.3 (W(\civnc) $\approx$ 5 $\pm 1$ \AA) with W(\oiii)$\sim$ 80 \AA\ (see
Fig. \ref{fig:civnc}). We do not expect a strong correlation physically, since emission of \civ\ is favored in the
innermost regions of the NLR, as shown below.

Even if RL sources often show prominent \civ\ cores of width
$\sim$ 2000 \kms, such emission has been ascribed to \civbc\
\cite{willsetal93} and to an intermediate line region
\cite{brothertonetal94b}. It is relatively na\"\i ve to believe
that \oiiiopt\ and \civnc\ should have the same width.
Collisionally excited lines such as \civ\ show a band of efficient
reprocessing running at constant $U$\ along a diagonal ridge in
the plane photon flux versus \ne. Emission is locally optimized
along this ridge which corresponds to $\log U \sim -1.5$ for \civ\
\cite{baldwinetal95}. Provided that a smooth density gradient
exists (i.e., that there is a significant amount of gas at \ne\
$\sim$10$^6$ \cm3), we expect \civ\ emission to be strong at
larger \ne\ (\civ\ emissivity is $\propto n_{\rm e}^2$ while
\oiiiopt\ becomes collisionally quenched). If velocity dispersion
increases with decreasing distance from the central continuum
source it is somewhat natural to expect FWHM(\civnc) $>$
FWHM(\oiiiopt). Detailed calculations
 based on these simple considerations indeed suggest  that 1000 \kms\ $\simlt$
FWHM(\civnc) $\simlt$ 2000 \kms\ for a variety of possible density-gradient laws \cite{netzer90,sulenticmarziani99}.
The same models  correctly predict that FWHM(\hbnc)$\sim$ 500 \kms $\simgt$ FWHM(\oiiiopt) $\ll$ FWHM(\civnc).  To
ascribe \civ\ core emission to the BLR (where \ne$\simgt 10^{10}$\cm3) seems  rather arbitrary. Since \civnc\ is
prominent in Pop.\ B sources but can be absent in Pop A., the wind that gives rise to the blue-shifted \civbc\
component may serve as a filter to the FUV ionizing radiation from the center that would otherwise reach the outer
self-gravitating parts of the disk, where \civnc\ may originate \cite{leighly04}.

\subsubsection{Is the Width of \civbc\ a Reliable \mbh\ Estimator?}

Estimating \mbh\ from the width of \civ\ is cumbersome. Failure to account for \civnc\ has dramatic consequences on
\mbh\ and Eddington ratio estimates (\S \ref{edd}). Dynamically, the HIL emitting gas is not in virial equilibrium at
least in a significant fraction of quasars (\S \ref{blr}; Fig. \ref{fig:civshift}). Assuming  virial equilibrium  for
Pop.\ B (as suggested by the absence of systematic line shifts and by the tentative similarity with the \hbbc\
profile) still leaves the problem of properly assessing  \civnc. Recent studies of the \civ\ profile using HST
archival spectra (\cite{baskinlaor04b,kuraszkiewiczetal02,kuraszkiewiczetal04,vestergaard02,warneretal04}) have
subtracted little or no \hbnc\ from the \civ\ line profiles. A comparison of  line widths and shift measures for \hb\
and \civ\ \cite{baskinlaor04b,warneretal04} shows that there are significant and systematic differences. An apparent
dichotomy occurs at FWHM(\hbbc)$\approx$ 4000 \kms\ if no \civnc\ is considered \cite{baskinlaor04b}: below 4000 \kms,
the \civ\ line is broader than \hbbc, but the reverse seems to hold when FWHM(\hbbc) $\simgt$ 4000 \kms\ where we
believe  that the NLR becomes important (Pop. B; \cite{bachevetal04}). This result is not necessarily  against the
view that \civ\ generally originates closer to the center than \hb\ since \civnc\ was not subtracted. In the case of
PG 0026+129, failure to subtract the NLR emission  yields spurious broad line parameter measures (e.g. FWHM $\sim$
8000 \kms\ instead of 1860 \kms; see Fig. \ref{fig:civnc}). We conclude that: (a) \mbh\ computations from \civ\ width
are wrong in the case the \civ\ profile shows large blueshifts i.e., for Pop.\ A sources; (b) failure to properly
correct for \civnc\ yields a large error in the mass estimate \cite{bachevetal04}. The \mbh\ error is especially large
for Pop. B sources; however, it must be remarked that \civnc\ can be strong also in Pop.\ A RQ sources, and that a
case-by-case analysis should be made.

\subsection{High-$z$\ Masses: Are They Really That Big? \label{big}}

If the empirical relationship between \rb\ and the source luminosity is used to obtain  \mbh\ at high $z$, the largest
\mbh\ are $\simgt 10^{10}$ \msol. Such values of \mbh\ suggest bulge mass $\simgt$ 10$^{13}$ \msol\ and $\sigma_\star
\simgt$\  700 \kms\ which are not observed at low $z$. Black holes with \mbh\ $ \simgt$ $ 3 \cdot 10^9$ \msol\ should
reside almost exclusively in high-redshift quasars \cite{netzer03}. This implication is in contrast with the
expectation that \mbh\ can only grow with cosmic time on timescales of the Universe present age, and inconsistent with
several suggested scenarios of black hole and galaxy formation. A solution of this dilemma may reside in the improper
use of \civ\ as a proxy of \hb\ to estimate \mbh.  A very good correlation has been recently found between \rb\ and
the specific continuum luminosity at 3000 \AA\  \cite{mclurejarvis02}. In a sample of objects with broad-line radii
determined from reverberation mapping the FWHM of \mgii\ and \hb\ are consistent with an exact one-to-one relation, as
expected if both \hbbc\ and \mgii\ are predominantly emitted at the same distance from the central ionizing source
within a factor of 2.5 \cite{mcluredunlop04,mclurejarvis02}.  The \mgii\ line seems to be systematically narrower than
\hbbc\ probably because \mgii\ (like \feii\ and the reverberating part of \hbbc) is emitted in the most optically
thick portion of the BLR. The FWHM(\mgii) can then be  used to estimate \mbh\ of quasars $0.25 \simlt z \simlt 2.5$\
via optical spectroscopy alone, and  it is in principle  even preferable to  FWHM(\hbbc). There are no strong
theoretical reasons or empirical evidence to doubt the assumption that the \mgii\ line broadening is virial, at least
no more than the ones known for \hbbc. FWHM(\mgii) typically indicates a factor of 5 times lower \mbh\ than \civ\
\cite{dietrichhamann04}.   It is interesting to point out that a mass   of 3 $\cdot 10^9$ \msol\ for the central black
hole in the  $z = 6.41$ quasar SDSS J114816.64+525150 (one of the most distant quasars known) has been estimated
applying the \mbh\ technique appropriate for a detected \mgii\  FWHM of 6000 \kms\ \cite{willottetal03}. This very
high luminosity quasar does not show extremely broad lines and does not require super-Eddington luminosity.

 Measurements of \hbbc\ at high $z \simgt 1$ can be achieved only through IR spectrometers. Instrumental capabilities lag behind the ones of
optical spectrometers, and lack of resolution as well as poor $S/N$\ can lead to huge overestimates of FWHM(\hbbc)
\cite{mcintoshetal99}, although excellent data for a few tens of sources are becoming available \cite{sulenticetal04}.

\subsection{Mass Estimates Along the E1 Sequence}

The previous results suggest a number of general remarks.

\begin{itemize}

\item The most accurate \mbh\ determinations are probably obtained
considering the reverberating part of emission lines
\cite{petersonetal04}. If \mbh\ estimates from the reverberating
component of \hbbc\  are compared to \civ\ \mbh\ estimates,
offsets of 1.0 dex or larger from a perfect one-to-one
relationship are possible \cite{vestergaard02}. \mbh\ derived from
\civbc\ has a probability of just 90\% to be within a factor 10
from the best reverberation \mbh\ estimates. For the \hb\
single-epoch  estimates, the probability  of getting \mbh\
accurate to within an order of magnitude is $\approx$ 95 \%, which
suggests that systematic sources of uncertainties and ambiguities
operate in the estimate of \mbh.

\item If an innermost region of optically thin (to the Lyman continuum) gas can be identified from the profile of \hb,
an estimate of the line emitting region size is relatively straightforward (\S \ref{thin}). However, the frequent
occurrence of asymmetries partly invalidates the virial assumption for the innermost line emitting regions in Pop.  B
sources.

\item Single epoch, single line \mbh\ determinations based on the
FWHM are valid in a statistical sense if the line is a LIL, like
\hb\ or \mgii. An estimation of the aspect angle $\theta$\ is a
first-hand necessity, as farther outlined below.

\item It is improper to use \civ\ for \mbh\ calculations is the
line is blueshifted or strongly asymmetric, as in Pop. A sources.
This seems to be the case of most high-$z$\ objects
\cite{richardsetal02}. The \mgii\ line may provide the best
estimator at high $z$.

\item Some special sources, like double peakers and black hole binaries may provide accurate \mbh, although  \rb\ and
\mbh\ determinations are still somewhat model dependent. Other sources -- like BAL QSOs -- may be favored if the
orientation angle can be constrained.

\item \oiiiopt\ may provide a suitable \mbh\ estimator only in RQ
sources, predominantly for spectral types A1 and Pop. B, if the
recessional velocity from\linebreak \oiiiopt\ agrees with the one deduced
from other narrow lines which are of low ionization.

\end{itemize}

\subsubsection{Orientation Effects}


As reviewed in the previous sections, there is growing evidence that the LILs are predominantly emitted in a flattened
system. LIL emission may be dominated by an extended, optically thick disk in Pop.\ A, while a fragmented disk may be
the main line emitter in Pop.\ B sources. Inclination relative to a flattened system causes a systematic underestimate
of the central mass. This effect is expected to be present in both RQ and RL sources. Considering the vertical
velocity dispersion for the line emitting gas at $\theta \rightarrow 0^\circ$ FWHM$(\theta=0)$, a suitable
parameterization could be $FWHM(\theta) \approx FWHM(\theta=0) + \Delta FWHM \cdot \sin \theta$, where
$FWHM(\theta=0)\approx$ 1000 \kms\ and 3000 \kms\ for RQ and RL sources respectively. $\Delta FWHM$\ can be estimated
from the FWHM  range from CD to LD sources (\cite{marzianietal01,sulenticetal03}; a similar approach has been followed
in Ref. \cite{mcluredunlop02}). Considering that $\theta$ may be in the range $few ^\circ \simlt \theta \simlt
45^\circ$, and distributed randomly, we have an average $<\theta> \approx 30^\circ$. This means that ignoring
orientation effects leads to a systematic underestimate of \mbh\ by $\Delta \log$ \mbh $\approx$ 0.6. The
underestimate may be a factor $\simgt$ 10 if the source is observed almost pole-on. Even if pole-on sources are the
rarest one in a randomly-oriented sample, errors of this amplitude may dramatically influence inferences on \mbh\ and
blur correlations involving \lm\ (which varies only by 2 -- 3 orders of magnitude in AGNs; \S \ref{edd}). It is also
important to consider that optical luminosity may not be appropriate in the \rb\ -- luminosity relationship for CD RL
AGNs ($\theta \rightarrow 0^\circ$) because
 jets may significantly contribute  to the optical continuum. A
relativistically boosted continuum  leads to an overestimation of \mbh. In such cases, it may be better to consider an
empirical relation between \rb\ and the \hbbc\
 luminosity \cite{wuetal04}.

\section{Black Hole Spin \&\ Observer's Orientation \label{spin}}

\subsection{Where Are Spinning Black Holes?}

The idea of a non rotating black hole is rather distressing because every known massive object in the universe (from
asteroids to neutron stars) does rotate. Conservation of angular momentum by the accreting matter should lead quickly
to maximally rotating black holes \cite{thorne74}. While \mbh\ is destined to increase on timescales comparable to the
present age of the Universe, a black hole angular momentum  can reverse and decrease through a diversity of mechanisms
\cite{gammieetal04,hughesblandford03}: the collapse of massive gas accumulations, gas accretion, capture of stellar
mass bodies, and successive mergers with other massive holes. A significant black hole specific angular momentum $a$\
is required for driving relativistic, radio-emitting jets \cite{blandfordznajek77} although a maximally rotating black
hole ($a/M \approx0.998 G/c$) is not necessary. In the E1 context, suggestions of  $ a \neq$ 0 come from time scale
variability arguments, especially from soft-X ray observations of NLSy1 sources. A variation in luminosity cannot
occur on an arbitrarily short time scale, a minimum physical limit being set by the light crossing time of the
emitting region. The emitting matter should have some opacity $\tau \approx n \sigma r \sim 1$, where $\sigma$\ is the
Thomson scattering cross-section. The time needed for radiation to diffuse out of a region of size $r$\ is therefore
$\Delta t \approx (1+\tau) r / c$ \cite{fabian79}. The change in luminosity $\Delta L$\ following an accretion event
with accretion rate $\Delta M/ \Delta t$ is $\Delta L = \eta \Delta M/\Delta t c^2$. Writing $\Delta M = \frac{4}{3}
\pi R^3 n m_{\rm p}$ where $n = \tau /\sigma R$\ and $m_{\rm p}$\ is the proton mass, and considering the travel time,
we have:

$$ \frac{\Delta L}{\Delta t} \lsim \frac{4}{3} \pi \frac{m_{\rm
p}}{\sigma} c^4 \eta \approx 2 \cdot 10^{41} \eta_{0.1} {\rm ergs~
s^{-1}}.$$

If changes of very large amplitude occur in such a short timescale
that the relation above is violated for $\eta \approx 0.1$, claims
have been laid that a rotating black hole is needed. This has been
the case for a handful of NLSy1 galaxies \cite{forsterhalpern96}.
In the view that some NLSy1s and at least some other Pop. A
sources are young/rejuvenated quasars
\cite{mathur00,sulenticetal00a,grupe04}, they may have experienced
one of few accretion events leading to a consistent increase of
the black hole angular momentum.

Two-fluid models for relativistic outflows include a fast,
relativistic beam  surrounded by a slower, possibly thermal,
outflow, with a mixing layer forming between the beam and the jet
(e.g., \cite{lobanovroland04}).  The mixing layer could  produce
the secondary absorption seen in several BAL sources (Fig.
\ref{fig:bal}). It is interesting to stress that the observed
\rfe\ for the BAL QSOs Mkn 231 and IRAS 07598+6508 would require
highly super-Eddington accretion unless the accreting object is a
rotating black hole. Another aspect favoring a maximally rotating
black hole is the remarkable behavior of the Fe K$\alpha$\
emission line in MCG --06--30--15 \cite{sulenticetal98}. This
source is pretty unremarkable optically; it is a Pop.\ A source
with modest \feiiopt\ emission. More circumstantial evidence
favoring rotating black holes is provided by the mass density of
supermassive black holes in the present-day universe, which agrees
with the quasar luminosity evolution and total power output if the
efficiency has been $\eta \simgt 0.1$, larger than that of a
non-rotating black hole \cite{haimanetal04}.

It is then intriguing that, at present, neither black hole spin nor morphology seem to be sufficient conditions to
explain radio loudness \cite{dunlopetal03}, disfavoring the idea that the RL and RQ dichotomy may be due to
systematically different black hole spins  \cite{wilsoncolbert95}. The physics of radio emission in the inner regions
of all quasars (RL and RQ) may be essentially the same, involving a compact, partially opaque core together with a
beamed jet \cite{barvainisetal04}. Other factors may be at play: a strong electromagnetic field, coupled with
differential rotation, could serve to convert rotational energy into kinetic energy of outflows \cite{camenzind04}. In
the context of optical/UV observations such effects are at present observationally unconfirmed, also because radio
loudness does not seem to have appreciable effects on the observed broad line spectrum \cite{marzianietal03b}. In
principle, we could expect that, if a viscous accretion disk is inclined in relation to the equatorial plane of a
rotating black hole, the differential precession will produce warps in the disk. The combined action of the
Lense-Thirring effect and the internal viscosity of the accretion disk forces the alignment between the angular
momenta of the rotating black hole and the accretion disk at $r \sim 1000$ \rg\ $\sim$ \rb. Some double peaked
emission line profiles may be produced by warped disks. The agreement between some peculiar \habc\ profiles and a toy
model is very good \cite{calvanietal97,bachev99}, but many other models may also be applicable without invoking a
warped disk \cite{storchibergmannetal03}.


\subsection{Can we Measure an Orientation Angle? \label{theta}}

\begin{figure}[htbp]
\centering
\includegraphics[width=0.5\textwidth]{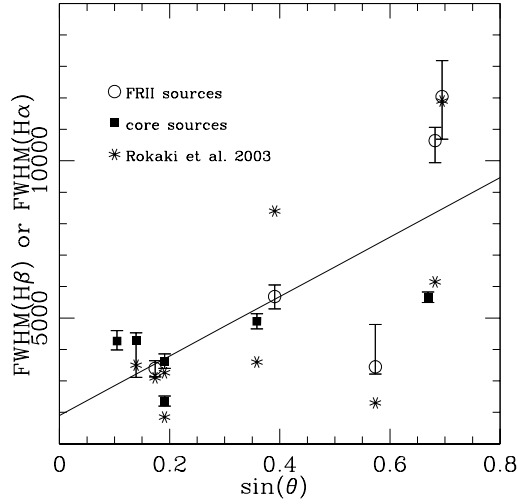}
\caption{Dependence of FWHM(\hbbc) on orientation angle $\theta$
for superluminal sources \cite{sulenticetal03} with data from
Refs. \cite{rokakietal03,marzianietal03b}.} \label{fig:orient}
\end{figure}

In radio-loud quasars, where a rough orientation indicator is available from the radio core-to-lobe ratio, there is
good evidence that  FWHM(\hbbc) shows an orientation dependence consistent with a flattened (or even disk-like)
geometry \cite{willsbrowne86,brotherton96}. There is a domain space separation between Fanaroff-Riley II (i.e., LD)
and CD sources in E1. The diagonal arrow in figure \ref{fig:e1radio} indicates the average  change in E1 position due
to change in orientation from LD to CD sources. As stressed earlier, \mbh\ can be underestimated if no correction is
applied to observed LIL widths \cite{oshlacketal02}. An analysis of the connection between \mbh\ and radio luminosity
in radio-selected flat-spectrum quasars shows that values of \mbh\ are not systematically lower than those of luminous
optically selected RL quasars, if a proper correction for the effects of inclination is applied
\cite{jarvismaclure02}.   It is also unlikely that there are sources radiating at \lledd $> 1 $ at $z \simlt 0.8$ (see
Fig. \ref{fig:blueoutliers}; \cite{marzianietal03b}) if a reasonable orientation correction is applied to line widths.
While a correction for the effect of varying $\theta$\ is feasible in a statistical sense, measuring $\theta$\ on a
source-by-source basis has proved to be very difficult.  Superluminal radio sources are an exception.   Radio
observations allow an estimate of the expected synchrotron self Compton flux relative to the observed X-ray flux.
Lorentz factor and $\theta$ can then be independently estimated \cite{ghisellinietal93}. Figure \ref{fig:orient} shows
$\theta$\ vs. FWHM(\hbbc)  for 11 superluminal sources, confirming a factor $\approx$ 3 effect on the \hbbc\ width at
extreme $\theta$\ values.

The observational properties of  some, extreme NLSy1s  may well be due to a pole-on orientation angle. A NLSy1 nucleus
has been found in the extensively studied eruptive BL Lac, 0846+51W1, out of a large sample of NLSy1  from the SDSS
\cite{zhouetal04}. The SDSS allowed the identification of another source with typical NLSy1 properties (strong
\feiiopt, undetectable \oiiiopt) that is definitely RL (\rk $\simgt$ 1000). The inverted radio spectrum and the very
high brightness temperature  derived from variation of the radio flux suggested the presence of a relativistic jet
beaming toward the observer \cite{zhouetal03}. It is  unclear whether this finding can be extended to (at least) some
RQ NLSy1s. Some NLSy1 might be considered the radio quiet equivalent of BL Lacs if they give rise to radio-silent
relativistic outflow. Even if evidence is accumulating about the existence of a relativistic jet in RQ sources (e.g.,
\cite{barvainisetal04}), RQ sources (especially if of Pop. A) are not optically violently variable. Although some
NLSy1 sources (those with largest \oiiiopt\ and \civbc\ blueshifts, as well as narrowest \hbbc\ and most strongly
variable soft X-ray excess) are likely to be oriented almost pole-on, it is still not clear whether we are observing a
beaming effect.

Orientation estimates for RQ sources may come from the correlation between \gs\ and soft-X ray variability amplitude,
although the poor understanding of the soft X-ray excess prevents the definition of an orientation indicator. Other
attempts at estimating $\theta$\  are rather model-dependent. A sample of several thousand quasars from the SDSS
confirms that HILs such \civ\ are significantly blueshifted with respect to LILs such as \mgii\ \cite{richardsetal02}.
Among the SDSS sources, \civ\ emission-line peaks have a range of shifts from a redshift of 500 \kms\ to blueshifts
well in excess of 2000 \kms\  compared to \mgii. The  anticorrelation between the shift of the \civ\ emission-line
peak and W(\civ) (Fig. \ref{fig:civshift}) is confirmed by the SDSS study. Composite quasar spectra as a function of
\civ\ shift suggest that the apparent shift of the \civ\ emission-line peak is not a shift so much as it is a lack of
flux in the red wing for the composite with the largest apparent shift \cite{richardsetal02}. An outflowing wind and
an optically thick disk  to act as a screen provide a tempting explanation. However, it is unlikely that purely
geometrical effects are at play in determining the width and shifts of \hbbc\ and \civbc\ alike \cite{marzianietal01},
even if a simple toy model shows that \civ\ shift amplitude is orientation dependent
 with the largest shifts  being produced by $\theta  \rightarrow 0^\circ$\ \cite{zamanovetal02}. W(\civ) is strongly
dependent on \lm\ (\cite{bachevetal04,baskinlaor04b}; see \S
\ref{beevol}),  as are the properties of  outflowing winds
\cite{murraychiang97,progaetal00}. To estimate $\theta$\ on a
source by source basis, we first need to understand how W(\civ)
and \civ\ shifts are affected by \lledd\ and $\theta$.

\begin{center}
\begin{figure}
\hspace{1.cm}\epsfxsize=10.5cm\epsfbox{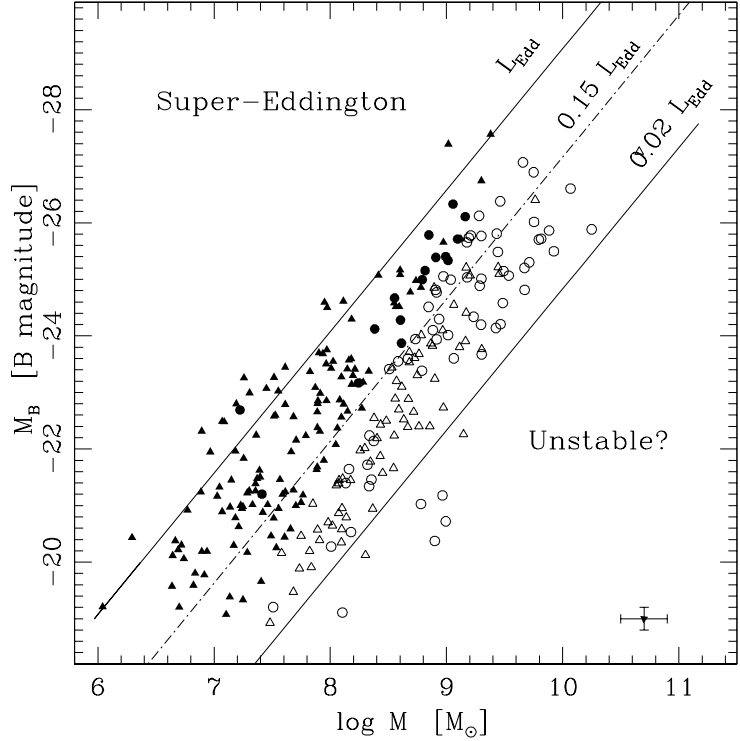} \caption{Mass
Luminosity relationship for a sample of $\approx$270 AGNs
\cite{marzianietal03b}.  Triangles and circles indicate
radio-quiet and radio-loud sources, respectively. Open symbols
indicate sources with FWHM(\hbbc) $\simgt$ 4000 \kms\ (Pop.\ B);
filled symbols sources with FWHM(\hbbc) $\simlt$ 4000 \kms\ (Pop.
A). The bars in the lower right corner indicate typical errors.
\label{fig:ml}}
\end{figure}
\end{center}

We mention here an interesting attempt that is, again, strongly
dependent on specific assumptions. Using the variable fraction of
the \hbbc, \hei, and \heii\ profiles to estimate \rb, and then
comparing the virial \mbh\ and the \mbh\ obtained assuming that
the variable-fraction shifts observed in Mkn 110 are due to
gravitational redshift, a measurement of the spin axis of the
central black hole is possible \cite{kollatschny03a}: $\theta
\approx 21 \pm 5^\circ$. Unfortunately, there is as yet no
convincing general evidence that the redward asymmetry seen in
\hbbc\   profiles is entirely due to gravitational redshift (\S
\ref{red}).

\section{The Mass - Luminosity  Diagram \label{ml}}

Fig.\ref{fig:ml} shows a plot of  \mb\  vs. \mbh\ for a combined sample of $\approx$ 280 AGNs which cover an absolute
B magnitude range $-20 \simgt$ \mb $\simgt -27$\ and an  \mbh\ interval $10^6 \simlt$ \mbh\ $\simlt 10^{10} $
(\cite{marzianietal03b}; cf. Refs. \cite{woourry02a,collinetal02}). The wide majority of sources are located between
$0.02 \simlt \rm L/L_{\rm Edd} \simlt 1.00$. RQ sources  show evidence for significant Malmquist bias while RL sources
show the opposite trend probably related to a bias towards selecting higher luminosity core-dominated sources likely
to be beamed.  We find that the Eddington limit defines an approximate upper boundary to the luminosity distribution
\cite{woourry02a}, indicating that there are no low-$z$\ AGNs radiating significantly above the Eddington limit (this
depends somewhat on the adopted $\rm H_0$\ value). Figure \ref{fig:ml} also suggests that there may be fewer high
(than low) mass  sources with \lm\ close to the Eddington limit. This might be a selection effect or an indication
that galaxies with  large \mbh\ may be unable to supply fuel at high \lm\ i.e., a limit on $\dot{M}$\
\cite{nicastroetal03}.  If a sample comprising more high luminosity objects ($L \simgt 10^{47}$ \ergss) is considered,
41\% of the objects in that sample with $L \simgt 10^{47}$ \ergss\ have super-Eddington ratios \cite{warneretal04}. We
caution however that correction for orientation and beaming should decrease \lledd. The lower envelope of the
luminosity distribution of Fig. \ref{fig:ml}\ may be due to a selection effect \cite{woourry02a}, or it may indicate
that only sources radiating at $0.01 \simlt (L_{\rm bol}/L_{\rm Edd}) \simlt 1.00 $ exhibit a stable BLR. The lower
limit may be connected with the absence or small size of a standard accretion disk  (\S \ref{popc}; see also Refs.
\cite{kollatschnybischoff02,nicastroetal03}). Redshift evolution of the black hole mass-luminosity ratio is found to
be much less than the $(1+z)^3$\ evolution seen in QSO luminosity function evolution \cite{corbettetal03}. This means
that quasars radiate in almost the same \lm\ range also at high-$z$.

\section{The Eddington Ratio as the Driver of the E1 Correlations\label{edd}}

A PCA reveals a first principal component accounting for 48\%\ of the spectrum-to-spectrum variance with just UV
emission line strengths, ratios and widths as input. This linear combination of the input variables depends on \lledd\
\cite{shangetal03,willsshang04}. A reanalysis of PG quasars confirms that E1  is driven predominantly by \lledd\ while
a second principal component is driven by accretion rate \cite{boroson02}. An UV based spectral PCA that uses the
whole spectrum in the rest wavelength interval 1170--2100 \AA\ find consistent results although the relative
importance of luminosity and \lledd\ is reversed i.e., \lledd\ appears in a second eigenvector accounting for 20\%\ of
the spectrum-to-spectrum variance \cite{yipetal04}. The E1 optical sequence and the \lledd\ trends have been recently
confirmed also for intermediate $z$ quasars \cite{netzeretal04,sulenticetal04,yuanwills03}. Comparison between high
and low-$z$ QSOs confirms that the inverse \feiiopt\ -- \oiiiopt\ relationship (i.e., E1) is indeed related to
\lledd\, rather than \mbh\ \cite{yuanwills03}.

Different \lm\ values explain the change in \feii\ and W(\hbbc) prominence  in the optical plane
\cite{marzianietal01,willsetal99,netzeretal04}.  The virial relationship can be rewritten in the form of FWHM(\hbbc)
$\propto$ (\lm)$^{-1}$\mbh$^{-0.4}$  using the \rb -- $\nu L\nu$\ correlation. The diagnostic ratios \siiii / \ciii\
and \aliii / \ciii\ provide an estimate of \ne. Since \ne\ correlates with FWHM(\hbbc)
\cite{willsetal99,marzianietal01} it is possible to write also the ionization parameter (and \rfe) as a function of
\lm\ and \mbh\ \cite{zamanovmarziani02}. The E1 ``elbow" sequence in the optical plane is reproduced by  varying
\lledd, with \lledd\ $\rightarrow 1$\ toward spectral type A3. Orientation can be modelled as a third parameter in
this scheme from the location of CD and LD RL QSOs \cite{sulenticetal03}. Model computations suggest that sources of
spectral type A1 are actually almost face-on, low \lledd\ sources \cite{marzianietal01}. This result explains why some
extremely low W(\hbbc) objects (the second extreme type mentioned in \S \ref{samples}) are observed in bin A1. Sources
of bin B1, A2, A3 are instead rather homogenous in terms of \lledd\ with \mbh\ and $\theta$\ acting as source od
scatter.  It has also been possible to show that the \hbbc\ different profile shapes of Pop. A and B sources are
linked to differences in \lledd\ \cite{marzianietal03b}.

\begin{figure}[htbp]
\centering
\includegraphics[width=0.5\textwidth]{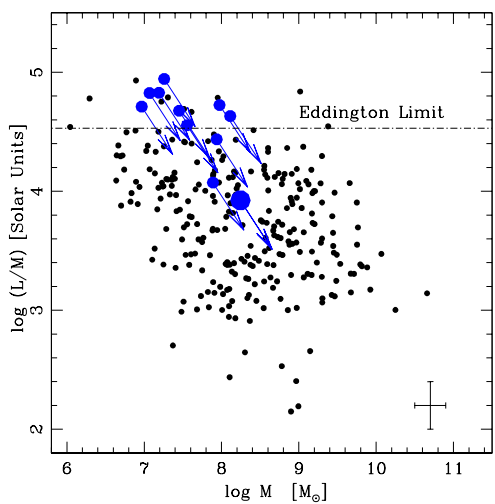}
\caption{Position of blue outliers (large blue spots) relative to other sources in our sample (filled circles). The
largest circle indicates the position of PKS 0736+01, the only known RL blue outlier. The dashed line indicates the
Eddington limit. Arrows indicate displacements of blue outliers if we apply an orientation correction 0.4 to the
derived masses \cite{marzianietal03a}.} \label{fig:blueoutliers}
\end{figure}

The soft X-ray spectral index correlates extremely closely with
Eddington ratio and FWHM(\hbbc)
\cite{grupe04,sulenticetal00a,willsshang04}. W(\civ) decreases
toward extreme population A (see a more detailed discussion in \S
\ref{be}, and Fig. \ref{fig:baldeff}, where a dependence of
W(\civ) on spectral type is clearly shown). Also, the
distributions of large \civbc\ blueshifts and blue outliers seem
to be governed by \lledd\
(\cite{marzianietal03b,baskinlaor04a,zamanovetal02};  Fig.
\ref{fig:blueoutliers}). Among type 1 AGNs with large blueshifted
\oiii, there is no correlation between the Eddington ratio and the
amount of \oiii\ blueshifts. However, the Eddington ratios of the
blue outliers are the highest among AGNs with the same \mbh\ in a
sample of 300 objects \cite{marzianietal03b}. These facts suggest
that high \lledd\ is a necessary condition  for \oiiiopt\ large
blueshifts.

Both NLSy1s galaxies and  BAL QSOs of the PG sample lie at the high \lm\ extreme
\cite{boroson02,marzianietal01,yuanwills03}, although these two  object classes are well separated in a second
eigenvector based on $\dot{M}$ \cite{boroson02}. Other measurements confirm this scenario.  BAL quasars show large
\civ\ {\em emission} blueshifts \cite{reichardetal03b,sulenticetal05}. Using the luminosity and \hbbc\ to derive \mbh\
and \md, both BAL and non-BAL QSOs at $z \sim$ 2 tend to have even higher \lledd\ than those at low $z$\
\cite{yuanwills03}.
Extreme Pop. A sources may represent the most intense accretors
with extreme BAL QSOs being observed at the largest inclination
angles possible for a given wind geometry. A large mass flow\ may
limit the viewing angle range of extreme Pop. A sources to be
$\approx$ 90$^\circ$\ (\S\ \ref{bal} \& Fig. \ref{fig:model}). It
is important to stress a possible difference between the most
extreme BALs (i.e., those with the highest terminal radial
velocity, $\sim$ 30000 \kms) and the generality of BAL sources.
While the most extreme BAL QSOs seem to be associated to highest
\lledd, the presence of a shallow BAL of moderate terminal
velocity is not necessarily a signature of high \lledd\ since the
terminal velocity correlates with \lledd\ if the wind giving rise
to the BALs is radiation driven
\cite{sulenticetal05,reichardetal03a,reichardetal03b}.

The separation between Pop.\ A  and Pop.\ B sources, which may
occur at \lledd\ $\approx$ 0.15 (Fig. \ref{fig:ml}) could be in
part explained by a change in accretion mode. We note that there
is no correlation between radio loudness \rk\ and  \lledd\ in a RL
sample \cite{guetal01b} although RL sources are differentiated in
terms of \lm\ from RQ sources \cite{marzianietal03b}. Most or all
of the weakly AGNs in nearby galaxies are RL, highly sub-Eddington
systems that are plausibly experiencing advection-dominated
accretion \cite{ho02}. The vast majority of these objects appear
to be radiating between 0.01 and 0.1 Eddington luminosity, and may
therefore be akin to Pop.\ B powerful quasars. Radiatively
inefficient accretion  at less than a few per cent of the
Eddington rate is unlikely to produce excess soft X ray emission
\cite{merlonietal03,narayan04}. Lower mass loss rate, and the
harder spectrum (which may decrease radiative acceleration) may
both contribute to the absence of a strong wind in Pop.\ B
sources.

\subsection{Are ``Double Peakers" a Peculiar Population? \label{popc}}

Recent work seems to confirm the accretion disk origin of the LILs for double-peakers. Inter-profile comparison
between Balmer lines and \civ\ (which, in a few cases, show a markedly different, single peaked Gaussian profile
\cite{eracleousetal04}) can be interpreted in the framework of the disk and wind paradigm. More \civ\ -- Balmer line
profile intercomparisons are needed to reach a firm conclusion also because the \civ\ profiles thus far observed
cannot be immediately ascribed to a wind. The LIL emitting portion of the disk to produce double-peaks as separated as
in Arp 102B is relatively modest, from $\sim$ 100 \rg\ to $\sim$ 500 \rg\ \cite{chenhalpern89}. At smaller radii, a
hot solution may be appropriate \cite{narayan04}, while at $\simgt$ 500 \rg\ self gravity may dominate.  Using the
customary empirical relation between the broad-line region size and optical continuum luminosity, \mbh\ and accretion
rate have been computed for 135 AGNs with double-peaked broad emission lines  from the SDSS from a survey of RL AGNs
\cite{wuliu04}. The inferred range of \mbh\ of double-peakers cover almost the full range of \mbh\ of type-1\ AGNs,
3$\cdot 10^{7}$ \msol\ to 5$ \cdot 10^9$ \msol and \lbol\ up to 10$^{46}$ \ergss \cite{wuliu04}. Typical Eddington
ratios are 0.001, much lower than the typical type-1 AGN; about 90\%\ of them show \lledd$\simlt 0.02$ (from the
distribution of \lledd\ shown by \cite{wuliu04}). Double-peaked sources lie close to the lower \lm\ boundary or below
in the mass-luminosity diagram.

The evidence reported does not suggest that double peakers are truly peculiar sources, i.e., an independent Population
C even if they lie out of the mass-luminosity sequence of Fig. \ref{fig:ml}. A two-component model assuming that the
line wings originate in an accretion disk with additional contribution from spherical emission region fits very well
the observed \hbbc\ profiles of Pop. B sources \cite{popovicetal04,stratevaetal03}. An evolutionary scheme from single
peaked Pop.\ B and double peakers is at least conceivable: double peakers show, on average, the lowest \lledd\ (and
probably also the lowest $\dot{M}$). They may be AGNs exhausting their fuel supply, with little gas reservoir, in
which only the innermost part of the accretion disk is left to produce broad emission lines.

\paragraph{``Naked" AGNs} Indirect arguments suggest that some ``true'' type 2
AGNs do exist \cite{laor03,nicastroetal03}.  They are AGNs without
a hidden BLR even in spectropolarimetric data \cite{tran01}. The
observed radius-luminosity relation for the BLR implies an
increasing line width with decreasing luminosity for a given black
\mbh. However, the upper limit to the observed width of broad
emission lines in AGNs at $\sim$ 25000 \kms\
\cite{sulenticetal00a} may reflect a physical limit above which
the BLR may not be able to survive \cite{laor03}. The
disappearance of the BLR has been linked to the critical radius at
which the disk changes from gas pressure dominated to radiation
pressure dominated, or to disk evaporation
\cite{czernyetal04,nicastro00}. For low enough accretion rates,
the critical radius becomes smaller than the innermost stable
orbit  of the accretion disk and the BLR may not form
\cite{nicastroetal03}. The intrinsic (i.e., unabsorbed) X-ray
luminosity and an estimation  of \mbh\ using the relationship
between nuclear mass and bulge luminosity in galaxies indicate
that all hidden BLR sources have accretion rates \md\ $\simgt$
10$^{-3}$. This is also approximately the limit at which
double-peakers are found \cite{wuliu04}. Naked AGNs may this be
observed if \md $\ll 10 ^{-3}$.


\subsection{On the Origin of Radio Loudness}

RL activity in low \mbh\ sources is observed, and there are RL
NLSy1s that radiate al $\log \rm L/M \approx 4.5$
\cite{marzianietal03b}. There are no appreciable effects (within
the limits set by  $S/N$) on the \hbbc\ profile attributable to
radio loudness.  Pop.\ B RL and RQ sources can have the same \mbh\
and \lm\ values. Therefore, {\em it is not unreasonable to
conclude that a similar range of \mbh\ and \lm\ is physically
possible for both RQ and RL sources} \cite{woourry02b}. However,
this does not mean that the mass function and the \lm\
distribution  are necessarily the same for RL and RQ sources. RL
and RQ sources are well separated  in terms of \mbh\ in the E1
space \cite{boroson02} at low $z$.  A robust inference from a
bootstrap analysis  is that the mass function and the conditional
probability of having certain \lm\ values at fixed \mbh\ are
different for the two AGN classes \cite{marzianietal03b}. With the
customary \mbh\ computation assumptions, \mbh\ in RL sources have
been found to be \mbh $\simgt 10^8$ \msol, while only a few
quasars may contain  smaller black holes \cite{guetal01b}. These
conclusions are in agreement with studies based on the SDSS
carried out using a sample of more than 6000 AGNs
\cite{mclurejarvis04}. RL sources were found to harbor
systematically more massive black holes than are the RQ quasars
with very high significance (see also Ref. \cite{dunlopetal03}).
It is important that RQ and RL samples have indistinguishable
distributions on the redshift-optical luminosity plane, excluding
the possibility that either parameter is responsible for the
observed \mbh\ difference \cite{marzianietal03b,mclurejarvis02}.
Quasars from the FIRST Bright Quasar Survey \cite{beckeretal95}
fill in the gap between the RL and RQ quasars in the radio vs.
optical luminosity plane, show a continuous variation of radio
luminosity with \mbh\ and no evidence for a discontinuity  that
may signal the turning on of powerful radio jets
\cite{lacyetal01}. We suggest that several factors drive the RQ/RL
separation in the E1 space: intrinsic mass function and \lm\
distribution differences, jet-related effects yielding \oiiiopt\
enhancement in RL sources, and sample selection criteria
\cite{laor00,boroson02,marzianietal03b}.  The range in radio
luminosity at a given \mbh\ is several orders of magnitude.
Additional  parameters other than \mbh\ and \lm \ like evolution
should be invoked to explain the quasar radio-loudness dichotomy
\cite{mclurejarvis04}.

\section{At the Origin and at the End \label{end}}

\begin{figure}[htbp]
\centering
\includegraphics[width=0.5\textwidth]{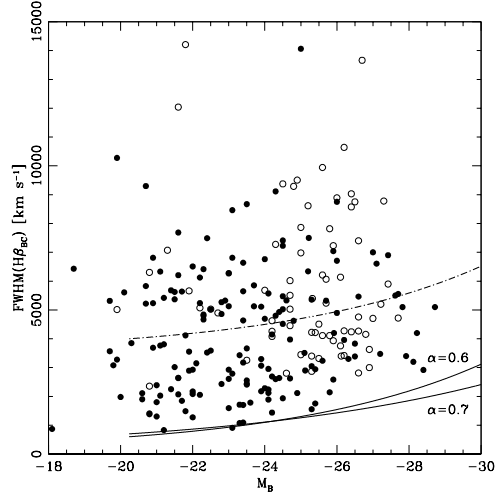}
\caption{FWHM(\hbbc) in \kms\ vs. absolute
magnitude \mb\ for two joined samples, one at low redshift
($z\simlt 0.8$, \cite{marzianietal03a}), and the other for
intermediate redshift quasars \cite{sulenticetal03}. Filled and
open circles indicate RQ and RL sources respectively. The
dependence on luminosity for the minimum FWHM(\hbbc) and for the
boundary between Pop.\ A and B sources is shown. See text for
details.}
\label{fig:hbevol}
\end{figure}

\subsection{Is the Emission Line Width Increasing with
Luminosity? \label{hbevol}}

A trend between  FWHM(\hbbc) and luminosity is  expected if the usual virial and \rb-$\nu L_\nu$\ reasonings are
applied. There is a well-defined lower boundary in the diagram \mbh\ vs. \mb\ (see Fig. \ref{fig:hbevol};
\cite{sulenticetal04}).  Here we make the assumption that low-redshift NLSy1 with the narrowest lines radiate very
close to the Eddington ratio. If we assume $\log $\lm\ $\approx4.5$, we obtain FWHM(\hbbc)$_{\rm min} \approx600$
\kms\ for $\log L = 11$\ in solar units and FWHM$_{\rm min}$(\hbbc)$ \propto 10^{(-0.08 M_{\rm B})}$. If we consider
the luminosity dependence of\linebreak FWHM(\hbbc)$_{\rm min}$, we see that the expected trend for $\alpha = 0.6$\
reproduces fairly well the FWHM(\hbbc) lower boundary as a function of \mb. A less pronounced trend, especially at
high luminosity, is expected for $\alpha=0.7$.  The presence of a  correlation between FWHM(\hbbc) and luminosity in a
survey may depend on sample and intrinsic dispersion of FWHM values in a narrow \mb\ range. One should consider that
{\em the expected luminosity dependence is very weak}: $\Delta$FWHM(\hbbc) increases by $\approx 1000$~\kms\ over an
increase of $\Delta$\mb~$\approx 10$, with FWHM(\hbbc)$_{\rm min}$ changing from 1000~\kms\ to 2000~\kms. In a narrow
\mb\ range, the FWHM(\hbbc) intrinsic scatter spans the range 1000--10000~\kms, therefore making any correlation
intrinsically very weak even if statistically significant.   A very large sample from the 2dF and 6dF redshift surveys
shows a weak increase of \hb\ line width with luminosity with a slope $\approx 0.15$--0.2 ($\Delta v$ vs.\ $\log L$),
very close to the one derived from our calculation \cite{corbettetal03}.

\subsection{Lack of Evolution in the \feii\ Spectrum?}

{\em An analysis of the spectra of 12 high-redshift quasars in the region of \mgii\ shows remarkable similarity with
the LBQS composite ($z \simlt$0.8; \cite{thompsonetal99})}. Near-infrared spectra of four QSOs located at $z> 6 $\
have been recently analyzed \cite{iwamuroetal04}. Two out of  four $z > 6$\ QSOs have significant \feiiuv. Spectra of
three additional sources indicate \feiiuv/\mgii\ similar to that observed in low-$z$ quasars
\cite{freudlingetal03,barthetal03}. The \feiiuv/\mgii\ ratio at high redshift should be a factor of 3 lower than for
low-redshift QSOs if the age of the universe at the earlier epoch is much less than 1 Gyr \cite{hamannferland99}. This
prediction is due to the delayed contribution of type Ia supernov\ae\  to the iron abundance. Apart from the
intriguing results for the highest $z$\ quasars, systematic examinations of the \feiiuv/\mgii\ intensity ratio with
redshift suggest that the ratio may indeed decline at $z \simgt 1.5$ \cite{iwamuroetal02,iwamuroetal04}. LBQS spectra
show a significant Baldwin effect in \feiiuv\ emission. Further analysis reveals that the primary correlation of iron
emission strength is probably with $z$, implying an evolutionary  effect \cite{greenetal01}. A very recent SDSS study
based on $\sim$ 16000 spectra in the reshift range from 0.08 to 5.41 shows that the dominant redshift effect is a
result of the evolution of the blended \feii\ emission (optical) and the Balmer continuum \cite{yipetal04}. This
redshift dependence can be explained by  the evolution of chemical abundance in the quasar environment, or by an
intrinsic change in the continuum itself \cite{yipetal04}. In another study \cite{vernerpeterson04} quasars were
grouped on the basis of their \feiiuv/\mgii\ intensity ratios. The fraction of quasars in each group with strong
\feiiuv/\mgii\ dropped significantly at $z \simgt 1.5$. To check whether there is a change in iron abundance at high
redshifts observations covering \feiiopt\ are essential.

\subsection{A Scheme of Physical Evolution for the AGNs \label{evolution}}

High-$z$\ quasars must have accreted at a relatively high rate in
the early stages of their evolution, if black holes of \mbh$\simgt
10^9$ \msol\ were already radiating at $z \approx 6.4$\
\cite{vestergaard04,warneretal04,willottetal03}.  From the low-$z$
data, it is tempting to understand the Eddington ratio  as the age
of an AGN within its total lifetime (i.e., considering the
totality of the time spent by a black hole as a ``switched on''
AGN): AGNs with steep X-ray spectra, strong \feii, and weak \oiii\
are AGNs in an early phase of their evolution, which proceeds
roughly from the lower left corner (\lledd\ $\rightarrow 1$) of
the optical plane of E1 toward the upper right corner (low \lm,
large \mbh). In this hypothetical scenario local NLSy1s are the
``seedlings'' of AGNs \cite{sulenticetal00a,grupe04}.  Several
results and correlations associated to the optical E1 can be
explained by an evolutionary sequence. RL  seem to be
systematically more massive than RQ sources. They also radiate at
lower \lledd. They show extended NLRs with large W(\oiiiopt). At
the other end we have ``blue outliers" which are Pop. A sources
with strong winds. Young AGNs  still possess a compact NLR; they
may have plenty of fuel and may show spectral evidence of
circum-nuclear Starbursts. More massive, evolved systems may have
less fuel and have had all the time to ionize interstellar clouds
in the bulge of their host galaxies as well as to produce radio
lobes exceeding the host galaxy size. Dying quasars have a limited
gas supply surrounding their continuum-emitting regions, and may
at best produce emission lines through the illumination of a last
``strip" or ring of optically thick material as it may be the case
of double-peakers. In this sense, the \lledd\ sequence of E1
becomes literally an age sequence.

\subsection{The Baldwin Effect \label{be}}

\begin{figure}
\hspace{-0.5cm} \epsfxsize=5cm \epsfysize=5cm \epsfbox{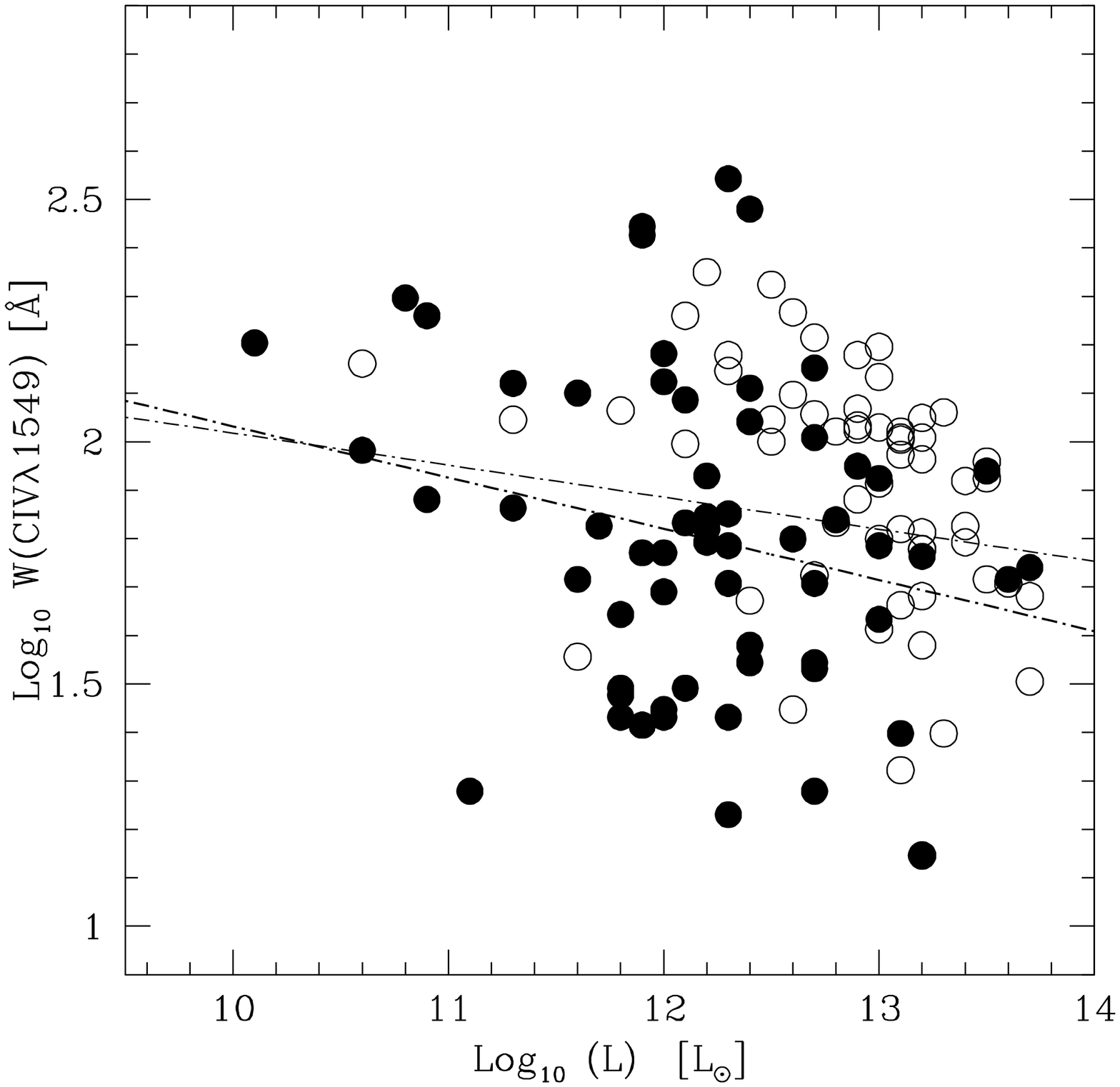}
\epsfxsize=5cm \epsfysize=5cm \epsfbox{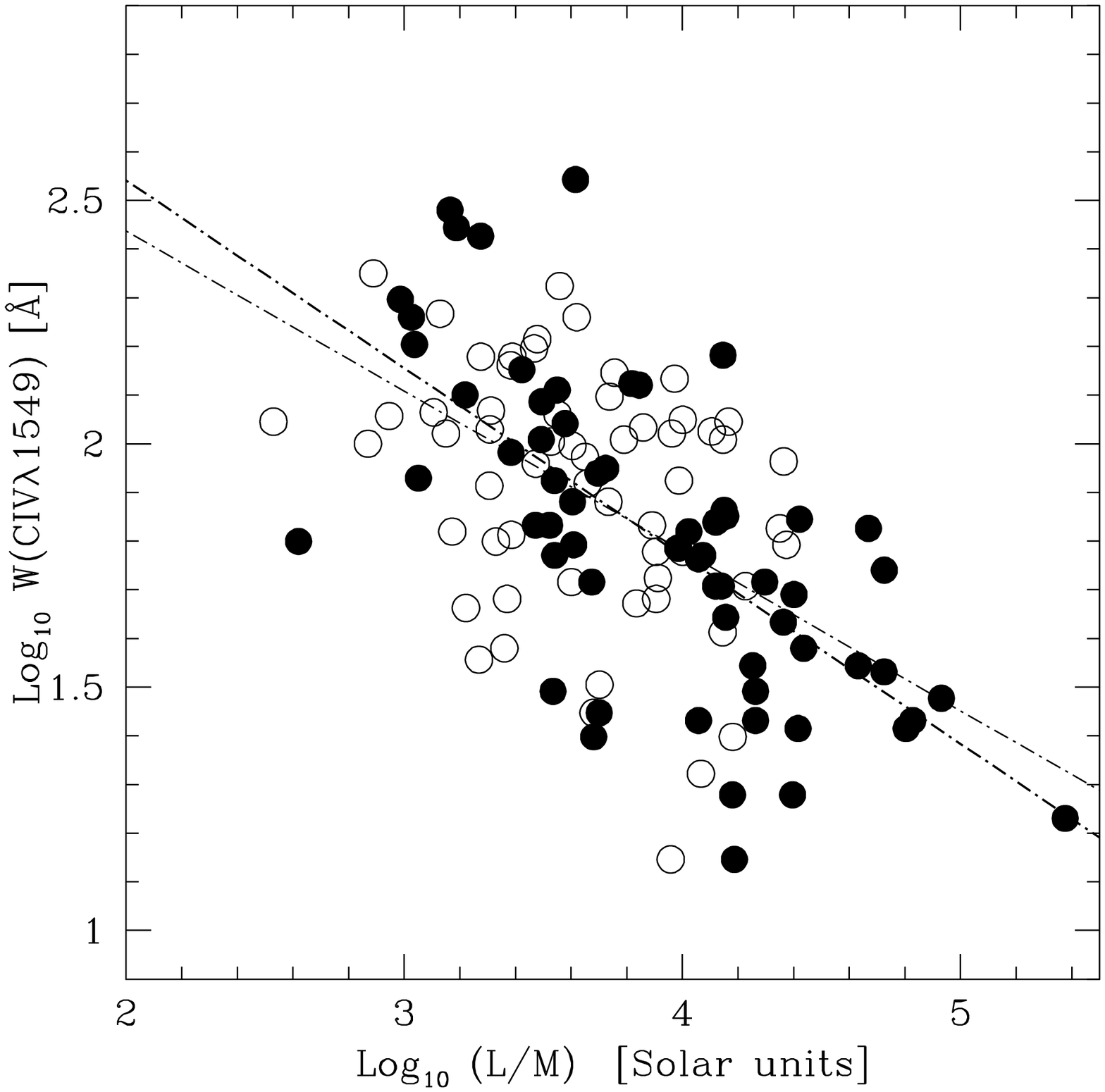}\epsfxsize=5cm
\epsfysize=5cm \epsfbox{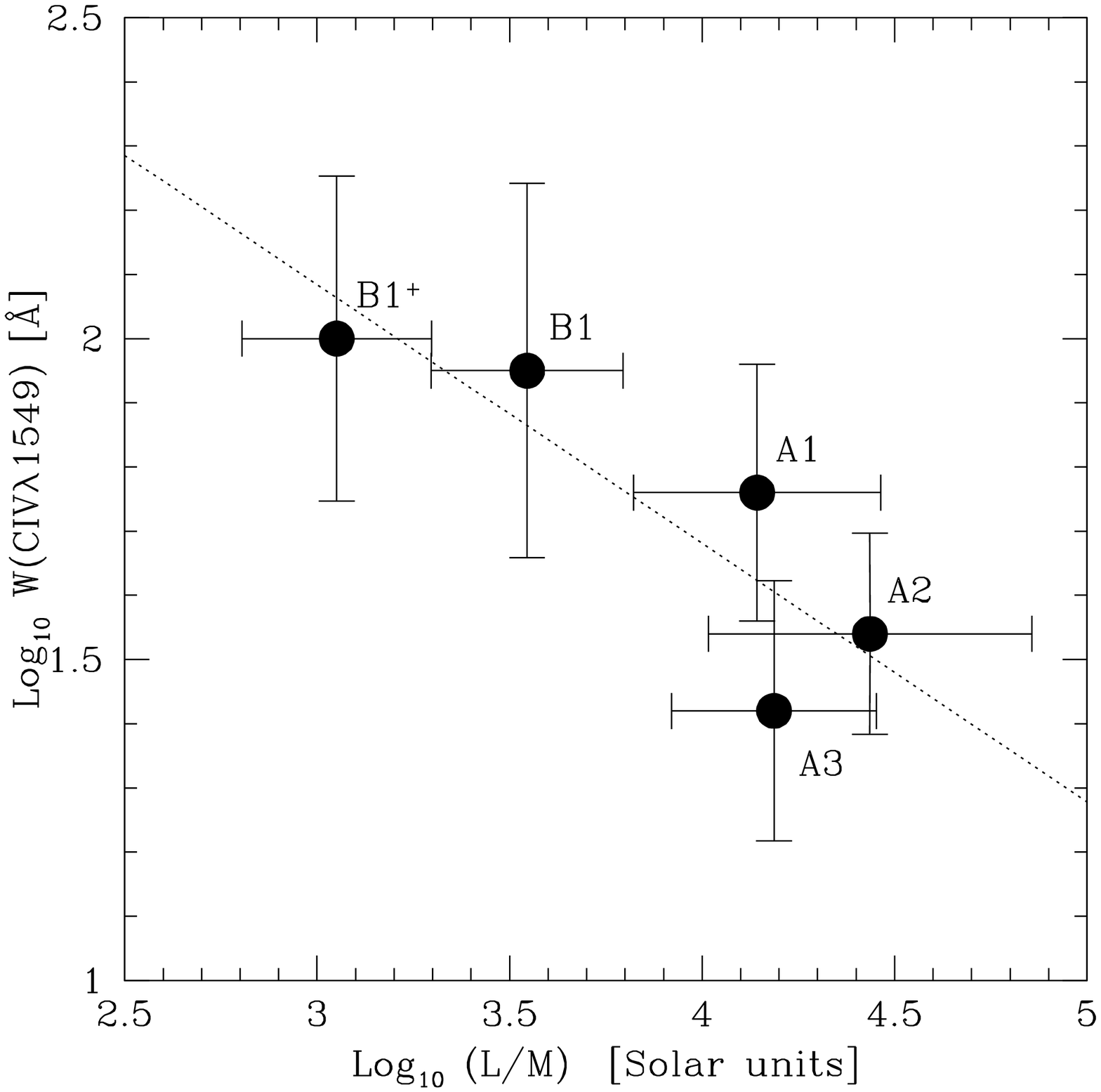} \caption{The Baldwin effects in
the HST based sample of Bachev et al.  \cite{bachevetal04}. Left
panel: dependence of rest-frame W(\civbc) on bolometric
luminosity. Middle panel: dependence of W(civbc) on the $L/M$
ratio (in solar units). Filled circles indicate RQ, open circles
RL sources. Right Panel, same as middle panel but with median
averages with the spectral types defined according to Ref.
\cite{sulenticetal02}.}\label{fig:baldeff}
\end{figure}

It is important to stress that the Baldwin effect is a very loose
correlation between specific luminosity and \civ\ equivalent
width. Claims and counter-claims of a Baldwin effect on the basis
of small samples (few tens of objects) are unreliable
\cite{sulenticetal00a}; the statistical weakness of the Baldwin
correlation implies that the effect becomes appreciable only if a
very large range in luminosity is considered,  4 -- 6 decades. As
pointed out by Sulentic et al. \cite{sulenticetal00a},  results
until mid-1999 have led to a standard scenario in which the
Baldwin effect occurs in all measurable HILs except \nv, and the
slope of the Baldwin relationship (i.e., W(\civ)$\propto
L_\nu^{\rm -a}$, \cite{baldwin77}) increases with ionization
potential (e.g. \cite{osmershields99}; see Ref.
\cite{hamannferland99} for the implications of the apparent lack
of any Baldwin effect for \nv). These results have been basically
confirmed by more recent studies based on large quasar samples
\cite{croometal02,dietrichetal02,richardsetal02}, LBQS
\cite{greenetal01} and  SDSS \cite{yipetal04} included.  A source
of dispersion involves low-luminosity objects that have low
W(\civ) $\sim 10 \div 30$ \AA\ since they tend to blur even more
the Baldwin relationship. These sources are NLSy1s that
preferentially occupy the lower left part of a $\log W$ vs. $\log
L$\ diagram.

Claims of a Baldwin effect on \mgii\ should be taken somewhat with
care \cite{yipetal04}, since the underlying \feiiuv\ emission can
be very strong as well as variable from source to source depending
on the BLR physical conditions, and the \feiiuv\ behavior has a
function of luminosity and redshift is still poorly understood.
Similar claims for W(\hbbc) have been rather erratic and
unconvincing due to small sample sizes and large intrinsic scatter
(Refs. \cite{mcintoshetal99,sulenticetal00a} found no trend; Ref.
\cite{netzeretal04,wilkesetal99} did).

The \oiiiopt\ lines seem to decrease in equivalent width with
increasing luminosity
\cite{dietrichetal02,netzeretal04,sulenticetal04,yipetal04}. In
the joint data of Refs. \cite{sulenticetal04} and
\cite{marzianietal03a} there is  definitely a significant
correlation. Even if a luminosity correlation may not be always
detectable (we stress again that the luminosity range must be very
large) weak or no \oiiiopt\ seems to be a common property of many
luminous AGNs \cite{netzeretal04,yuanwills03}. The reason why
\oiiiopt\ seem to experience a Baldwin effect is presently not
clear. It is possible that the \oiiiopt\ luminosity is limited by
the  physical size of the NLR \cite{netzeretal04}, or that we are
seeing an evolutionary effect at high-$z$, making high-$z$\
sources more similar to low-$z$ Pop. A AGNs.

\subsubsection{The Baldwin Effect as an Evolutionary Effect
\label{beevol}}

A recent analysis of 80 PG quasars and one of $\approx$ 120 HST archive spectra indicates a strong correlation of
W(\civ) with some of the emission parameters which define E1. Since \lledd\ drives the E1 correlations, \lledd\ may be
the primary physical parameter behind the Baldwin effect for \civ\ (\cite{bachevetal04,baskinlaor04a,calvanietal04};
Fig. \ref{fig:baldeff}). Also, high-$z$\ quasars show large \civ\ blueshifts like Pop.\ A sources at low-$z$
\cite{richardsetal02,sulenticetal04,sulenticetal05}.  Near-IR, low resolution spectra of eight of the most distant
quasars  (in the range $4.9 <z < 6.4$)\ show that half of these quasars are characterized by deep, broad and
blueshifted absorption features typical of BAL quasars \cite{maiolinoetal04}. Similarities and close association
between NLSy1, BAL QSOs, strong \feii\ emission, and low W(\civ) led to the suggestions that NLSy1s might be Seyfert
galaxies in their early stage of evolution and as such may be low-redshift, low-luminosity analogues of high-redshift
quasars \cite{mathur00,sulenticetal00a}. The \civ\ profiles show decreasing  equivalent widths with increasing
luminosity and the line profiles become similar to the ``trapezoidal"  shapes often observed in low-$z$ NLSy1s
\cite{warneretal04}. However, a direct comparison of composite spectra of NLSy1 and  $z \sim $ 4 quasars through a PCA
suggests that high-$z$ quasi-stellar objects do not show a strong preference toward NLSy1 behavior
\cite{constantinshields03}. Since this result is sample dependent and the mass-luminosity  diagram of quasars can be
similar up to $z \sim 2.5$\ \cite{corbettetal03}, it is still debatable whether evolution of \lledd\ with $z$,
selection effects or a combination of both may lead to the Baldwin effect \cite{sulenticetal00a}. After all, we are
not observing other evolutionary effects even at $z \approx 6$. What seems unlikely is that the driving parameter of
the Baldwin effect is \mbh\ \cite{warneretal03}. Computing \mbh\ employing FWHM(\civ) and assuming virial motions does
not demonstrate that \mbh\ is driving the Baldwin effect because it is assumed to be \mbh $\propto L^{0.7}$, and this
assumption leads, in part, to circular reasoning.

\subsection{Quasars as Standard Candles}

The Supernova Cosmology Project results supporting a cosmological
constant $\Lambda \neq$ 0\ are still highly  uncertain. The
expected effect is just a few tens of magnitude at $z\approx 0.6$,
and goes down to 0 at $z\approx1.5$. A robust verification of a
cosmological model with relative energy density  $\rm \Omega_{\rm
M} \approx 0.28$ and $\Omega_\Lambda \approx$ 0.72 would require
accurate measurements of standard candles in the range $0.6 \simlt
z \simlt 3$, as it can be seen from the Hubble diagram with type
Ia supernov\ae\ \cite{riessetal01,riessetal04}.

It is  clear from the previous discussion that, {\em  at high $z$, we are observing quasars that can be very similar
to the AGNs we are observing at low $z$, in terms of line width, \feii\ prominence, HIL equivalent widths.} Trends
with \lm\ are much better defined than trends with luminosity. Luminosity effects remain weak and prone to sample bias
(see, for example, Fig. \ref{fig:baldeff}, where the Baldwin effect is not significant while W(\civ) shows a
significant correlation with \lledd). Furthermore, \lledd\ seems to affect many parameters related to physics and
structure of the BLR, from LIL widths to HIL shifts and ionization degree. It makes therefore sense to use E1
correlations to estimate \lm. If  \mbh\ is known with reasonable accuracy, it may become possible to retrieve
$z$-independent information on \lbol. In principle, this can be achieved through two methods:
\begin{itemize}
\item a multivariate analysis which isolates the best linear
combination of variables which are dependent only on \lm\ and
\mbh\ \cite{willsshang04};

\item an extension of the E1 diagram to high luminosity and high
$z$ \cite{marzianietal03d}.
\end{itemize}

Instruments like the infrared spectrometers mounted on 10-m sized telescopes make possible observations of
intermediate redshift quasars (1 $\simlt z\simlt$ 2.2) with $S/N$\ and resolution sufficient to apply the same data
analysis procedure used for optical spectra of sources with $z\simlt 1$. If these data are available, it is possible
to measure the E1 parameters and to build an extension of ``optical" E1 diagram. To fully exploit the E1 diagram,
orientation needs to be estimated on an object-by-object basis. In the case of a radiation pressure driven wind, the
\civ\ shift and profile shape should be sensitive to both \lm\ and $\theta$. Therefore, \civ\ measurements may provide
an independent estimate of orientation, and hence a 3D observational parameter space to map into a 3D physical space
($\theta$, \lm, \mbh) which will ultimately yield a redshift-independent estimate of the luminosity (if an accurate
low-$z$\ calibration can be defined). Merging results on \mbh\ and on \lbol\ from the previous analysis, it is
reasonable to infer that, if $\theta$ is known within $\pm 5^\circ$, errors could be $\sim$ 60\%\ on \mbh, 80\% on
\lm. If errors are distributed randomly, an uncertainty as high as $\Delta m \approx 1$\ could be farther reduced if
samples of sources within a spectral type are averaged, yielding valuable data points for the Hubble diagram.

\section{Conclusion}

The main purpose of this paper was to point out how it is becoming possible to measure accretion  parameters
(especially \mbh\ and \lm) of AGNs.  We highlighted major results and sources of uncertainty, as well as the
connections between \mbh\ and \lledd, the structure of the broad line emitting region, and physical evolution. If we
compare our present knowledge to the one of about 10 years ago, we can see an impressive progress. \mbh\
determinations through the virial assumption have become widely popular, and have been applied to quasars at all known
$z$, up to 6.4, even if several sources of uncertainty limit strongly the accuracy by which \mbh\ is known for
individual objects. It has been possible to show that many line properties and trends (including the Baldwin effect)
correlate with \lledd, and the the BLR structure is strongly affected by \lledd\ through E1. There is no doubt that we
have a much better framework with a strong observational support to understand the BLR structure, minority objects
like ``double-peakers" and BAL QSOs, and source evolution. Improvements on \mbh\ and \lm\ estimates are possible
through new multivariate analysis including orientation sensitive parameters, or through disk and wind BLR models
accounting for the combined effects on  line parameters of orientation and physics. We may be able to achieve a
satisfactory vision of quasar structural evolution (not to mention boundaries on cosmography) well before the emitting
regions may be resolved through space interferometry missions (e.g., Ref. \cite{unwinetal02}).

\bigskip

{\small We thank Dr. Massimo Calvani for a careful reading of the manuscript.}



\begin{thebibliography}{999}

\bibitem{antonucci93} Antonucci, R.\ 1993, ARAAp, 31, 473
\bibitem{aokietal04} Aoki, K., Kawaguchi, T., \& Ohta, K.\ 2004, ArXiv Astrophysics e-prints, astro-ph/0409546
\bibitem{bachev99} Bachev, R.\ 1999, \textit{AAp}, \textbf{348}, 71
\bibitem{bachevetal04} Bachev, R., Marziani, P., Sulentic, J.~W., Zamanov, R., Calvani, M., \& Dultzin-Hacyan, D.\ 2004, \textit{ApJ}, \textbf{617}, 171
\bibitem{baldwin77} Baldwin, J.~A.\ 1977, \textit{ApJ}, \textbf{214}, 679
\bibitem{baldwinetal04} Baldwin, J.~A., Ferland, G.~J., Korista, K.~T., Hamann, F., \& LaCluyz{\' e}, A.\ 2004, \textit{ApJ}, \textbf{615}, 610
\bibitem{baldwinetal95} Baldwin, J., Ferland, G., Korista, K., \& Verner, D.\ 1995, \textit{ApJL}, \textbf{455}, L119
\bibitem{bardeenpetterson75}  Bardeen, J.M., \& Petterson, J.A., 1975, \textit{ApJL}, \textbf{195}, 65
\bibitem{barthetal03} Barth, A.~J., Martini, P., Nelson, C.~H., \& Ho, L.~C.\ 2003, \textit{ApJl}, \textbf{594}, L95
\bibitem{barvainisetal04} Barvainis, R., Lehar, J., Birkinshaw, M., Falke, H., \& Blundell, K.~M.\ 2004, ArXiv Astrophysics e-prints, astro-ph/0409554
\bibitem{baskinlaor04a} Baskin, A.~\& Laor, A.\ 2004, \textit{MNRAS}, \textbf{350}, L31
\bibitem{baskinlaor04b} Baskin, A.~\& Laor, A.\ 2004, ArXiv Astrophysics e-prints, astro-ph/0409196
\bibitem{beckeretal95} Becker, R.~H., White, R.~L., \& Helfand, D.~J.\ 1995, \textit{ApJ}, \textbf{450}, 559
\bibitem{bianzhao02} Bian, W., \& Zhao, Y.\ 2002, \textit{AAp}, \textbf{395}, 465
\bibitem{bicknell02} Bicknell, G.~V.\ 2002, \textit{New Astronomy Review}, \textbf{46}, 365
\bibitem{blandford90} Blandford, R.~D., Netzer, H., Woltjer, L., Courvoisier, T.~J.-L., \& Mayor, M.\ 1990, \textit{Saas-Fee Advanced Course} \textbf{20}.~Lecture Notes 1990.~Swiss Society for Astrophysics and Astronomy, XII, 280 pp.~97 figs..~ Springer-Verlag Berlin Heidelberg New York,
\bibitem{blandfordznajek77} Blandford, R.~D., \& Znajek, R.~L.\ 1977, \textit{MNRAS}, \textbf{179}, 433
\bibitem{bolleretal96}  Boller, Th., Brandt, W.N., Fink, H.H., 1996, \textit{AAp}, \textbf{305}, 53 (BBF96)
\bibitem{bolleretal97}  Boller, Th., Brandt, W.N., Fabian, A.C. \& Fink, H.H., 1997, \textit{MNRAS}, \textbf{289},393-405
\bibitem{bolleretal02} Boller, T., Gallo, L.~C., Lutz, D., \& Sturm, E.\ 2002, \textit{MNRAS}, \textbf{336}, 1143
\bibitem{boroson02} Boroson, T.~A.\ 2002, \textit{ApJ}, \textbf{565}, 78
\bibitem{borosongreen92} Boroson T.A., Green R.F., 1992, \textit{ApJS} \textbf{80}, 190
\bibitem{borosonmeyers92} Boroson, T.~A.~\& Meyers, K.~A.\ 1992, \textit{ApJ}, \textbf{397}, 442
\bibitem{borosonoke87} Boroson, T.~A., \& Oke, J.~B.\ 1987, \textit{PASP}, \textbf{99}, 809
\bibitem{bottorffetal97} Bottorff, M., Korista, K.~T., Shlosman, I., \& Blandford, R.~D.\ 1997, \textit{ApJ}, \textbf{479}, 200
\bibitem{braitoetal04} Braito, V., et al.\ 2004, \textit{AAp}, \textbf{420}, 79
\bibitem{breeveldetal01} Breeveld, A.~A., Puchnarewicz, E.~M., \& Otani, C.\ 2001, \textit{MNRAS}, \textbf{325}, 772
\bibitem{brinkmannetal04} Brinkmann, W., Papadakis, I. E., Ferrero, E., 2004, \textit{AAp}, \textbf{414},107
\bibitem{brotherton96} Brotherton, M.~S.\ 1996, \textit{ApJS}, \textbf{102}, 1
\bibitem{brothertonetal94b} Brotherton, M.~S., Wills, B.~J., Steidel, C.~C., \& Sargent, W.~L.~W.\ 1994, \textit{ApJ},\linebreak \textbf{423}, 131
\bibitem{brothertonetal94a} Brotherton, M.~S., Wills, B.~J., Francis, P.~J., \& Steidel, C.~C.\ 1994, \textit{ApJ}, \textbf{430}, 495
\bibitem{calvanietal04} Calvani, M., Marziani, P., Bachev, R., Sulentic, J.~W., Zamanov, R.~K., \& Dultzin-Hacyan, D.\ 2004, \textit{Memorie della Societa Astronomica Italiana Supplement}, \textbf{5}, 223
\bibitem{calvanietal97} Calvani, M., Marziani, P., \& Sulentic, J.\ 1997, \textit{Memorie della Societa Astronomica Italiana}, \textbf{68}, 93
\bibitem{camenzind04} Camenzind, M.\ 2004, ArXiv Astrophysics e-prints, astro-ph/0411573
\bibitem{camenzindandcrockenberger92}  Camenzind, M., \& Krockenberger, M. A. 1992, \textit{AAp}, \textbf{255}, 59
\bibitem{capettietal96} Capetti, A., Axon, D.~J., Macchetto, F., Sparks, W.~B., \& Boksenberg, A.\ 1996, \textit{ApJ}, \textbf{469}, 554
\bibitem{caproniabraham04}  Caproni A., Abraham Z., 2004, \textit{ApJ}, \textbf{602}, 625
\bibitem{chenhalpern89} Chen, K., \& Halpern, J.~P.\ 1989, \textit{ApJ}, \textbf{344}, 115
\bibitem{chiangmurray96} Chiang, J., \& Murray, N.\ 1996, \textit{ApJ}, \textbf{466}, 704
\bibitem{collieretal01} Collier, S., et al.\ 2001, \textit{ApJ}, \textbf{561}, 146
\bibitem{collinetal02} Collin, S., Boisson, C., Mouchet, M., Dumont, A.-M., Coup{\' e}, S., Porquet, D., \& Rokaki, E.\ 2002, \textit{AAp}, \textbf{388}, 771
\bibitem{collinhure01}  Collin, S., \& Huré, J.-M. 2001, \textit{AAp}, \textbf{372}, 50
\bibitem{collinjoly00} Collin, S., \& Joly, M.\ 2000, \textit{New Astronomy Review}, 44, 531
\bibitem{collinsouffrinetal88} Collin-Souffrin, S., Dyson, J.~E., McDowell, J.~C., \& Perry, J.~J.\ 1988, \textit{MNRAS},\linebreak \textbf{232}, 539
\bibitem{comastrietal92}  Comastri , A., Setti, G., Zamorani, G., Elvis, M., Giommi, P., Wilkes, B.J., \& McDowell. J.C. 1992, \textit{ApJ}, \textbf{384}, 62
\bibitem{constantinshields03} Constantin, A., \& Shields, J.~C.\ 2003, \textit{PASP}, \textbf{115}, 592
\bibitem{conwayetal95}  Conway, J. E., \& Wrobel, J. M. 1995, \textit{ApJ}, \textbf{439}, 98
\bibitem{corbettetal03} Corbett, E.~A., et al.\ 2003, \textit{MNRAS}, \textbf{343}, 705
\bibitem{corbin97} Corbin, M.~R.\ 1997, \textit{ApJ}, \textbf{485}, 517
\bibitem{corbinboroson96} Corbin, M.~R., \& Boroson, T.~A.\ 1996, \textit{ApJS}, \textbf{107}, 69
\bibitem{crenshawetal96}  Crenshaw, D.M., et al. 1996, \textit{ApJ}, \textbf{470}, 322
\bibitem{crenshawetal02} Crenshaw, D.~M., et al.\ 2002, \textit{ApJ}, \textbf{566}, 187
\bibitem{crescietal04} Cresci, G., Maiolino, R., Marconi, A., Mannucci, F., \& Granato, G.~L.\ 2004, \textit{AAp}, \textbf{423}, L13
\bibitem{croometal02} Croom, S.~M., et al.\ 2002, \textit{MNRAS}, \textbf{337}, 275
\bibitem{czernyetal01} Czerny, B., Niko{\l}ajuk, M., Piasecki, M., \& Kuraszkiewicz, J.\ 2001, \textit{MNRAS},\linebreak \textbf{325}, 865
\bibitem{czernyetal04} Czerny, B., R{\' o}za{\' n}ska, A., \& Kuraszkiewicz, J.\ 2004, \textit{AAp}, \textbf{428}, 39
\bibitem{daviesetal04} Davies, R.~I., Tacconi, L.~J., \& Genzel, R.\ 2004, \textit{ÀpJ}, \textbf{613}, 781
\bibitem{dietrichetal03} Dietrich, M., Hamann, F., Appenzeller, I., \& Vestergaard, M.\ 2003, \textit{ApJ}, \textbf{596}, 817
\bibitem{dietrichetal02} Dietrich, M., Hamann, F., Shields, J.~C., Constantin, A., Vestergaard, M., Chaffee, F., Foltz, C.~B., \& Junkkarinen, V.~T.\ 2002, \textit{ApJ}, \textbf{581}, 912
\bibitem{dietrichhamann04} Dietrich, M., \& Hamann, F.\ 2004, \textit{ApJ}, \textbf{611}, 761
\bibitem{dongetal04} Dong, X., Zhou, H., Wang, T., Wang, J., Li, C., \& Zhou, Y.\ 2004, ArXiv Astrophysics e-prints, astro-ph/0411171
\bibitem{dultzinhacyanetal00} Dultzin-Hacyan, D., Marziani, P., \& Sulentic, J.~W.\ 2000, R\textit{evista Mexicana de Astronomia y Astrofisica Conference Series}, \textbf{9}, 308
\bibitem{dunlopetal03} Dunlop, J.~S., McLure, R.~J., Kukula, M.~J., Baum, S.~A., O'Dea, C.~P., \& Hughes, D.~H.\ 2003, \textit{MNRAS}, \textbf{340}, 1095
\bibitem{elvis00} Elvis, M.\ 2000, \textit{ApJ}, \textbf{545}, 63
\bibitem{elvisetal94} Elvis, M., et al.\ 1994, \textit{ApJS}, \textbf{95}, 1
\bibitem{eracleousetal97}  Eracleous, M., et al. 1997, \textit{ApJ}, \textbf{490}, 216.
\bibitem{eracleoushalpern03} Eracleous, M., \& Halpern, J.~P.\ 2003, \textit{ApJ}, \textbf{599}, 886
\bibitem{eracleousetal04} Eracleous, M., Halpern, J.~P., Storchi-Bergmann, T., Filippenko, A.~V., Wilson, A.~S., \& Livio, M.\ 2004, ArXiv Astrophysics e-prints, astro-ph/0404506
\bibitem{fabian79} Fabian, A.~C.\ 1979, R\textit{oyal Society of London Proceedings Series A}, \textbf{366}, 449
\bibitem{falckeetal98} Falcke, H., Wilson, A.~S., \& Simpson, C.\ 1998, \textit{ApJ}, \textbf{502}, 199
\bibitem{ferland00} Ferland, G.~J.\ 2000, \textit{Revista Mexicana de Astronomia y Astrofisica Conference Series}, \textbf{9}, 153
\bibitem{ferraresemerritt00} Ferrarese, L., \& Merritt, D.\ 2000, \textit{ApJL}, \textbf{539}, L9
\bibitem{floydetal04} Floyd, D.~J.~E., Kukula, M.~J., Dunlop, J.~S., McLure, R.~J., Miller, L., Percival, W.~J., Baum, S.~A., \& O'Dea, C.~P.\ 2004, \textit{MNRAS}, \textbf{355}, 196
\bibitem{forsterhalpern96} Forster, K., \& Halpern, J.~P.\ 1996, \textit{ApJ}, \textbf{468}, 565
\bibitem{freemanetal01}  Freeman, P. E.,  Doe, S., Siemiginowska, A. 2001, SPIE Proceedings vol.4477, 76
\bibitem{francisetal01} Francis, P.~J., Drake, C.~L., Whiting, M.~T., Drinkwater, M.~J., \& Webster, R.~L.\ 2001, \textit{Publications of the Astronomical Society of Australia}, \textbf{18}, 221
\bibitem{francisetal91} Francis, P.~J., Hewett, P.~C., Foltz, C.~B., Chaffee, F.~H., Weymann, R.~J., \& Morris, S.~L.\ 1991, \textit{ApJ}, \textbf{373}, 465
\bibitem{freudlingetal03} Freudling, W., Corbin, M.~R., \& Korista, K.~T.\ 2003, \textit{ApJL}, \textbf{587}, L67
\bibitem{galloetal04} Gallo, L.C., Boller, Th., Tanaka, Y., Fabian, A.C., Brandt, W.N., Welsh, W.F., Anabuki, N. \& Haba, Y., 2004, \textit{MNRAS}, \textbf{347}, 269-276
\bibitem{gammieetal04} Gammie, C.~F., Shapiro, S.~L., \& McKinney, J.~C.\ 2004, \textit{ApJ}, \textbf{602}, 312
\bibitem{garciabarretoetal95} Garcia-Barreto, J.~A., Franco, J., Guichard, J., \& Carrillo, R.\ 1995, \textit{ApJ}, \textbf{451}, 156
\bibitem{gaskell96}  Gaskell, C. M. 1996, \textit{ApJ}, \textbf{464}, L107
\bibitem{gaskelletal04} Gaskell, C.~M., Goosmann, R.~W., Antonucci, R.~R.~J., \& Whysong, D.~H.\ 2004, \textit{ApJ}, \textbf{616}, 147
\bibitem{ghisellinietal93} Ghisellini, G., Padovani, P., Celotti, A., \& Maraschi, L.\ 1993, \textit{ApJ}, \textbf{407}, 65
\bibitem{gierlinskidone04}  Gierlinski \& Done, C., 2004, \textit{MNRAS}, \textbf{349}, L7
\bibitem{giveonetal99} Giveon, U., Maoz, D., Kaspi, S., Netzer, H., \& Smith, P.~S.\ 1999, \textit{MNRAS}, \textbf{306}, 637
\bibitem{greenetal01} Green, P.~J., Forster, K., \& Kuraszkiewicz, J.\ 2001, \textit{ApJ}, \textbf{556}, 727
\bibitem{grupe04} Grupe, D.\ 2004, \textit{AJ}, \textbf{127}, 1799
\bibitem{grupeetal99} Grupe, D., Beuermann, K., Mannheim, K., \& Thomas, H.-C.\ 1999, \textit{AAp}, \textbf{350}, 805
\bibitem{grupeetal98} Grupe, D., Beuermann, K., Thomas, H.-C., Mannheim, K., \& Fink, H.~H.\ 1998, \textit{AAp}, \textbf{330}, 25
\bibitem{grupeetal01} Grupe, D., Thomas, H.-C., \& Beuermann, K.\ 2001, \textit{AAp}, \textbf{367}, 470
\bibitem{guetal01b} Gu, M., Cao, X., \& Jiang, D.~R.\ 2001, \textit{MNRAS}, \textbf{327}, 1111
\bibitem{guetal01a} Gu, Q., Dultzin-Hacyan, D., \& de Diego, J.~A.\ 2001, R\textit{evista Mexicana de Astronomia y Astrofisica}, \textbf{37}, 3
\bibitem{haimanetal04} Haiman, Z., Quataert, E., \& Bower, G.~C.\ 2004, \textit{ApJ}, \textbf{612}, 698
\bibitem{halpernetal96} Halpern, J.~P., Eracleous, M., Filippenko, A.~V., \& Chen, K.\ 1996, \textit{ApJ}, \textbf{464}, 704
\bibitem{hamannferland99} Hamann, F., \& Ferland, G.\ 1999, \textit{ARAAp}, \textbf{37}, 487
\bibitem{hartigbaldwin86} Hartig, G.~F.~\& Baldwin, J.~A. 1986, \textit{ApJ}, \textbf{302}, 64
\bibitem{hesetal93} Hes, R., Barthel, P.~D., \& Fosbury, R.~A.~E.\ 1993, \textit{Nature}, \textbf{362}, 326
\bibitem{ho02} Ho, L.~C.\ 2002, \textit{ApJ}, \textbf{564}, 120
\bibitem{horneetal04} Horne, K., Peterson, B.~M., Collier, S.~J., \& Netzer, H.\ 2004, \textit{PASP}, \textbf{116}, 465
\bibitem{hughesblandford03} Hughes, S.~A.~\& Blandford, R.~D.\ 2003, \textit{ApJl}, \textbf{585}, L101
\bibitem{imanishiwada04} Imanishi, M., \& Wada, K.\ 2004, \textit{ApJ}, \textbf{617}, 214
\bibitem{iwamuroetal04} Iwamuro, F., Kimura, M., Eto, S., Maihara, T., Motohara, K., Yoshii, Y., \& Doi, M.\ 2004, \textit{ApJ}, \textbf{614}, 69
\bibitem{iwamuroetal02} Iwamuro, F., Motohara, K., Maihara, T., Kimura, M., Yoshii, Y., \& Doi, M.\ 2002, \textit{ApJ}, \textbf{565}, 63
\bibitem{jarvismaclure02} Jarvis, M.~J., \& McLure, R.~J.\ 2002, \textit{MNRAS}, \textbf{336}, L38
\bibitem{joly91} Joly, M.\ 1991, \textit{AAp}, \textbf{242}, 49 
\bibitem{joly87} Joly, M.\ 1987, \textit{AAp}, \textbf{184}, 33
\bibitem{kaspietal00} Kaspi, S., Smith, P.~S., Netzer, H., Maoz, D., Jannuzi, B.~T., \& Giveon, U.\ 2000, \textit{ApJ}, \textbf{533}, 631
\bibitem{kidgeretal92} Kidger, M., Takalo, L., Sillanp\aa\aa, A.: 1992, \textit{AAp}, \textbf{264}, 32
\bibitem{klimeketal04} Klimek, E.S., Gaskell, C.M. \& Hedrick, C.H., 2004, \textit{ApJ}, \textbf{609}, 69-79
\bibitem{kollatschny03b} Kollatschny, W.\ 2003, \textit{AAp}, \textbf{412}, L61
\bibitem{kollatschny03a} Kollatschny, W.\ 2003, \textit{AAp}, \textbf{407}, 461
\bibitem{kollatschnybischoff02} Kollatschny, W., \& Bischoff, K.\ 2002, \textit{AAp}, \textbf{386}, L19
\bibitem{kollatschnyetal01} Kollatschny, W., Bischoff, K., Robinson, E.L., Welsh, W.F. \& Hill, G.J., 2001, \textit{AAp}, \textbf{379}, 125-135
\bibitem{koristaetal97} Korista, K., Baldwin, J., Ferland, G., \& Verner, D.\ 1997, \textit{ApJS}, \textbf{108}, 401
\bibitem{koristagoad04} Korista, K.~T., \& Goad, M.~R.\ 2004, \textit{ApJ}, \textbf{606}, 749
\bibitem{koristaetal93} Korista, K.~T., Voit, G.~M., Morris, S.~L., \& Weymann, R.~J.\ 1993, \textit{ApJS}, \textbf{88}, 357
\bibitem{krichbaumetal04} Krichbaum, T.~P., et al.\ 2004, ArXiv Astrophysics e-prints, astro-ph/0411487
\bibitem{krongoldetal01} Krongold, Y., Dultzin-Hacyan, D., \& Marziani, P.\ 2001, \textit{AJ}, \textbf{121}, 702
\bibitem{krongoldetal03} Krongold, Y., Nicastro, F., Brickhouse, N.~S., Elvis, M., Liedahl, D.~A., \& Mathur, S.\ 2003, \textit{ApJ}, \textbf{597}, 832
\bibitem{kuraszkiewiczetal04} Kuraszkiewicz, J.~K., Green, P.~J., Crenshaw, D.~M., Dunn, J., Forster, K., Vestergaard, M., \& Aldcroft, T.~L.\ 2004, \textit{ApJS}, \textbf{150}, 165
\bibitem{kuraszkiewiczetal02} Kuraszkiewicz, J.~K., Green, P.~J., Forster, K., Aldcroft, T.~L., Evans, I.~N., \& Koratkar, A.\ 2002, \textit{ApJS}, \textbf{143}, 257
\bibitem{krolik01} Krolik, J.~H.\ 2001, \textit{ApJ}, \textbf{551}, 72
\bibitem{lacyetal01} Lacy, M., Laurent-Muehleisen, S.~A., Ridgway, S.~E., Becker, R.~H., \& White, R.~L.\ 2001, \textit{ApJl}, \textbf{551}, L17
\bibitem{laor00} Laor, A.\ 2000, \textit{ApJL}, \textbf{543}, L111
\bibitem{laor01} Laor, A.\ 2001, \textit{ApJ}, \textbf{553}, 677
\bibitem{laor03} Laor, A.\ 2003, \textit{ApJ}, \textbf{590}, 86
\bibitem{lehtovaltonen96} Lehto, H. J., Valtonen, M. J., 1996, \textit{ApJ} \textbf{460}, 207
\bibitem{leighly99} Leighly, K.~M.\ 1999, \textit{ApJS}, \textbf{125}, 317
\bibitem{leighly04} Leighly, K.~M.\ 2004, \textit{ApJ}, \textbf{611}, 125
\bibitem{leighlymoore04} Leighly, K.~M., \& Moore, J.~R.\ 2004, \textit{ApJ}, \textbf{611}, 107
\bibitem{lewisetal04} Lewis, K.~T., Eracleous, M., Halpern, J.~P., \& Storchi-Bergmann, T.\ 2004, ArXiv Astrophysics e-prints, astro-ph/0404342
\bibitem{liparietal93} Lipari, S., Terlevich, R., \& Macchetto, F.\ 1993, \textit{ApJ}, \textbf{406}, 451
\bibitem{lobanovroland04} Lobanov, A.~P., \& Roland, J.\ 2004, ArXiv Astrophysics e-prints, astro-ph/0411417
\bibitem{maiolinoetal04} Maiolino, R., Oliva, E., Ghinassi, F., Pedani, M., Mannucci, F., Mujica, R., \& Juarez, Y.\ 2004, \textit{AAp}, \textbf{420}, 889
\bibitem{marzianisulentic93} Marziani, P., \& Sulentic, J.~W.\ 1993, \textit{ApJ}, \textbf{409}, 612
\bibitem{marzianietal93} Marziani, P., Sulentic, J.~W., Calvani, M., Perez, E., Moles, M., \& Penston, M.~V.\ 1993, \textit{ApJ}, \textbf{410}, 56
\bibitem{marzianietal96} Marziani P., Sulentic J.W., Dultzin-Hacyan D., Calvani M., Moles M., 1996, \textit{ApJS} \textbf{104}, 37
\bibitem{marzianietal01} Marziani P., Sulentic J.W., Zwitter T., Dultzin-Hacyan D., Calvani M., 2001, \textit{ApJ}, \textbf{558}, 553 (M01)
\bibitem{marzianietal03a} Marziani P., Sulentic J. W., Zamanov R., Calvani M., Dultzin-Hacyan D., Bachev R., Zwitter T., 2003a, \textit{ApJS}, \textbf{145}, 199 (M03)
\bibitem{marzianietal03d} Marziani, P., Sulentic, J.~W., Zamanov, R., Calvani, M., Della Valle, M., Stirpe, G., \& Dultzin-Hacyan, D.\ 2003, \textit{Memorie della Societa Astronomica Italiana Supplement}, \textbf{3}, 218
\bibitem{marzianietal03c} Marziani, P., Zamanov, R., Sulentic, J.~W., Dultzin-Hacyan, D., Bongardo, C., \& Calvani, M.\ 2003b, \textit{ASP Conf.~Ser.~290: Active Galactic Nuclei: From Central Engine to Host Galaxy}, \textbf{229}
\bibitem{marzianietal03b} Marziani P., Zamanov R., Sulentic J. W., Calvani M., 2003c, \textit{MNRAS}, \textbf{345}, 1133 S.\ 2000, \textit{MNRAS}, \textbf{314}, L17
\bibitem{mathur00} Mathur, S.\ 2000, \textit{MNRAS}, \textbf{314}, L17
\bibitem{mathuretal01} Mathur, S., Kuraszkiewicz, J., \& Czerny, B.\ 2001, \textit{New Astronomy},\textbf{ 6}, 321
\bibitem{mcintoshetal99} McIntosh, D.~H., Rieke, M.~J., Rix, H.-W., Foltz, C.~B., \& Weymann, R.~J.\ 1999, \textit{ApJ}, \textbf{514}, 40
\bibitem{mcluredunlop01} McLure, R.~J., \& Dunlop, J.~S.\ 2001, \textit{MNRAS}, \textbf{327}, 199
\bibitem{mcluredunlop02} McLure, R.~J., \& Dunlop, J.~S.\ 2002, \textit{MNRAS}, \textbf{331}, 795
\bibitem{mcluredunlop04} McLure, R.~J., \& Dunlop, J.~S.\ 2004, \textit{MNRAS}, \textbf{352}, 1390
\bibitem{mclurejarvis02} McLure, R.~J., \& Jarvis, M.~J.\ 2002, \textit{MNRAS}, \textbf{337}, 109
\bibitem{mclurejarvis04} McLure, R.~J., \& Jarvis, M.~J.\ 2004, \textit{MNRAS}, \textbf{353}, L45
\bibitem{merlonietal03} Merloni, A., Heinz, S., \& di Matteo, T.\ 2003, MNRAS, \textbf{345}, 1057
\bibitem{milleretal00} Miller, H.R. Ferrara, E.C., McFarland, J.P., Wilson, J.W., Daya, A.B., Fried, R.E., 2000, \textit{New Astron. Rev.}, \textbf{44}, 539
\bibitem{moranetal96} Moran, E.~C., Halpern, J.~P., \& Helfand, D.~J.\ 1996, \textit{ApJS}, \textbf{106}, 341
\bibitem{murraychiang97} Murray, N., \& Chiang, J.\ 1997, \textit{ApJ}, \textbf{474}, 91
\bibitem{murtaghheck87} Murtagh, F.~\& Heck, A.\ 1987,\textit{ Astrophysics and Space Science Library, Dordrecht: Reidel}, 1987
\bibitem{mushotzkietal93} Mushotzki, R.F., Done, C., Pounds, K.A., 1993, \textit{ARAAp}, \textbf{31}, 717
\bibitem{nagaoetal01} Nagao, T., Murayama, T. \& Taniguchi, Y., 2001, \textit{ApJ}, \textbf{546}, 744-758
\bibitem{narayan04} Narayan, R.\ 2004, ArXiv Astrophysics e-prints, astro-ph/0411385
\bibitem{nelsonetal04} Nelson, C.~H., Green, R.~F., Bower, G., Gebhardt, K., \& Weistrop, D.\ 2004, \textit{ApJ}, \textbf{615}, 652
\bibitem{nelson00} Nelson, C.~H.\ 2000, \textit{ApJl}, \textbf{544}, L91
\bibitem{netzeretal04} Netzer, H., Shemmer, O., Maiolino, R., Oliva, E., Croom, S., Corbett, E., \& di Fabrizio, L.\ 2004, \textit{ApJ}, \textbf{614}, 558
\bibitem{newmanetal97} Newman, J.~A., Eracleous, M., Filippenko, A.~V., \& Halpern, J.~P.\ 1997, \textit{ApJ},\linebreak \textbf{485}, 570
\bibitem{nicastro00}     Nicastro, F.\ 2000, \textit{ApJL}, \textbf{530}, L65
\bibitem{nicastroetal03} Nicastro, F., Martocchia, A., \& Matt, G.\ 2003, \textit{ApJL}, \textbf{589}, L13
\bibitem{onkenetal04} Onken, C.~A., Ferrarese, L., Merritt, D., Peterson, B.~M., Pogge, R.~W., Vestergaard, M., \& Wandel, A.\ 2004, \textit{ApJ}, \textbf{615}, 645
\bibitem{oshlacketal02} Oshlack, A.~Y.~K.~N., Webster, R.~L., \& Whiting, M.~T.\ 2002, \textit{ApJ}, \textbf{576}, 81
\bibitem{osmershields99} Osmer, P.~S., \& Shields, J.~C.\ 1999, \textit{Astronomical Society of the Pacific Conference Series}, \textbf{162}, 235
\bibitem{netzer03} Netzer, H.\ 2003, \textit{ApJL}, \textbf{583}, L5
\bibitem{netzer90} Netzer, H. 1990, in Blandford, R.~D., Netzer, H., Woltjer, L., Courvoisier, T.~J.-L., \& Mayor, M.\ 1990, \textit{Saas-Fee Advanced Course} \textbf{20}.~Lecture Notes 1990.~\textit{Swiss Society for Astrophysics and Astronomy, XII}, \textbf{280} pp.~97 Springer-Verlag Berlin Heidelberg New York,
\bibitem{ostoreroetal04} Ostorero, L.; Villata, M.; Raiteri, C. M. 2004, \textit{AAp}, \textbf{419}, 913.
\bibitem{osterbrockpogge85} Osterbrock, D.E., Pogge, R.W., 1985, \textit{ApJ}, \textbf{297}, 166
\bibitem{padovanietal04} Padovani, P., Allen, M.~G., Rosati, P., \& Walton, N.~A.\ 2004, \textit{AAp}, \textbf{424}, 545
\bibitem{pageetal04} Page, K.~L., Reeves, J.~N., O'Brien, P.~T., Turner, M.~J.~L., \& Worrall, D.~M.\ 2004, \textit{MNRAS}, \textbf{353}, 133
\bibitem{pauletal05}  Paul C.,  Brotherton M.S.,  Diamond-Stanic A.),  Vanden Berk D., Canalizo G. 2005, BAAS, in press
\bibitem{peterson93} Peterson, B.~M.\ 1993, \textit{PASP}, \textbf{105}, 247
\bibitem{peterson97} Peterson, B.~M.\ 1997, An introduction to active galactic nuclei,  Cambridge, New York Cambridge University Press, 1997
\bibitem{peterson04} Peterson, B.~M.\ 2004, ArXiv Astrophysics e-prints, astro-ph/0404539
\bibitem{petersonetal04} Peterson, B.~M., et al.\ 2004, \textit{ApJ}, \textbf{613}, 682
\bibitem{petersonetal00} Peterson, B.M., et al. 2000, \textit{ApJ}, \textbf{542}, 161
\bibitem{petersonetal02} Peterson, B.~M., et al.\ 2002, \textit{ApJ}, \textbf{581}, 197 
\bibitem{petersonhorne04} Peterson, B.~M.~\& Horne, K.\ 2004, ArXiv Astrophysics e-prints, astro-ph/0407538
\bibitem{petersonetal89} Peterson, B.M., Korista, K.T., \& Wagner, R.M., 1989, \textit{AJ}, \textbf{98}, 100
\bibitem{petersonetal98} Peterson, B.~M., Wanders, I., Horne, K., Collier, S., Alexander, T., Kaspi, S., \& Maoz, D.\ 1998, \textit{PASP}, \textbf{110}, 660
\bibitem{picaetal88} Pica, A.J., Smith, A.G., Webb, J.R., Leacock, R.J., Clements, S., \& Gombola, P.P. 1988, \textit{AJ}, \textbf{96}, 1215
\bibitem{piconcellietal04} Piconcelli, E., Jiménez-Bailon, E., Guainazzi, M., et al., 2004, \textit{MNRAS}, \textbf{351}, 161
\bibitem{popovicetal04} Popovi{\' c}, L.~{\v C}., Mediavilla, E., Bon, E., \& Ili{\' c}, D.\ 2004, \textit{AAp}, \textbf{423}, 909
\bibitem{popovic03} Popovi{\' c}, L.~{\v C}.\ 2003, \textit{ApJ}, \textbf{599}, 140
\bibitem{porquetetal04} Porquet, D., Reeves, J.~N., O'Brien, P., \& Brinkmann, W.\ 2004, \textit{AAp}, \textbf{422}, 85
\bibitem{progaetal00} Proga, D., Stone, J.~M., \& Kallman, T.~R.\ 2000, \textit{ApJ}, \textbf{543}, 686
\bibitem{poundsetal03a} Pounds, K.~A., King, A.~R., Page, K.~L., \& O'Brien, P.~T.\ 2003, \textit{MNRAS}, \textbf{346}, 1025
\bibitem{poundsetal03b} Pounds, K.~A., Reeves, J.~N., King, A.~R., Page, K.~L., O'Brien, P.~T., \& Turner, M.~J.~L.\ 2003, \textit{MNRAS}, \textbf{345}, 705
\bibitem{poundsetal03c} Pounds, K.~A., Reeves, J.~N., Page, K.~L., Edelson, R., Matt, G., \& Perola, G.~C.\ 2003, \textit{MNRAS}, \textbf{341}, 953
\bibitem{pursimoetal00} Pursimo, T., Takalo, L.O., Sillanp\aa\aa, A., Kidger, M., Lethto, H.J., Heidt, J., Charles, P.A., Aller, H., Aller, M., Beckmann, V., and 61 coauthors, 2000, \textit{AApS}, \textbf{146}, 141.
\bibitem{reevesturner00} Reeves, J. N., \& Turner, M. J. L., 2000, \textit{MNRAS}, \textbf{316}, 234
\bibitem{reichardetal03a} Reichard, T.~A., et al.\ 2003, \textit{AJ}, \textbf{125},1711
\bibitem{reichardetal03b} Reichard, T.~A., et al.\ 2003, \textit{AJ}, \textbf{126}, 2594
\bibitem{richardsetal02} Richards, G.~T., Vanden Berk, D.~E., Reichard, T.~A., Hall, P.~B., Schneider, D.~P., SubbaRao, M., Thakar, A.~R., \& York, D.~G.\ 2002, \textit{AJ}, \textbf{124}, 1
\bibitem{richardsetal03} Richards, G.~T., et al.\ 2003, \textit{AJ}, \textbf{126}, 1131
\bibitem{riessetal01} Riess, A.~G.~et al.\ 2001, \textit{ApJ}, \textbf{560}, 49
\bibitem{riessetal04} Riess, A.~G., et al.\ 2004, \textit{ApJ}, \textbf{607}, 665
\bibitem{rodriguezardilaviegas03} Rodr{\'{\i}}guez-Ardila, A., \& Viegas, S.~M.\ 2003, \textit{MNRAS}, \textbf{340}, L33
\bibitem{rodriguezpascualetal97} Rodr\'\i guez-Pascual, P.M., Mas-Hesse, J.M., Santos-Lleo, M., 1997, \textit{AAp}, \textbf{327}, 72
\bibitem{rokakietal03} Rokaki, E., Lawrence, A., Economou, F., \& Mastichiadis, A.\ 2003, \textit{MNRAS},\linebreak \textbf{340}, 1298
\bibitem{romanoetal02} Romano, P., Turner, T.~J., Mathur, S., \& George, I.~M.\ 2002, \textit{ApJ}, \textbf{564}, 162
\bibitem{rossfabian93} Ross, R. R., \& Fabian, A. C. 1993, \textit{MNRAS}, \textbf{261}, 74
\bibitem{saslawetal94} Saslaw, W. C., Waltonen, M. J., \& Aarseth, S. J. 1974, \textit{ApJ}, \textbf{190}, 253
\bibitem{schneideretal02} Schneider, D.~P., et al.\ 2002, \textit{AJ}, \textbf{123}, 567
\bibitem{scottetal04} Scott, J.~E., Kriss, G.~A., Brotherton, M., Green, R.~F., Hutchings, J., Shull, J.~M., \& Zheng, W.\ 2004, \textit{ApJ}, \textbf{615}, 135
\bibitem{sdss} http://www.sdss.org
\bibitem{shakurasunyaev73} Shakura, N.~I., \& Sunyaev, R.~A.\ 1973, \textit{AAp}, \textbf{24}, 337
\bibitem{shangetal04} Shang, Z., et al.\ 2004, ArXiv Astrophysics e-prints, astro-ph/0409697
\bibitem{shangetal03} Shang, Z., Wills, B.~J., Robinson, E.~L., Wills, D., Laor, A., Xie, B., \& Yuan, J.\ 2003, \textit{ApJ}, \textbf{586}, 52
\bibitem{shieldsetal95} Shields, J.~C., Ferland, G.~J., \& Peterson, B.~M.\ 1995, \textit{ApJ}, \textbf{441}, 507
\bibitem{stepanianetal03} Stepanian, J.~A., et al.\ 2003, \textit{ApJ}, \textbf{588}, 746
\bibitem{stratevaetal03} Strateva, I.~V., et al.\ 2003, \textit{AJ}, \textbf{126}, 1720
\bibitem{shapovalovaetal01} Shapovalova, A.I., Burenkov, A.N., Carrasco, L., Chavushyan, V.H., Doroshenko, V.T., Dumont, A.M., Lyuty, V.M., Valdés, J.R., Vlasuyk, V.V., Bochkarev, N.G., 2001, \textit{AAp}, \textbf{376}, 775.
\bibitem{shapovalovaetal04} Shapovalova, A.~I., et al.\ 2004, \textit{AAp}, \textbf{422}, 925
\bibitem{shields96} Shields, G.~A.\ 1996, \textit{ApJL}, \textbf{461}, L9
\bibitem{shieldsetal03} Shields, G.~A., Gebhardt, K., Salviander, S., Wills, B.~J., Xie, B., Brotherton, M.~S., Yuan, J., \& Dietrich, M.\ 2003, \textit{ApJ}, \textbf{583}, 124
\bibitem{sigutetal04} Sigut, T.~A.~A., Pradhan, A.~K., \& Nahar, S.~N.\ 2004, \textit{ApJ}, \textbf{611}, 81
\bibitem{sillanpaaetal96} Sillanp\aa\aa, A., et al. 1996, A \& A, in press
\bibitem{sillanpaaetal88} Sillanp\aa\aa, A., Haarala S., Valtonen M. J., Sundelius B., Byrd G. G., 1988, \textit{ApJ},\linebreak \textbf{325}, 628.
\bibitem{storchibergmannetal03} Storchi-Bergmann, T., et al.\ 2003, \textit{ApJ}, \textbf{598}, 956
\bibitem{sulenticetal05} Sulentic, J.~W., Dultzin-Hacyan, D., Marziani, P., C. Bongardo, V. Braito, M. Calvani, Zamanov, R., 2005, RevMexAAp, submitted
\bibitem{sulenticmarziani99} Sulentic, J.~W.~\& Marziani, P.\ 1999, \textit{ApJL}, \textbf{518}, L9
\bibitem{sulenticetal98} Sulentic, J.~W., Marziani, P., \& Calvani, M.\ 1998, \textit{ApJL}, \textbf{497}, L65
\bibitem{sulenticetal95} Sulentic, J.~W., Marziani, P., Zwitter, T., \& Calvani, M.\ 1995, \textit{ApJL}, \textbf{438}, L1
\bibitem{sulenticetal00a} Sulentic, J.~W., Marziani, P., \& Dultzin-Hacyan, D.\ 2000a, \textit{ARA\&A}, \textbf{38}, 521
\bibitem{sulenticetal02} Sulentic, J.~W., Marziani, P., Zamanov, R., Bachev, R., Calvani, M., \& Dultzin-Hacyan, D.\ 2002, \textit{ApJ}, \textbf{566}, L71
\bibitem{sulenticetal00b} Sulentic, J.~W., Marziani, P., Zwitter, T., Dultzin-Hacyan, D., \& Calvani, M.\ 2000b, \textit{ApJ}, \textbf{545}, L15
\bibitem{sulenticetal00c} Sulentic, J.~W., Marziani, P., Zwitter, T., Dultzin-Hacyan, D., \& Calvani, M.\ 2000, \textit{ApJL}, \textbf{545}, L15
\bibitem{sulenticetal04} Sulentic, J.~W., Stirpe, G.~M., Marziani, P., Zamanov, R., Calvani, M., \& Braito, V.\ 2004, ArXiv Astrophysics e-prints, astro-ph/0405279
\bibitem{sulenticetal03} Sulentic J. W., Zamfir S., Marziani P., Bachev R., Calvani M., Dultzin-Hacyan D., 2003, \textit{ApJ}, \textbf{597}, L17
\bibitem{sulenticetal90} Sulentic, J.~W., Zheng, W., Calvani, M., \& Marziani, P.\ 1990, \textit{ApJL}, \textbf{355}, L15
\bibitem{terasrantavaltaoja95} Teräsranta, H., Valtaoja, E.: 1995, \textit{Private Communication}
\bibitem{thompsonetal99} Thompson K. L., Hill G. J., Elston R. 1999. \textit{ApJ}, \textbf{515}, 487
\bibitem{thorne74} Thorne, K.~S.\ 1974, \textit{ApJ}, \textbf{191},507
\bibitem{toleaetal02} Tolea, A., Krolik, J.~H., \& Tsvetanov, Z.\ 2002, \textit{ApJL}, \textbf{578}, L31
\bibitem{tran01} Tran, H.~D.\ 2001, \textit{ApJL}, \textbf{554}, L19
\bibitem{turneretal99} Turner, T.J., George, I.M., Nandra, K., \& Turcan, D., 1999, \textit{ApJ}, \textbf{524}, 667
\bibitem{turnshek84} Turnshek, D.~A.\ 1984, \textit{ApJ}, \textbf{280}, 51
\bibitem{turnsheketal97} Turnshek, D.~A., Monier, E.~M., Sirola, C.~J., \& Espey, B.~R.\ 1997, \textit{ApJ}, \textbf{476}, 40
\bibitem{ulrichetal97} Ulrich, M.-H., Maraschi, L., \& Urry, C.M., 1997, \textit{ARAAp}, \textbf{35}, 445
\bibitem{unwinetal02} Unwin, S.~C., Wehrle, A.~E., Jones, D.~L., Meier, D.~L., \& Piner, B.~G.\ 2002, \textit{Publications of the Astronomical Society of Australia}, \textbf{19}, 5
\bibitem{valtonenetal99} Valtonen, M. J.; Lehto, H. J.; Pietilä, H., 1999, \textit{AAp}, \textbf{342}, 29.
\bibitem{vandenberketal01} Vanden Berk, D.~E., et al.\ 2001, \textit{AJ}, \textbf{122}, 549
\bibitem{vaughanetal99} Vaughan, S., Reeves,J., Warwick, R., \& Edelson, R.\ 1999, \textit{MNRAS}, \textbf{309}, 113
\bibitem{verneretal03} Verner, E., Bruhweiler, F., Verner, D., Johansson, S., \& Gull, T.\ 2003, \textit{ApJL},\linebreak \textbf{592}, L59
\bibitem{verneretal04} Verner, E., Bruhweiler, F., Verner, D., Johansson, S., Kallman, T., \& Gull, T.\ 2004, \textit{ApJ}, \textbf{611}, 780
\bibitem{vernerpeterson04} Verner, E.~M., \& Peterson, B.~A.\ 2004, \textit{ApJL}, \textbf{608}, L85
\bibitem{verneretal99} Verner, E.~M., Verner, D.~A., Korista, K.~T., Ferguson, J.~W., Hamann, F., \& Ferland, G.~J.\ 1999, \textit{ApJS}, \textbf{120}, 101
\bibitem{veroncettyveron03} V{\'e}ron-Cetty, M.-P. ~ \& V{\' e}ron, P.\ 2003, \textit{AAp}, \textbf{412}, 399
\bibitem{veroncettyetal01} V{\' e}ron-Cetty, M.-P., V{\'e}ron, P., \& Gon{\c c}alves, A.~C.\ 2001, \textit{AAp}, \textbf{372}, 730
\bibitem{veroncettyetal04} V{\' e}ron-Cetty, M.-P., Joly, M., \& V{\' e}ron, P.\ 2004, \textit{AAp} \textbf{417}, 515
\bibitem{vestergaard02} Vestergaard, M.\ 2002, \textit{ApJ}, \textbf{571}, 733
\bibitem{vestergaard04} Vestergaard, M.\ 2004, \textit{ApJ}, \textbf{601}, 676
\bibitem{vestergaardwilkes01} Vestergaard, M., \& Wilkes, B.~J.\ 2001, \textit{ApJS}, \textbf{134}, 1
\bibitem{vicenteetal95} Vicente, L, et al., 1995, in "Extragalactic Radio Sources", IAU symposium 175, Bologna.
\bibitem{villataetal99} Villata, M., \& Raiteri, C. M. 1999, \textit{AAp}, \textbf{347}, 30
\bibitem{wampler86} Wampler, E.~J.\ 1986, \textit{AAp}, \textbf{161}, 223
\bibitem{wandeletal99} Wandel, A., Peterson, B.~M., \& Malkan, M.~A.\ 1999, \textit{ApJ}, \textbf{526}, 579
\bibitem{wangetal96} Wang, T., Brinkmann, W., Bergeron, J., 1996, \textit{AAp}, \textbf{309}, 81
\bibitem{wangzhang03} Wang, T., \& Zhang, X.\ 2003, \textit{MNRAS}, \textbf{340}, 793
\bibitem{warneretal03} Warner, C., Hamann, F., \& Dietrich, M.\ 2003, \textit{ApJ}, \textbf{596}, 72
\bibitem{warneretal04} Warner, C., Hamann, F., \& Dietrich, M.\ 2004, \textit{ApJ}, \textbf{608}, 136
\bibitem{webbetal88} Webb, J.R., Smith, A.G., Leacock, R.J., Fitzgibbons, G.L., Gombola, P.P., \& Shepherd, D.W. 1988, \textit{AJ}, \textbf{95}, 374
\bibitem{webbetal00} Webb, W. \& Malkan, M., 2000, \textit{ApJ}, \textbf{540}, 652-677
\bibitem{weymannetal91} Weymann R.J., Morris S.L., Foltz C.B., Hewett P.C., 1991,      \textit{ApJ} \textbf{373}, 23
\bibitem{whangzhang03} Wang, T., \& Zhang, X.\ 2003, \textit{MNRAS}, \textbf{340}, 793
\bibitem{whittle92} Whittle, M.\ 1992, \textit{ApJS}, \textbf{79}, 49
\bibitem{wilkeselvis87} Wilkes, B. J., \& Elvis, M. 1987, \textit{ApJ}, \textbf{323}, 243
\bibitem{wilkesetal99} Wilkes, B.~J., Kuraszkiewicz, J., Green, P.~J., Mathur, S., \& McDowell, J.~C.\ 1999, \textit{ApJ}, \textbf{513}, 76
\bibitem{willottetal03} Willott, C.~J., McLure, R.~J., \& Jarvis, M.~J.\ 2003, \textit{ApJL}, \textbf{587}, L15
\bibitem{willsbrowne86} Wills, B.~J., \& Browne, I.~W.~A.\ 1986, \textit{ApJ}, \textbf{302}, 56
\bibitem{willsetal85} Wills, B.~J., Netzer, H., \& Wills, D.\ 1985, \textit{ApJ}, \textbf{288}, 94
\bibitem{willsetal93} Wills, B.~J., Brotherton, M.~S., Fang, D., Steidel, C.~C., \& Sargent, W.~L.~W.\ 1993, \textit{ApJ}, \textbf{415}, 563
\bibitem{willsetal99} Wills, B.~J., Brandt, W.~N., \& Laor, A.\ 1999, \textit{ApJL}, \textbf{520}, L91
\bibitem{willsshang04} Wills, B.~J., \& Shang, Z.\ 2004, \textit{Advances in Space Research}, \textbf{34}, 2584
\bibitem{wilsoncolbert95} Wilson, A.~S., \& Colbert, E.~J.~M.\ 1995, \textit{ApJ}, \textbf{438}, 62
\bibitem{woourry02a} Woo, J., \& Urry, C.~M.\ 2002, \textit{ApJ}, \textbf{579}, 530
\bibitem{woourry02b} Woo, J., \& Urry, C.~M.\ 2002, \textit{ApJ}, \textbf{581}, L5
\bibitem{wuliu04} Wu, X., \& Liu, F.~K.\ 2004, \textit{ApJ}, \textbf{614}, 91
\bibitem{wuetal04} Wu, X.-B., Wang, R., Kong, M.~Z., Liu, F.~K., \& Han, J.~L.\ 2004, \textit{AAp}, \textbf{424}, 793
\bibitem{yipetal04} Yip, C.~W., et al.\ 2004, \textit{AJ}, \textbf{128}, 2603
\bibitem{youngetal99}  Young, A.J., Crawford, C.S., Fabian, A.C., Brandt, W.N., O'Brien, P.T., 1999, \textit{MNRAS}, \textbf{304}, 4
\bibitem{yuanwills03} Yuan, M.~J.~\& Wills, B.~J.\ 2003, \textit{ApJL}, \textbf{593}, L11
\bibitem{zamanovetal02} Zamanov, R., Marziani, P., Sulentic, J.~W., Calvani, M., Dultzin-Hacyan, D., \& Bachev, R.\ 2002, \textit{ApJL}, \textbf{576}, L9
\bibitem{zamanovmarziani02} Zamanov, R.~\& Marziani, P.\ 2002, \textit{ApJL}, \textbf{571}, L77
\bibitem{zhengetal97} Zheng, W., Kriss, G.~A., Telfer, R.~C., Grimes, J.~P., \& Davidsen, A.~F.\ 1997, \textit{ApJ}, \textbf{475}, 469
\bibitem{zhouetal03} Zhou, H., Wang, T., Dong, X., Zhou, Y., \& Li, C.\ 2003, \textit{ApJ}, \textbf{584}, 147
\bibitem{zhouetal04} Zhou, H., Wang, T., Dong, X., Li, C., \& Zhang, X.\ 2004, ArXiv Astrophysics e-prints, astro-ph/0411252


\end{thebibliography}
\end{document}